# Physics of Spin Glass Freezing and Paired Cluster Model of High-$T_c$ Superconductivity

J. K. Srivastava*

*Tata Institute of Fundamental Research, Mumbai-400005, India*

(11 August 2005; revised (v3) 25 November 2005)

We present here a new mechanism of high-$T_c$ (critical temperature) superconductivity. This new model is able to explain almost all the HTSC (high-$T_c$ superconductivity) properties, like high $T_c$, origin and nature of NSPG (normal state pseudogap), anomalous superconducting state (SS) energy gap (SSEG) properties, absence of NMR spin relaxation rate coherence peak, existence of NSPG below $T_c$, nature of high-$T_c$ magnetic superconductivity, external magnetic field (H) dependence of the NSPG formation temperature $T^*$, etc. The coexistence of NSPG and SS BCS energy gap (EG), below $T_c$, has been predicted by the model before its experimental observation, showing that the observed SSEG is a superposition of NSPG and SS BCS EG. The model, called the paired cluster (PC) model, is also able to explain the pseudogap critical concentration, vortex core pseudogap and possible stripe phase in high-$T_c$ superconductors. We do not find any experimental observation which is not in favour of the model. Their results follow as a natural consequence of the physical picture contained in the model. Several such results are discussed.

## A. INTRODUCTORY REMARKS

This article (extended paper) presents *a new mechanism of high-$T_c$ superconductivity*. The mechanism is based on the physics of spin glass freezing and the magnetically frustrated nature of high-$T_c$ (cuprate) superconductors. Consequently in the earlier parts of the article the concept of spin glass freezing is discussed and in the later parts the new mechanism of high-$T_c$ superconductivity (HTSC) is presented in detail alongwith its application to different HTSC phenomena and comparison with other existing HTSC mechanisms bringing out the improved nature of the present mechanism. Accordingly this article is divided into four parts. The part I introduces the physics of spin glass freezing, part II describes the cluster phase transition (CPT) model (proposed by us) of spin glass systems and its extension to HTSC which results in a new HTSC mechanism, called paired cluster (PC) model by us. This part also discusses the PC model explanation of the pseudogap, the behaviour of superconducting state energy gap and several other HTSC properties. Part III deals with the specific topics of pseudogap critical concentration, vortex core pseudogap, coexistence of pseudogap and superconducting sate energy gap below $T_c$ and stripe phase. The parts II and III use a three dimensional (3D) electronic density of states (DOS) for pseudogap explanation. The part IV examines the validity of the PC model pseudogap and related predictions for a two dimensional (2D) system where the electronic DOS has a van Hove singularity in its DOS vs. $E_{el}$ (electron energy) curve. This part assumes importance since several cuprates have more 2D nature than 3D nature. These parts are presented in the same sequence in which the model has been developed. This has been done purposely to give the reder a feeling about the way the model has been developed. Thus some concepts of Part I may be more refined in Part II and so on. It is therefore necessary to read the entire article before arriving at any particular conclusion. Finally, the overall conclusion of the present study is that the PC model is capable of explaining the HTSC properties. The faith in the model grows as one observes that several predictions of the model made much earlier, like the coexistence of pseudogap and superconducting state energy gap below $T_c$, are being found true by the experiments now (Appendix 1). Other remarkable successes of the model too are described at different places in this article. Part V discusses the PC model explanation of high-$T_c$ magnetic superconductivity.

## B. DETAILS

## I. INTRODUCTION TO SPIN GLASS MAGNETISM

# I.1. INTRODUCTION

Whereas in a normal magnetically ordered system (ferro-, ferri-, antiferro- magnet), spins are aligned in a parallel (up, up or down, down) manner (ferromagnetic alignment, Fig. 1(a)) or antiparallel (up, down) manner (antiferro-, ferri-magnetic alignment, Fig. 1(b)), in a spin glass (SG) system there is no well defined direction (up or down) for the spins on the global scale (*i.e.* when seen for the entire lattice). Instead the spins are frozen (pointing) in random directions (Fig.1(c)). Thus though, unlike paramagnetic spins, each spin has a well defined direction, there is no net spin direction for the entire lattice (sublattice) as such. Such a random spin freezing occurs in the lattice due to the presence of 'frustration' (*i.e.* simultaneous presence of competing exchange interactions, *e.g.* both the positive and negative exchange interactions (exchange integral, $J_{ij}$, + and −), at a given spin site) and 'disorder' (*i.e.* random location of magnetic (diamagnetic) ions in the lattice). The net magnitude and sign of frustration at a spin site (defined more quantitatively later on) decides the direction of that particular spin and the disorder causes global spin direction randomness.

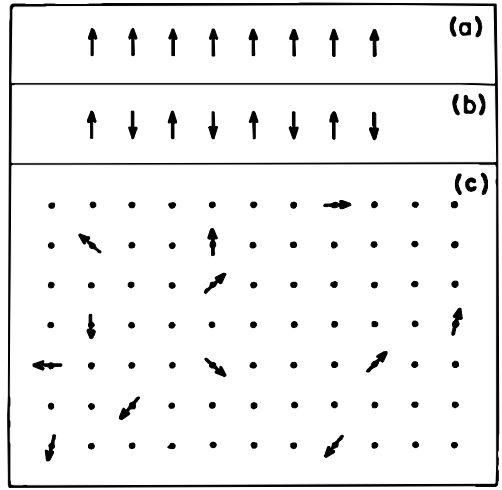

Fig.1. Schematic representation of spin arrangement in a ferromagnetic lattice (a), antiferromagnetic lattice (b) and spin glass lattice (c).

In a metallice lattice containing magnetic ions, the RKKY interaction exists [1] which gives rise to oscillatory conduction electron polarisation (Fig. 2 (a)). In magnetically dilute alloys, this interaction predominates. Due to this the sign of $J_{ij}$ for two spins i and j depends on their separation as these spins interact primarily through the polarized conduction electron cloud. For example in Fig.2 (a) there is a magnetic spin in up direction at the origin O which is polarising the conduction electron cloud ($n_\uparrow$ = number of conduction electrons with spins in up direction and $n_\downarrow$ = number of conduction electrons with spins down). Due to this polarisation, a spin at the point a will point in the down direction and a spin at the point b will point in the up direction. Thus at a particular spin site, say at the site of the spin b ($\equiv$ spin i), $J_{ij}$ is positive if the separation, $r_{ij}$, between the spin b and any other interacting spin is equal to $r_A$ (or its multiples) and $J_{ij}$ is negative if $r_{ij} = r_B$ (or its multiples). In magnetically dilute alloys, like Au - ~ 1-20 at. % Fe where SG behaviour has been observed [2,3], there is a random distribution of $r_{ij}$. Consequently both the positive and the negative $J_{ij}$, $(J_{ij})_+$ and $(J_{ij})_-$, are present at the spin b. If $N_{\uparrow+}[N_{\downarrow+}]$ =number of up (↑)[down (↓)] j spins producing an average $(J_{ij})_+$ per spin at the site of the spin b and $N_{\uparrow-}[N_{\downarrow-}]$ = number of up (↑−)[down(↓)] j spins producing an average $(J_{ij})_-$ per spin at the spin b and $N_{\uparrow+}[N_{\downarrow+}] | (J_{ij})_+ | \sim N_{\uparrow-}[N_{\downarrow-}] | (J_{ij})_- |$, spin b gets frustrated and instead of pointing in up or down direction points in a random direction. Depending on the relative magnitudes of $N_{\uparrow+}[N_{\downarrow+}] | (J_{ij})_+ |$ and $N_{\uparrow-}[N_{\downarrow-}] | (J_{ij})_- |$ at each spin site i, each i will have its own random direction resulting in a SG ordering [ $\equiv$ SG disordering] (global spin direction randomness (global disordering)) like that of Fig.1(c). It can be easily seen that the same result is obtained when $N_{\uparrow+} | (J_{ij})_+ | [N_{\uparrow-} | (J_{ij})_- | ] \sim N_{\downarrow+} | (J_{ij})_+ | [ N_{\downarrow-} | (J_{ij})_- | ]$. It may be noted that in dilute magnetic alloys,

very close to the spin glass temperature, $T_{SG}$, the polarising spin directions are either random (when system enters SG state from paramagnetic state on cooling) or parallel (when system enters SG state from ferromagnetic state on cooling).

In a nonmetallic SG lattice, where magnetic ions have superexchange and direct exchange interactions, frustration arises due to the presence of diamagnetic ions (lattice vacancies) or in some specific cases, like $FeCoCrO_4$ [4], due to the presence of magnetic ions of different magnetic moments. In Fig.2( (b) - (e) ) we describe different situations which arise in a nonmetallic SG lattice. Fig.2 (b) discusses the situation when the nearest neighbour (nn) interaction is antiferromagnetic and the next nn (nnn) interaction is also antiferromagnetic. Since nn interaction is stronger than the nnn interaction, neighbouring spins get aligned in antiparallel directions as is shown in the top row of Fig.2 (b). In the bottom row of Fig.2 (b), we reproduce the top row again but now with one of the spins replaced by a diamagnetic ion (shown by a dot at the lattice point). We now focus our attention on the bottom row spin marked by the thick arrow. Whereas the spin on the left of the marked spin (MS) wants the MS to point downwards, owing to the antiferromagnetic nn interaction, the spin on the right side of MS wants the MS to point upwards owing to the antiferromagnetic nnn interaction. As such MS will point downwards because of the stronger nature of the nn interaction. But if the number of next nearest neighbours separated by diamagnetic ions ($n_{nnn}$) is larger than the number of nearest neighbours ($n_{nn}$) in the lattice, it is possible to have a situation where MS receives opposite instructions of comparable strength for its alignment from nearest neighbours and the diamagnetic ion separated next nearest neighbours; (*i.e.* $n_{nn} |(J_{ij})_{nn}| \sim n_{nnn} |(J_{ij})_{nnn}|$, assuming each nn produces an average $(J_{ij})_{nn}$ at MS site and each diamagnetic ion separated nnn produces an average $(J_{ij})_{nnn}$ at the MS site). In such a situation MS will get frustrated and point in a random direction. Again if the diamagnetic ions are randomly located in the lattice, each frustrated spin will point in a different direction depending on the extent of frustration (*i.e.* the relative magnitude of the opposite instructions it receives at its site from the surrounding spins). Such a situation exists in systems like $FeMgBO_4$ [5].

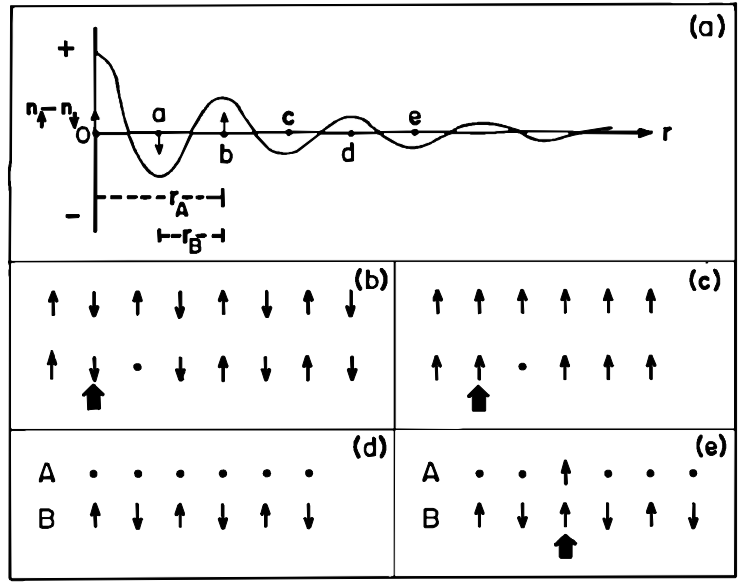

Fig.2. Schematic representation of the occurrence of frustration in a metallic lattice (a) and in a nonmetallic lattice ((b) - (e)). Details are described in the text.

In Fig.2 (c) we discuss another situation where nn interaction is ferromagnetic and the nnn interaction is antiferromagnetic; (such a situation exists in systems like $Eu_xSr_{1-x}S$ (x ~ 0.5) [6]). Because of the ferromagnetic nn interaction, in the top row of Fig.2(c) all the spins point in parallel directions. In the bottom row of Fig.2(c) we reproduce the same top row but now with one of the spins replaced by a diamagnetic ion (shown by a dot). We now focus our attention on the thick arrow marked spin (MS). The spin on the left of MS wants the MS to point upwards owing to ferromagnetic nn interaction and the spin on the right of MS wants the MS to point downwards owing to antiferromagnetic nnn interaction. Thus as explained above in the discussion of Fig.2 (b), the frustration is produced in the lattice and the SG ordering (random spin freezing) can occur if the diamagnetic ions are randomly located in the lattice.

The situations described in Figs. 2(b) and 2 (c) are the only two situations, involving competing nearest neighbours and higher order neighbours, in which theoretically, as discussed above, a SG freezing (random spin freezing) can occur. Experimentally also only in these two circumstances, nonmetallic systems have been found to show a SG behaviour. The other situations like (a) ferromagnetic nn interaction and ferromagnetic nnn interaction or (b) antiferromagnetic nn interaction and ferromagnetic nnn interaction, as can be easily seen from the discussion given above, can not produce a SG ordering. Experimentally also no system has yet been found where any of these two latter situations ((a) or (b)) exists and a SG ordering also occurs. However examples are known, in systems like $Rb_2Mn_xCr_{1-x}Cl_4$ (x ~ 0.5) [6], where competing exchange interactions are produced at a spin site by the nearest neighbours alone causing SG ordering.

The spinel ferrites [7], which are an important class of material, belong to the category where the situation discussed in Fig.2 (b) exists. In this type of materials we have A (tetrahedral) and B (octahedral) cation sites. Each A site cation has 12 B site cations as nearest neighbours and each B site cation has 6 A site cations plus 6 B site cations as its nearest neighbours. However these 6 B site nn cations act as next nearest neighbours since $|J_{AB}| >> |J_B|$; ($J_B \equiv J_{BB}$). Both the nn interaction ($J_{AB}$) and the nnn interactions ($J_A$ ($\equiv J_{AA}$) and $J_B$) are antiferromagnetic. Fig.2 (d) shows a situation where all the A-site ions are diamagnetic and only the B-site has magnetic ions which are antiparallely aligned owing to the negative sign of $J_B$. Such a situation exists in systems like $(Zn^{2+})_A[Fe_2^{3+}]_BO_4$ or $(Ga^{3+})_A[Ni^{2+}Cr^{3+}]_BO_4$. In Fig.2(e) we reproduce the lattice of Fig.2(d) but now with one of the A-site diamagnetic ions replaced by a magnetic ion (spin). We now focus our attention on the thick arrow marked B-site spin (MS). Whereas the B-site spins on the left and right of MS want it to point upwards owing to the negative $J_B$, the A-site spin on the top of MS wants it to point downwards due to negative $J_{AB}$. Though $|J_{AB}|$ is larger than $|J_B|$, for the MS the number of nn A spins is smaller than the number of nnn B spins. Thus, as discussed before, MS can get frustrated and depending on the diamagnetic ion concentration in the lattice, a random location of these ions can occur on the A-site giving rise to a SG ordering (Fig.1(c)). A system where such a situation occurs is $(Fe_x^{3+}Ga_{1-x}^{3+})_A[Ni^{2+}Cr^{3+}]_BO_4$ [8].

Out of various theories developed for understanding SG systems, the replica symmetric mean field theory, also called infinite range Sherrington-Kirkpatrick (SK) model [9], is able to explain several of the SG system's results. The SK model was developed for an Ising lattice using the Hamiltonian,

$$H = -\sum_{(ij)} J_{ij} S_i S_j - H \sum_i S_i, \qquad (1)$$

where $S_i = \pm 1$, i,j = 1 to N, N= Number of spins in the lattice, H= external magnetic field and the summation is over all the distinct i, j spin pairs ($\Sigma_{(ij)} = \frac{1}{2} \Sigma_{i \neq j}$, $\frac{N(N-1)}{2}$ terms). In this model, unlike to what happens in normal magnetically ordered systems, the exchange integral $J_{ij}$ does not have unique + or − value but has a Gaussian distribution (Fig.3) given by,

$$p(J_{ij}) = \frac{1}{J\sqrt{2\pi}} \exp[-\frac{(J_{ij} - J_0)^2}{2J^2}]. \qquad (2)$$

Thus $J_0$ is the mean of the distribution and $J \propto \Delta$, the width of the distribution (and accordingly its physically acceptable sign is always +).

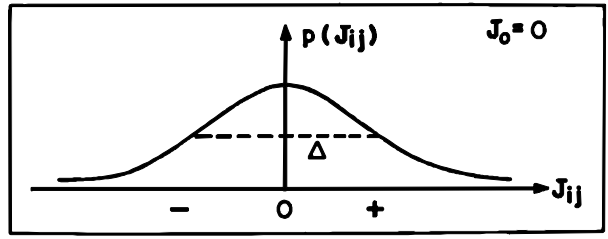

Fig.3. Gaussian distribution of exchange integral, $J_{ij}$, in a spin glass lattice. $J_0$ = mean and $\Delta$ = width of the distribution. $p(J_{ij})$ = probability for a particular $J_{ij}$ at any given spin site.

Figure 3 represents a special situation where $J_0 = 0$ and thus $p(J_{ij})$ has equal magnitude for $+|J_{ij}|$ and $-|J_{ij}|$. In this case both the positive and negative exchange interactions are present in equal strength at a given spin site. As will be discussed later, in such a situation a system becomes SG. Various other situations are also possible; they are discussed in more detail later on. For example for $J_0$ (+ or −) > $\Delta$, the system becomes reentrant (sometimes also called as reentrant SG) where on cooling the lattice paramagnetic state (P) $\xrightarrow{T_C, T_N}$ ferromagnetic state (F)/ ferrimagnetic state (F)/ antiferromagnetic state (AF) $\xrightarrow{T_{SG}}$ SG state transitions occur; these transitions are reversible in temperature *i.e.* on warming SG $\xrightarrow{T_{SG}}$ F $\xrightarrow{T_C}$ P transitions are obtained ($T_C$ = Curie temperature, $T_N$ = Néel temperature and $T_{SG}$ = spin glass temperature). In the $J_0 = 0$ case, only P $\xleftrightarrow{T_{SG}}$ SG transition occurs. Thus by adjusting the parameters $J_0$ and $J$, theoreticians are able to explain many of the SG system's results. Experimentally $J_0$ and $J$ depend on the concentration and distribution of the magnetic and diamagnetic ions in the lattice.

Defining $J_0 = \tilde{J}_0/N$ and $J^2 = \tilde{J}^2/N$, the free energy of the system (average, per spin) can be written as,

$$f = \frac{<F>_d}{N} = -kT(N)^{-1} <\ln Z>_d,$$

where Z is the partition function and $<...>_d$ means an average over the disorder *i.e.* over $p(J_{ij})$. Since theoretically it is difficult to obtain $<\ln Z>_d$, SK used the replica trick, introduced by Edwards and Anderson for SG calculation [1], $\ln x = \lim_{n\to 0} (x^n - 1)/n$, n being the number of replicas. Thus,

$$f = -kT \lim_{n\to 0, N\to\infty} (Nn)^{-1}\{<Z^n>_d - 1\}.$$

Assuming that all the replicas are symmetric, $Z^n = \prod_{\alpha=1}^{n} Z_\alpha$. Thus,

$$f = kT[(\frac{\tilde{J}_0 M^2}{2kT}) - (\frac{\tilde{J}^2(1-q)^2}{4(kT)^2}) - (2\pi)^{-1/2} \int_{-\infty}^{\infty} dz \exp(-\frac{1}{2}z^2) \ln(2\cosh\theta)], \qquad (3)$$

where $\theta = (\tilde{J}_0 M + \tilde{J} q^{1/2} z + H)/kT$, $M = <<S_i>_T>_d = <\sum_i <S_i>_T>_d/N$, $q = <<S_i>_T^2>_d = <\sum_i <S_i>_T^2>_d/N$ and $<...>_T$ means thermal average (*i.e.* average over the Boltzmann spin population distribution). The susceptibility, $\chi$, is obtained from the free energy using the relation, $\chi = -(\partial^2 f/\partial H^2)_{H\to 0}$.

M is the lattice magnetisation and q is called the Edwards-Anderson (EA) parameter. Fig.4((a) - (c)) shows the temperature (T)-, field (H)- and $\tilde{J}_0$- dependence of $\chi$ and M. Fig.4(d) shows the phase diagram (for H=0) obtained using the above given equations and the following criteria. In P, M=0, q=0. In F, M ≠ 0, q ≠ 0 and in the SG state, M=0, q ≠ 0. The phase diagram (Fig.4(d)) is symmetric for the positive and negative halves of the x-axis. Thus on changing the temperature, the system undergoes $P \xleftrightarrow{T_{SG}}$ SG transition for $\tilde{J}_0/\tilde{J} \leq 1$ and $P \xleftrightarrow{T_C} F \xleftrightarrow{T_{SG}}$ SG transition for $\tilde{J}_0/\tilde{J} > 1$. This latter ($\tilde{J}_0/\tilde{J} > 1$) behaviour, as mentioned above, is called reentrant SG behaviour (or sometimes only the reentrant behaviour). At each transition temperature, $T_C$ or $T_{SG}$, $\chi$ shows a cusp.

The Ising model of SK has been extended to a general vector system by Gabay and Toulouse (GT) [10]. They use the Hamiltonian,

$$H = -\sum_{(ij)} J_{ij} \sum_\mu S_{i\mu} S_{j\mu} - H \sum_i S_{i1}, \qquad (4)$$

where $\mu = 1...m$, m = number of spin components (m=3) and $\sum_{\mu=1}^{m} S_\mu^2 = m$; ($\mu=1$ ($\equiv z$) is the longitudinal (L) direction and $\mu = 2,3$ ($\equiv x,y$) are the transverse (T′) directions). Following the procedure of SK, it can be seen that the free energy in this case is given by,

$$f = \frac{\tilde{J}_0}{2} \sum_\mu (M_\mu^\alpha)_0^2 + \frac{\tilde{J}^2}{4kT} \sum_\mu \{(q_\mu^\alpha)_0^2 - (q_\mu^{(\alpha\beta)})_0^2\} - \frac{kT}{(2\pi)^{m/2}} \int_{-\infty}^{\infty} (\prod_\mu dt_\mu) \exp(-\frac{1}{2} \sum_\mu t_\mu^2) \ln Q, \qquad (5)$$

where,

$$Q = \sqrt{m} (2\pi)^{(m-1)/2} (|a|_{m-1})^{(3-m)/2} \int_{-\sqrt{m}}^{\sqrt{m}} dS_1 \exp[a_1 S_1 + b_1 S_1^2 + b_{\mu>1}(m - S_1^2)](m - S_1^2)^{(m-3)/4}$$

$$\times I_{(m-3)/2}(|a|_{m-1} \sqrt{m - S_1^2}),$$

$I_\nu(z)$ = modified Bessel function of the first kind,

$$b_{\mu>1} = b_2 \text{ or } b_3, |a_{m-1}| = \sqrt{a_2^2 + a_3^2},$$

$$a_\mu = \frac{\tilde{J}_0}{kT}(M_\mu^\alpha)_0 + \frac{\tilde{J}}{kT} t_\mu (q_\mu^{(\alpha\beta)})_0^{1/2} + \frac{H}{kT} \delta_{1,\mu},$$

$$b_\mu = \frac{\tilde{J}^2}{2(kT)^2}[(q_\mu^\alpha)_0 - (q_\mu^{(\alpha\beta)})_0],$$

$(M_1^\alpha)_0 = M_L = <<S_{i1}>_T>_d, (M_2^\alpha)_0 = (M_3^\alpha)_0 = M_{T'},$

$(q_1^{(\alpha\beta)})_0 = q_L = <<S_{i1}>_T^2>_d, (q_2^{(\alpha\beta)})_0 = (q_3^{(\alpha\beta)})_0 = q_{T'},$

$(q_\mu^\alpha)_0 = <<S_{i\mu}^2>_T>_d,$

$(q_1^\alpha)_0 = 1 + (m-1)x$ and $(q_2^\alpha)_0 = (q_3^\alpha)_0 = 1-x$.

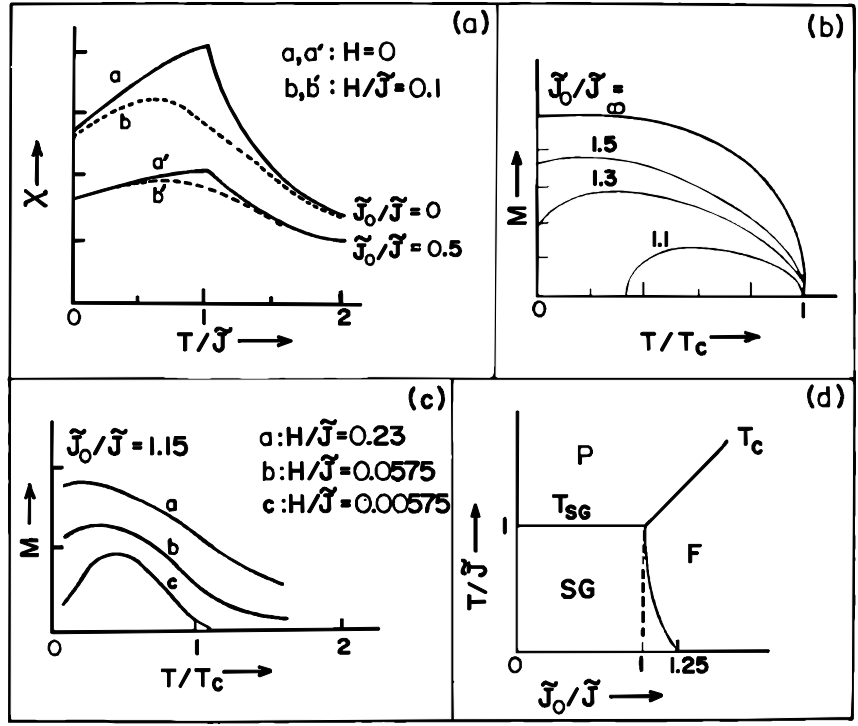

Fig. 4. Results of the Sherrington-Kirkpatrick (SK) model. Details are described in the text. Fig.(d) gives the SK model spin glass phase diagram. For Figs. (b) and (d), external magnetic field, H, =0.

Thus M and q of the SK model have now three components; $\mu = 1$ gives the L component and $\mu = 2, 3$ give the T′ components. In addition a new parameter $(q_\mu^\alpha)_0$ appears; x is called the quadrupolar parameter and is zero for H = 0.

Figure 5(a) shows the phase diagram obtained for the vector system (H=0). The following criteria have been used for obtaining this phase diagram. For P, $M_L = M_{T'} = q_L = q_{T'} = 0$. For SG state, $M_L = M_{T'} = 0$, $q_L \neq 0$, $q_{T'} \neq 0$. For F, $M_L \neq 0$, $M_{T'} = 0$, $q_L \neq 0$, $q_{T'} = 0$. For the mixed phase M1, $M_L \neq 0$, $M_{T'} = 0$, $q_L \neq 0$, $q_{T'} \neq 0$ and for the mixed phase M2, $M_L = 0$, $M_{T'} = 0$, $q_L \neq 0$, $q_{T'} \neq 0$. Thus in the mixed phase M1, L spin components are magnetically ordered but the T′ components are frozen in a SG configuration (*i.e.* are frozen in random directions). Some people also use the term 'semi spin glass' for the mixed phase M1 and 'spin glass' for the mixed phase M2 since, as in the SG phase, in this latter phase all the three spin components ($\mu = 1, 2, 3$) are frozen in SG configuration [11]. $\chi$ shows cusp (peak) at all the three transition temperatures ($T_C$, $T_{M1}$ and $T_{M2}$) [2,3]. Fig.5(b) shows the field dependence of $T_{M1}$ and $T_{M2}$. The two curves follow the equations given below [10,12].

$T_{M1}(H) = T_{M1}(0) - A_{GT} H^2$,

$T_{M2}(H) = T_{M2}(0) - A_{AT} H^{2/3}$.

$T_{M1}$ *vs.* H curve is called the GT line and the $T_{M2}$ *vs.* H curve is called the de Almeida - Thouless (AT) line. In the SK model (Ising case), $T_{SG}$ *vs.* H follows the AT line [9,12]. For $\tilde{J}_0 \neq 0$, $A_{GT}$ and $A_{AT}$ get modified and the curves do not meet at H = 0 in Fig.5(b) plot [10] when $\tilde{J}_0 / \tilde{J} > 1$. In Fig.5 (b), $\vec{H}$ provides the quantisation (z) direction and the GT line – AT line meeting point at H=0 is the $T_{SG}$ of Fig.5 (a) (where the quantisation direction is provided by the Weiss field, $\vec{H}_W$).

### I.2. NEAR SEPARATE FREEZING OF A- AND B- SITE MOMENTS

From the above discussion it is clear that in a $\chi$ vs. T measurement in any system, one can normally expect at the most three $\chi$-peaks occurring respectively at $T_C$ ($T_N$), $T_{M1}$ and $T_{M2}$ (double reentrant behaviour). In one of our group $\chi$-measurements in a spinel ferrite $Co_2TiO_4$ (cation distribution $(Co^{2+})_A[Co^{2+}Ti^{4+}]_BO_4$), we however observed five peaks on cooling the system down to 1.7K through $T_C$ [13]. The neutron diffraction and magnetisation (M) measurements carried out by earlier workers have shown this system to have a reentrant behaviour. $\chi$ vs. T measurements done in $Co_{1.2}Zn_{0.8}TiO_4$ and $Co_2SnO_4$ showed one and three $\chi$-peaks respectively (Figs.6 -8) which can be easily understood on the basis of the discussion given before as indicating the $T_{SG}$ peak in $Co_{1.2}Zn_{0.8}TiO_4$ and the $T_C$, $T_{M1}$ and $T_{M2}$ peaks in $Co_2SnO_4$. In order to understand the $Co_2TiO_4$ result, it is necessary to extend the GT vector model to a two sublattice system. The Hamiltonian for such a system will be,

$$H = -\sum_{(ij)} J_{ij}^A \sum_\mu S_{i\mu}^A S_{j\mu}^A - \sum_{(ij)} J_{ij}^B \sum_\mu S_{i\mu}^B S_{j\mu}^B - \sum_{i,j} J_{ij}^{AB} \sum_\mu S_{i\mu}^A S_{j\mu}^B - H^A \sum_i S_{il}^A - H^B \sum_i S_{il}^B , \quad (6)$$

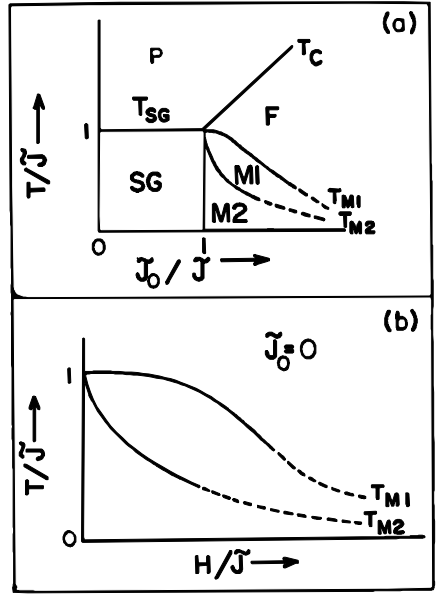

Fig.5. Gabay - Toulouse (GT) vector model spin glass phase diagram (a) and magnetic field (H) dependence of spin glass temperature, $T_{SG}$, in GT model (b). Symbols are described in the text. In Fig.(a), H=0.

where $H^A$ and $H^B$ are the components of $\vec{H}$ present at the A- and B- site spins respectively. The free energy for such a system is very complicated and is described in Appendix 2 [13]. We give here only the results of our calculation. For the specific case of $H^A = H^B = H$, $\chi = -(\partial^2 f / \partial H^2)_{H\to 0}$. We have done calculation for this case for the following two situations.

(a) $\tilde{J}_0^A = \tilde{J}_0^B = \tilde{J}_0^{AB} = 0$, $\tilde{J}^A > \tilde{J}^B$.

(b) $\tilde{J}^A = \tilde{J}^B = \tilde{J}^{AB} = 0$, $|\tilde{J}_0^A| > |\tilde{J}_0^B|$.

It can be seen from the phase diagram (Fig.5(a)) that the situation (a) aims in finding the behaviour of χ-peak at $T_{SG}$ and the situation (b) concerns with the χ-peak behaviour at $T_C$. The results of the calculation are given in Fig.9. For the situation (a) (Fig.9(a)), we see one χ-peak at $T_{SG}$ when $\tilde{J}^{AB} > (\tilde{J}^A \tilde{J}^B)^{1/2}$ i.e. when the coupling between the two sublattices is strong. For completely uncoupled sublattices ($\tilde{J}^{AB} = 0$), we get two χ-peaks situated respectively at $T_{SG}^A = \tilde{J}^A$ and $T_{SG}^B = \tilde{J}^B$ which indicates that the two sublattices freeze separately. For the intermediate case of weak soupling, $\tilde{J}^{AB} < (\tilde{J}^A \tilde{J}^B)^{1/2}$, we see in

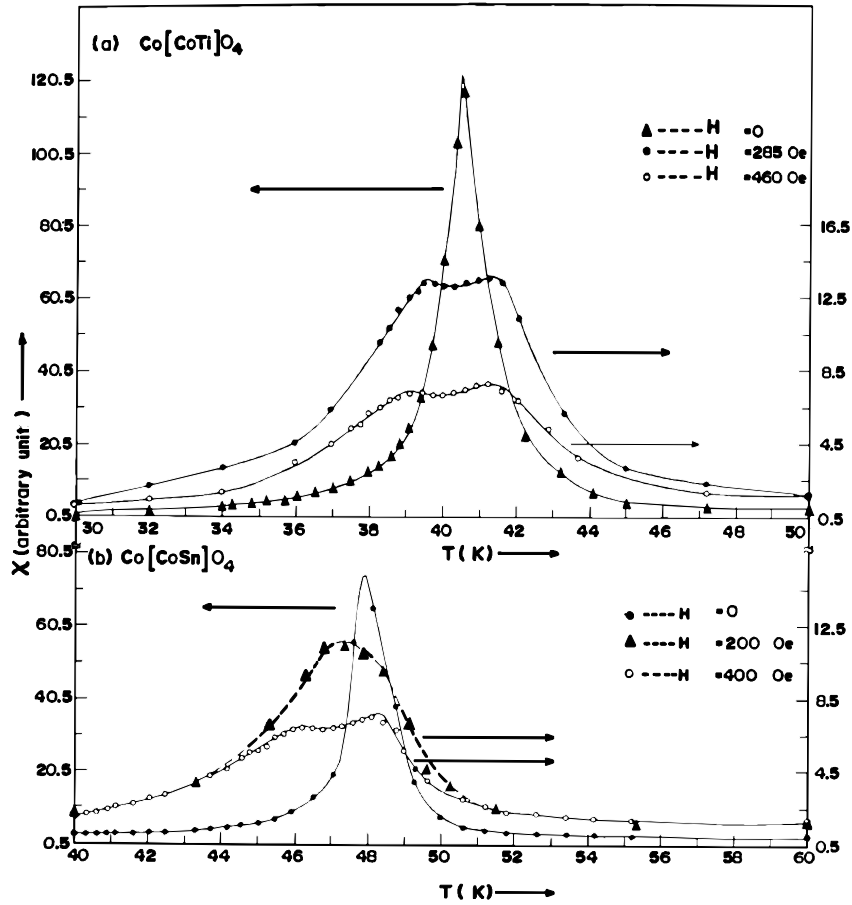

Fig.6. Temperature (T) dependence of the a.c. susceptibility, $\chi_{ac}$, in (a) $Co_2TiO_4$ and (b) $Co_2SnO_4$ for some typical values of the external d.c. field, H, (longitudinal to a.c. field) for $30 \leq T \leq 60$ K.

Fig.9(a) [ii], a strong χ-peak at ~ $T_{SG}^A$ and a broad, weak amplitude χ-peak (a hump) at ~ $T_{SG}^B$ indicating a near separate freezing of the A- and B- site spins [13]. Similar result is obtained for the situation (b) (when $|\tilde{J}_0^{AB}| > (|\tilde{J}_0^A||\tilde{J}_0^B|)^{1/2}$, $\tilde{J}_0^{AB} = 0$ and $|\tilde{J}_0^{AB}| < (|\tilde{J}_0^A||\tilde{J}_0^B|)^{1/2}$) (Fig.9(b)). From this we conclude that for each of the phase diagram transition temperatures, $T_C$, $T_{M1}$ and $T_{M2}$, it is possible to get two χ-peaks, one of them a strong peak and the other a hump, when the coupling between the two sublattices is weak. The stronger peak represents the true transition and the hump represents a pseudo-transition [13]. Thus as many as six χ-peaks can be observed in a system having weak A-B coupling. $Co_2TiO_4$ seems to represent such a case. Some other examples of a separate A-, B- site SG behaviour are also known [14].

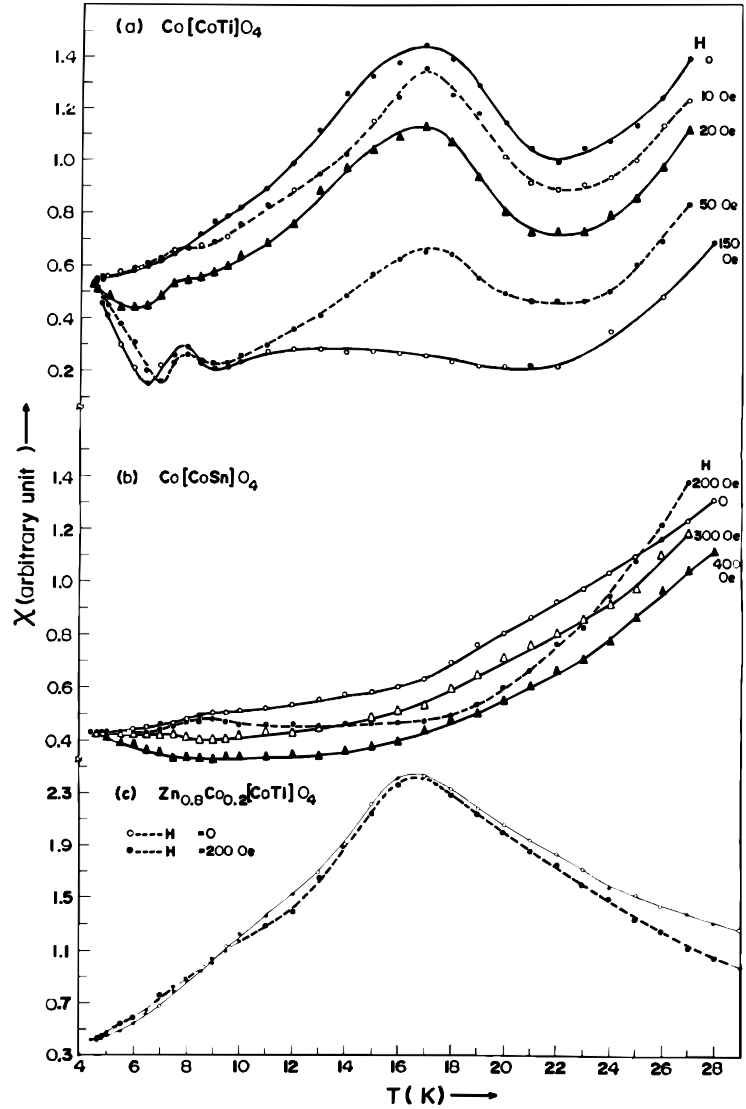

Fig.7. Temperature dependence of $\chi_{ac}$ in (a) $Co_2TiO_4$, (b) $Co_2SnO_4$ and (c) $Co_{2-x}Zn_xTiO_4$ (x = 0.8) for some typical H for $4 \leq T \leq 30$ K.

## I.3. HISTORY, AGE AND TIME DEPENDENCE OF SPIN GLASS MAGNETISATION

In a normal magnetically ordered system, where the domain or sublattice magnetisation follows a Brillouin function temperature dependence, the observed M vs. T behaviour depends on the type of magnetic ordering [15]. In an antiferromagnet, the sublattice magnetisations are exactly equal and opposite, and the applied field, H, (needed for M measurement) produces a very small net observable M. In a ferrimagnet, the observed M is a sum of unbalanced sublattice magnetisations and thus M vs. T can have a variety of shapes. In a ferromagnet, the observed M vs. T follows a Brillouin function provided H is large

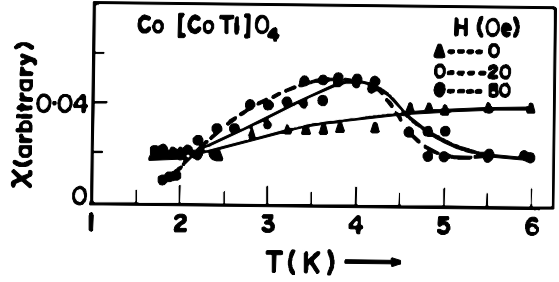

Fig.8. Temperature dependence of $\chi_{ac}$ in $Co_2TiO_4$ for some typical H for $1 \leq T \leq 6$ K.

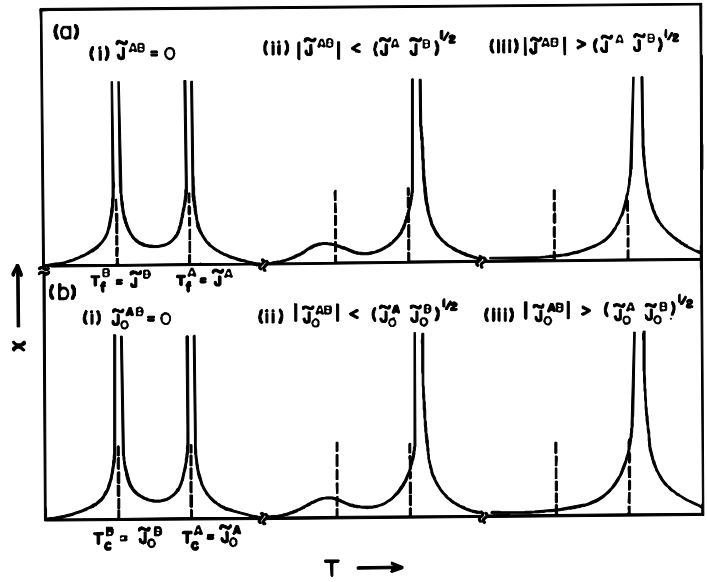

Fig.9. Results of the two-sublattice calculation.

enough to align the domains. However in all these cases, the M vs. T curve obtained is normally the same whether H is applied after cooling the system to the lowest measurement temperature (zero field cooled (ZFC) state) or in the paramagnetic state (P) itself (T > $T_C$, $T_{SG}$ (for the SG case discussed below)) (called the field cooled (FC) state since the system in this case gets cooled through $T_C$, $T_{SG}$ to the lowest T in a magnetic field) (Fig.10(a)). In SG systems, on the other hand, the ZFC and FC M vs. T curves do not overlap i.e. there is a history dependence. For systems showing P $\xleftrightarrow{T_{SG}}$ SG transition, the ZFC and FC curves get separated (i.e. magnetic irreversibility, $M_{irr}$, occurs) below $T_{SG}$ (Fig.10(b)). Generally for the $M_{irr}$ measurement, the system is first cooled to the lowest T. Then a small H is applied and M vs. T curve recorded with T increasing upto T > $T_C$, $T_{SG}$; this is ZFC curve. With the same H present, the temperature is now decreased and M vs. T recorded again; this is FC curve. Other variations of this procedure also exist [16]. Sometimes it is possible to record the FC curve with T increasing also (H present), after field cooling the system to the lowest T from T > $T_C$, $T_{SG}$, provided various time dependences of FC M (discussed below) are negligibly small during the measurement period. Fig.10(c) shows $M_{irr}$ data for a double reentrant system. $M_{irr}$, which is weak between $T_{M1}$ and $T_{M2}$, becomes strong for T < $T_{M2}$ [10]. In actual practice, normally the $T_{M1}$ and $T_{M2}$ positions are confirmed by observing the presence of the a.c. susceptibility ($\chi_{ac}$) peaks at these positions and also by studying the H dependence of these $\chi_{ac}$-peak positions [13,14]. Fig.11 shows the history dependence observed [17,18] in $Co_{0.5}Zn_{0.5}Fe_2O_4$ (double reentrant system) and $Co_xZn_{1-x}FeCrO_4$, x ≅ 0.5, (another double reentrant system (Sec. I.5, II.2 )). The $\chi_{ac}$ peaks are also shown in the

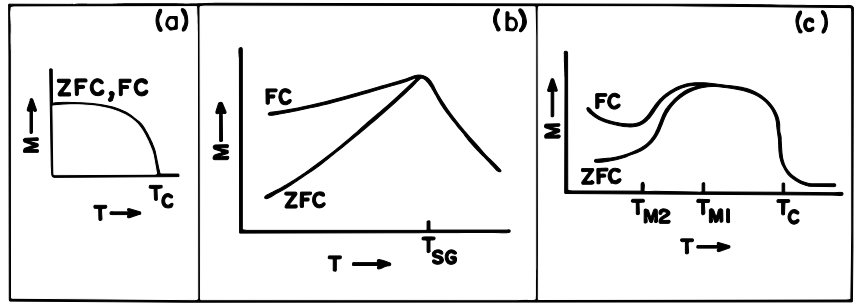

Fig.10. Temperature dependence of magnetisation (M) in a normal magnetically ordered system (a), in a spin glass (b) and in a double reentrant system (c). Symbols are described in the text; (FC= field cooled system, ZFC = zero field cooled system).

figure. In $Co_{0.5}Zn_{0.5}Fe_2O_4$ only two $\chi_{ac}$ peaks are seen because $T_C$ and $T_{M1}$ are very close. This is confirmed by the nature of the IRM (isothermal (or ZFC) remanent magnetisation) *vs.* T curve recorded for this system (Fig.12 ); (for recording IRM *vs.* T, the system is zero field cooled from T > $T_C$, $T_{SG}$ to the desired low T. A field H is then applied at that T and removed at the same T. The remanent magnetisation is then measured as a function of temperature). Fig.12 inset shows M *vs.* H behaviour at T = 291K (curve (a)) and T = 77K (curve (b)). Contrary to what is observed in normal magnetically ordered systems, in Fig.12 the M saturation is difficult at lower tempeature indicating the presence of low temperature SG freezing in $Co_{0.5}Zn_{0.5}Fe_2O_4$ [19] ;(T = 77K < $T_{M2}$, $T_{M2}$ ~ 250K (Fig.12)).

Apart from history dependence the SG M also shows age dependence. This latter dependence is also called waiting time, $t_W$, dependence. This means that in the ZFC case the value of M depends on the time for which one waits before applying H at the desired low T [20]. Similarly both the ZFC and FC M values depend on the cooling rate of the system [21]. In addition for a fixed T,H,$t_W$ and cooling rate, both the ZFC and FC magnetisations show a time (*i.e.* observation time), t, dependence [22]. However this t-dependence is more pronounced for the ZFC M ($M_{ZFC}$). The FC M ($M_{FC}$) shows very small t dependence [20, 22, 23]. $M_{ZFC}$ actually drifts with time towards the $M_{FC}$ value which is supposed to be nearer to the equilibrium value [23]. Similarly IRM and TRM (thermoremanent (or FC remanent) magnetisation) also show $t_W$, t and cooling rate dependence [20–24]; (TRM is measured by field cooling the system to the desired low T and then removing the field, H, after a certain $t_W$ ( ≥ 0) for TRM *vs.* t measurement). Fig.13 schematically shows the $t_W$- , t- dependence of $M_{FC}$, $M_{ZFC}$ [20] and Fig.14 shows the M *vs.* t curve observed in ZFC $Co_2TiO_4$ system (T = 30K, H = 2 kOe) [25].

It is possible to understand these dependences of M on the basis of the description given above (Fig.1). Fig.15(a) shows few possible spin configurations for some ferromagnetically coupled spins. It is obvious that all these spin configurations in space (*i.e.* A to E) can be obtained from any single spin configuration (A or B or C or D or E) by just rotating the quantisation axis and thus they are the same (*i.e.* not different from each other). Same is true for the different configurations of a set of antiferromagnetically coupled spins. This is not the case for a SG system, however, which can have several independent (distinct) spin configurations. For example, let us see the two configurations given in Fig.15(b) and 15(c). It is clear that these two configurations can not be matched with each other by quantisation axis rotation of any kind. Thus a SG system has many ground states [26]. The SG spin configuration drifts, from one possible configuration (one ground state) to the other, towards some equilibrium configuration giving rise to history, $t_W$, t and cooling rate dependence of SG M [23,27]. Experiments indicate that the number of SG ground states increases with decreasing T [28].

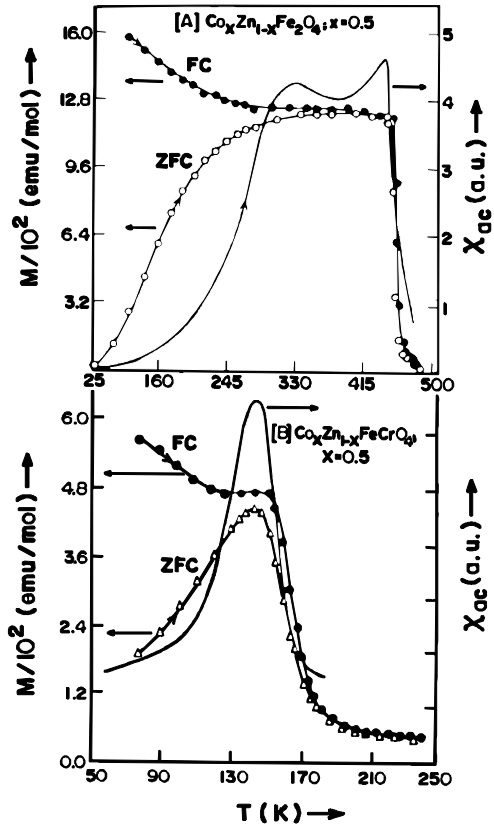

Fig.11. Temperature dependence of M and $\chi_{ac}$ in $Co_{0.5}Zn_{0.5}Fe_2O_4$ (A) and $Co_xZn_{1-x}FeCrO_4$ ($x \cong 0.5$) (B). For ZFC - FC measurement, H = 50 Oe in (A) and 30 Oe in (B); a.u.= arbitrary unit.

Three types of time functions have been used to fit the M *vs.* t data. For $M_{ZFC}$, they can be written as follows [23,29].

(i) Algebraic : $M_{ZFC} = M_0 - A\, t^{-\alpha}$.

(ii) Logarithmic: $M_{ZFC} = M_0 - M' [\beta' - (1-n)\, lnt]$ (for certain t interval).

(iii) Stretched exponential: $M_{ZFC} = M_0 - M' \exp [-(t/\tau_P)^{1-n}]$.

$0 \leq n \leq 1$; $(1/\tau_P) \propto \exp(-\gamma\, T_{SG}/T)$; $\alpha, \beta', \gamma$ =constants.

Even after carefully taking care of H, T, $t_W$ and cooling rate, different SG systems show different types of t behaviour, $M_{ZFC}$ *vs.* t fitting with one of the above three expressions. Attempts have been made to understand this aspect [20–24,27]. In some cases, a combination of the above expressions has also been tried for fitting the experimental data [24,30]. TRM *vs.* t has also been studied in detail [24,31]. It may be clarified that all the SG experimental results are reproducible.

## I.4. ANISOTROPIC SPIN GLASS BEHAVIOUR

In several SG systems, the spin components along different crystallographic axes do not freeze in SG configuration simultaneously at one temperature. Such systems are called anisotropic SG systems [23,32–34]. Examples are known for planar SG systems [32] and also for uniaxially anisotropic SG systems [14,23,33,34]. In well known uniaxially anisotropic SG $Fe_2TiO_5$ [33,34] (orthorhombic; lattice parameters: a= 9.79Å, b= 9.93 Å and c= 3.72 Å), the $Fe^{3+}$ - spin component along the c - axis (L component) freezes at $T_{LF} \sim 51K$ (L freezing from P) whereas the spin components along the a, b - axes freeze at $T_{TF} \sim 9K$ (T' freezing). The phase diagram for such a system has been calculated [35], following the procedure described before for the isotropic SG systems, using the Hamiltonian,

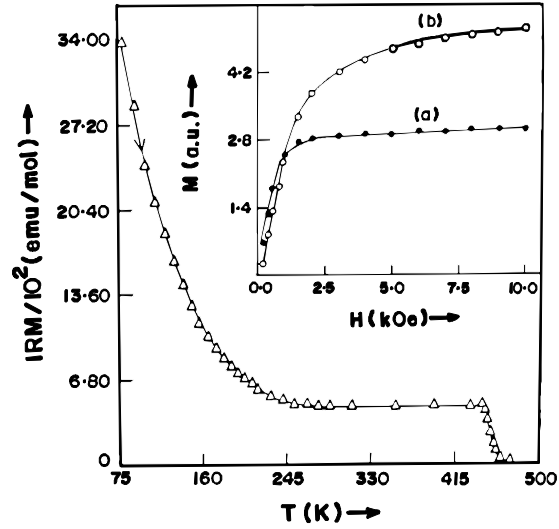

Fig.12. Temperature dependence of isothermal remanent magnetisation (IRM) and magnetic field, H, dependence of magnetisation, M, (T = 291K (a) and 77K (b)) in $Co_{0.5}Zn_{0.5}Fe_2O_4$. For IRM measurement, H = 15 kOe has been applied to the zero field cooled sample at 77K and then reduced to zero at the same temperature.

$$H = -\sum_{(ij)} J_{ij} \sum_{\mu} S_{i\mu} S_{j\mu} - H \sum_i S_{i1} - D \sum_i (S_{i1})^2, \qquad (7)$$

where D is the uniaxial crystal field anisotropy constant. Fig.16 (a) shows the phase diagram obtained ($\tilde{J}_o$ =H=0). For $Fe_2TiO_5$, D > 0 [34] and L freezing is observed first as one cools the lattice. It may be noted that whereas in Figs.4(d) and 5((a),(b)) the applied field, $\vec{H}$, or the Weiss field, $\vec{H}_W$, provides the quantisation direction (μ=1 direction), in Fig.16 (a) the uniaxial anisotropy axis (say the c-axis) of the crystal provides the μ=1 (quantisation) direction.

Single crystal measurements bring out the uniaxially anisotropic SG property of $Fe_2TiO_5$ very clearly [33]. Even the powder measurements, representing average crystallographic behaviour, show small amplitude $\chi_{ac}$-peak at $T_{LF} \sim 51K$ and a broad $\chi_{ac}$-maximum at $T_{TF} \sim 9K$ (Fig.17 ). $T_{TF}$ is seen more clearly in the $M_{irr}$ data (Fig.18 ) which also shows an arrow at ~ 61K indicating susceptibility, $\chi$, ( $\equiv \chi_{ac}$ or d.c. susceptibility, $\chi_{dc}$ (= M/H)) cross over temperature [36], $T_{cr}$, above which $\chi$ follows a Curie-Weiss behaviour (Fig.18, inset marked (iii)). In Fig.18 (inset marked (ii)) $M_{irr}$ data is shown for H = 1 kOe. The three arrows in this figure respectively indicate, starting from the highest temperature side, $T_{cr}$, $T_m$ and $T_{Mirr}$. $T_m$ is the temperature at which M ( and so also $\chi$) shows a maximum and $T_{Mirr}$ is the temperature at which $M_{irr}$ starts on cooling; ($T_{Mirr} \equiv T_{LF}$). As shown in Fig.16(b), all these temperatures are expected to meet at H = 0 [37] which gives $T_{LF}(0)$, the correct (*i.e.* undisturbed system's) $T_{LF}$; ($T_{LF}$ (0) = 51.5K [34] ). $T_{LF}$ *vs.* H follows the AT line and $T_{TF}$ *vs.* H follows the GT line [34].

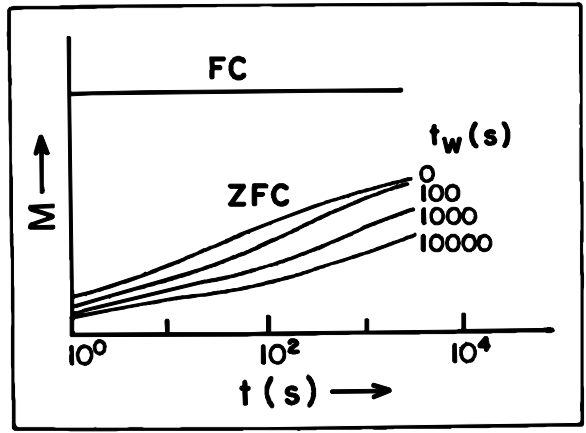

Fig.13. Typical time (t) dependence of M for the ZFC and FC states of the system. $t_W$ = waiting time before the field, H, is applied to the ZFC system for M measurement. For the FC state, H is applied at T > $T_{SG}$, $T_C$ (paramagnetic state) where $t_W$ dependence does not exist.

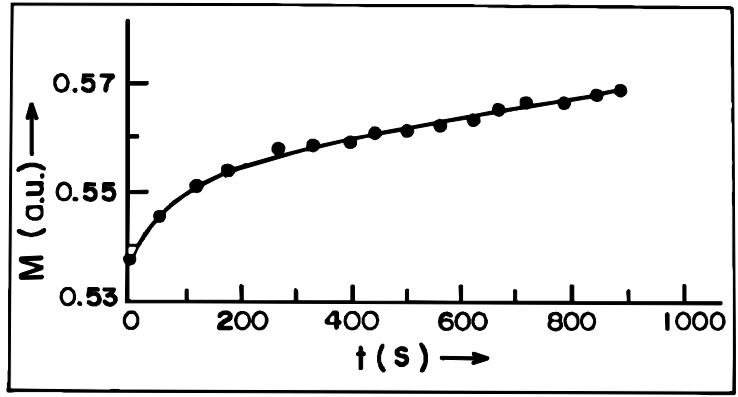

Fig.14. Time (t) dependence of M in ZFC $Co_2TiO_4$ at T=30K and H=2 kOe.

In the discussion given so far, the Dzyaloshinsky-Moriya (DM) interaction term ($\vec{d}_{ij} \cdot [\vec{S}_i \times \vec{S}_j]$) [38] has not been included in the Hamiltonian expressions. This term can couple the L and T′ irreversibilities and affect the H dependence of transition temperatures [33, 34, 38]

### I.5. CLUSTER PHASE TRANSITION

Whatever we have described so far, *i.e.* the spin (or the spin components) freezing abruptly in random directions at well defined temperatures $T_{M1}$ ($T_{TF}$) or $T_{M2}$ ($T_{LF}$) or $T_{SG}$, is the phase transition model. Several arguments have been given in support of this model [23, 27, 31, 39–41]. The static and dynamic scalings have been found to exist for M, $\chi_{ac}$, TRM and various characteristic times [31, 39, 40]. Fig.19 shows the scaling behaviour of M seen in anisotropic SG $Fe_2TiO_5$, yielding $T_{LF}$ (0) = 51.5 K [34]; $\tau$ = | T- $T_{LF}$ (0) | /$T_{LF}$ (0); $T_{LF}$ (0) ≡ $T_{LF}$ (H = 0). In several papers, the order of phase transition has also been discussed [1, 9, 10, 27, 41].

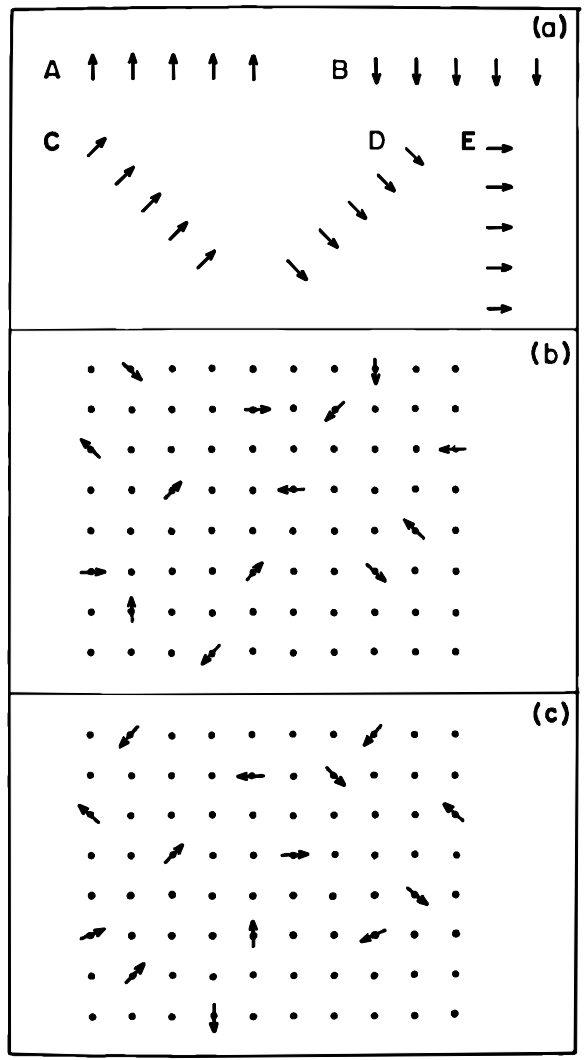

Fig.15. Schematic representation of few spin configurations in a ferromagnetic lattice (a) and in a spin glass lattice ((b) and (c)).

Apart from the phase transition model, another model, called 'continuous spin freezing model' or 'cluster by cluster freezing model', has also been used by several workers to describe the SG systems' results [27, 42, 43]. According to this model, the spin freezing does not occur abruptly at one well defined temperature ($T_{SG}$ or $T_{M1}$ or $T_{M2}$) but occurs continuously as the lattice is cooled and also instead of individual spins, it is the spin clusters which take part in the random freezing. Thus instead of spins freezing in random directions, we have spin clusters frozen in random directions (Fig.20(a)). The clusters are formed owing to the frustration existing in the lattice [44] and are separated from each other by the diamagnetic ions (lattice vacancies) or few frustrated spins (Fig. 20(a)) or may be just immersed in a magnetically ordered network of spins [43]. There is a cluster size distribution and as one cools the lattice,. the spin freezing occurs cluster by cluster where the biggest cluster freezes first and the smallest in the last (at the lowest temperature). These clusters are different from superparamagnetic clusters [45] as they interact with each other and the matrix, and therefore the fluctuation frequency, $\nu_F$, of the cluster magnetisation instead of following the Arrhenius law ($T = T_f$), $\nu_F = \nu_0 \exp(-E_A/kT_f)$, generally follows the Vogel-Fulcher law $\nu_F = \nu_0 \exp[-E_A /k(T_f - T_0)]$ [42]; $E_A$ = anisotropy energy barrier for cluster magnetisation fluctuation, $T_f$ is defined below and $T_0$ is a characteristic constant. Several SG properties, including magnetic viscosity and H dependence of transition temperatures (Fig.5 (b)), have been explained on the basis of this model [42, 46]. Support to this model comes from the fact that χ-peak (cusp) position (the cusp temperature) observed in a $\chi_{ac}$ vs. T measurement in SG systems has been found to be dependent on the frequency, $\nu_{ac}$, of the applied a.c. field [42] (Fig.20(b)). Therefore the cusp temperature is generally not denoted by $T_{SG}$ but by $T_f$ and called the freezing temperature. According to the cluster by cluster freezing model, in the $\chi_{ac}$ vs. T experiment only those clusters respond to the applied a.c. field for which $\nu_{ac} = \nu_F$. For different size clusters, this condition will be satisfied at different temperatures since $E_A \propto$ cluster volume; ($E_A$ depends on cluster shape also since it is proportional to the anisotropy constant, $K_A$, too [42] ). Thus each $T_f$ (*i.e.* each peak position in Fig.20 (b)) corresponds to the clusters of one particular size and accordingly the shift of $T_f$ with $\nu_{ac}$ indicates the freezing of different size clusters as T is changed. Depending on the mean and width of cluster size distribution, $T_f$ *vs.* $\nu_{ac}$ is more pronounced in some systems than in others. It may be mentioned here that in most of the SG experiments, $\chi_{ac}$ has been used for χ-investigations and not the $\chi_{dc}$ (= M/H) which normally does not show a sharp cusp (peak) at the SG transition temperatures owing to the large H needed for M measurement.

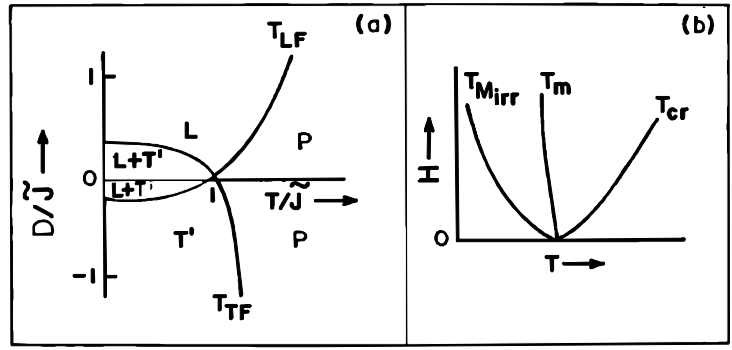

Fig.16. Phase diagram for an anisotropic spin glass (assuming $\tilde{J}_o = H = 0$) (a) and H dependence of various transition temperatures (b). Symbols are described in the text. D = 0 describes isotropic Heisenberg (vector) system, D = ∞ the Ising system and D = –∞ the planar system (Fig.a).

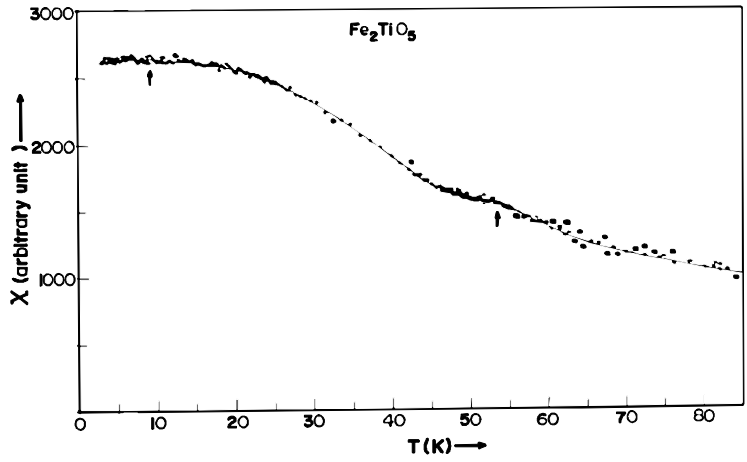

Fig.17. Temperature dependence of $\chi_{ac}$ in $Fe_2TiO_5$. Arrows indicate the transition temperatures $T_{LF}$ and $T_{TF}$ described in the text; $T_{LF} > T_{TF}$.

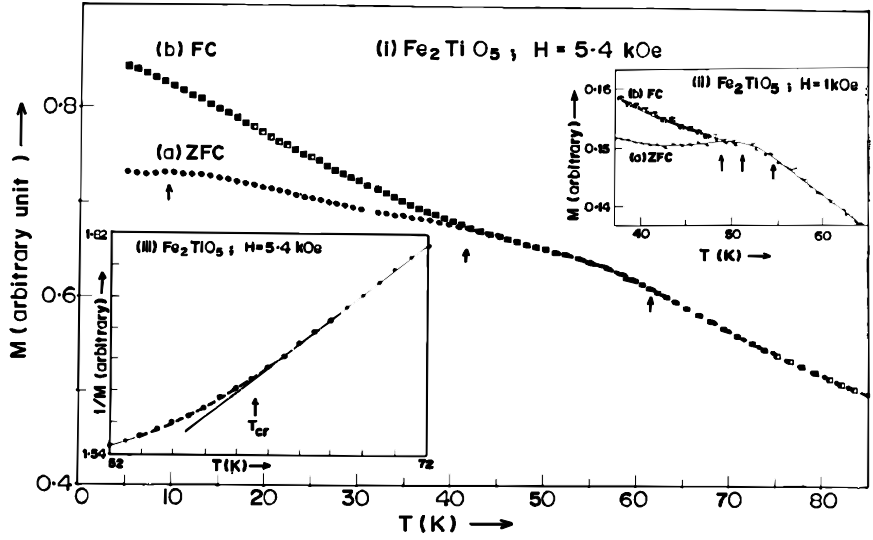

Fig.18. Temperature dependence of ZFC and FC magnetisation (M) in $Fe_2TiO_5$ for H=5.4 kOe and 1 kOe. Inset marked (iii) shows the definition of the susceptibility, χ ( ≡ $\chi_{ac}$ , $\chi_{dc}$ (=M/H)), cross over temperature, $T_{cr}$, above which χ (in this case $\chi_{dc}$ ) vs. T shows a Curie-Weiss behaviour. The arrows marked in the M(ZFC-FC) vs. T curve for H=5.4 kOe, starting from the higher temperature side, indicate $T_{cr}$, $T_{LF}$ and $T_{TF}$ respectively. The arrows shown in the inset marked (ii) (H=1 kOe) indicate, starting from the higher temperature side, $T_{cr}$, $T_m$ and $T_{Mirr}$ ( ≡ $T_{LF}$). Symbols are described in the text.

It is not yet settled that out of the two models described above which one represents the SG systems more correctly [27]. According to the research workers believing in phase transition model, the shift of $T_f$ with $\nu_{ac}$ does not have much significance since $T_f$ does not represent $T_{SG}$ which has to be obtained in the $\chi_{ac}$ vs. T experiments from a dynamic scaling argument [40]. This needs measurement of both the real ($\chi'$) and imaginary ($\chi''$) parts of $\chi_{ac}$ [40]. $T_{SG}$ obtained by such a scaling method agrees with the extrapolated value $T_f(\nu_{ac} = 0)$. Thus it is possible that the presence of a.c. field drives the SG system to a new, $\nu_{ac}$ dependent, spin configuration resulting in the shift of the cusp temperature with change in $\nu_{ac}$ [34].

From our group experiments in several insulating SG systems [47] and from other people's results in insulating and metallic SG systems [48], we have arrived at a new model for the SG freezing, the cluster phase transition model, which is somewhat intermediate to the above two models and looks more appropriate for SG systems. On the basis of our model we have modified the SG phase diagrams of Figs. 4(d), 5(a) and 16(a), and the modified phase diagrams are given in Fig.21 ; ($T_{CF}$ = cluster formation temperature, CF = cluster formation). Thus according to our model though the clusters are formed owing to the frustration existing in the lattice, they are formed in the material's otherwise paramagnetic state itself below a certain temperature $T_{CF} > T_C$ or $T_{SG}$ or $T_{LF}$ or $T_{TF}$. Above $T_{CF}$, the system is truly paramagnetic and χ follows true Curie-Weiss behaviour [34]. Also though the clusters are present the SG freezing does not occur cluster by cluster (the smallest cluster freezing in the last during cooling) but owing to the cluster-cluster interaction the components ($\vec{S}'_x, \vec{S}'_y, \vec{S}'_z$) of the effective cluster spin $\vec{S}'$ (cluster magnetisation $M_c = g'\mu_B S'$, $g'$ = the cluster g-factor, $\mu_B$ = Bohr magneton) of all size clusters freeze simultaneously at well defined temperatures, $T_C$ or $T_{M1}$ or $T_{M2}$ or $T_{SG}$ or $T_{LF}$ or $T_{TF}$; at $T_C$, $\vec{S}'_z$ is ordered and $\vec{S}'_x, \vec{S}'_y = 0$ (on average ( uniform precession) ), at $T_{M1}$, $\vec{S}'_z$ is ordered and $\vec{S}'_x, \vec{S}'_y$ are frozen in SG configuration, etc. Thus the SK model or GT model calculations still hold good but with the individual spin $\vec{S}$ replaced by $\vec{S}'$. Above $T_{CF}$ (Fig.21), $\tilde{J}_0, \tilde{J}$ represent parameters for interaction between the individual spins ($\vec{S}_i, \vec{S}_j$) while below $T_{CF}$ they represent parameters for interaction between the effective cluster spins ($\vec{S}'_i, \vec{S}'_j$). Even in the infinite range model (SK or GT), clusters are formed owing to the Gaussian nature of interaction which does not allow uniform interaction strength and sign for all the interacting ($\vec{S}_i, \vec{S}_j$) pairs. Unlike $T_{M1}$ or $T_{M2}$ or $T_{SG}$, which decrease with H, $T_{CF}$ vs. H is complex (Sec. II.2).

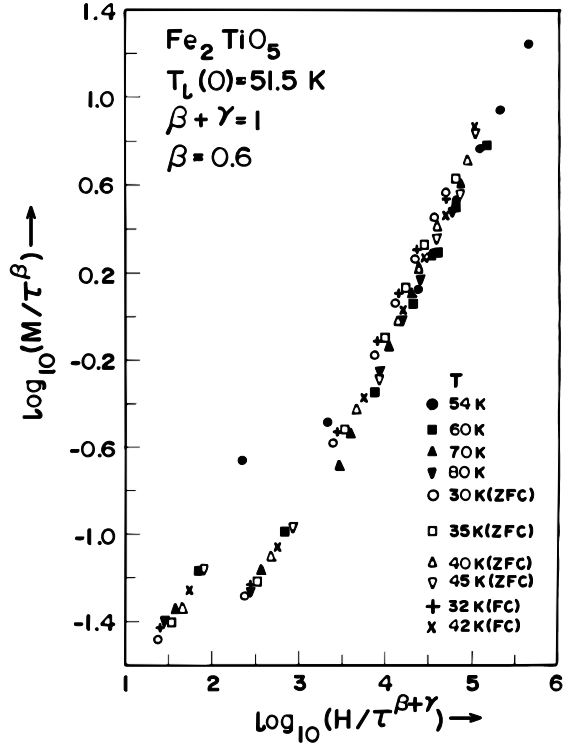

Fig.19. Static scaling observed in $Fe_2TiO_5$. $\tau = |T - T_{LF}(0)|/T_{LF}(0)$; $T_{LF}(0) \equiv T_{LF}(H=0)$.

Figure 22 shows the magnetic hysteresis curve recorded for anisotropic SG $Fe_2TiO_5$ [34]. Even at T ~ 84K (T > $T_{LF}$) M - H behaviour is nonlinear indicating that the system is not truly paramagnetic even at such a high temperature as compared to $T_{LF}$ ($T_{LF}$ = 51.5K for H = 0 and goes down with increasing H). At T ~ 84K, a small remanence (a very small hysteresis loop around H = 0) is also found to be present confirming the above statement. Further confirmation is obtained from the system's M vs. H, M/H vs. H, M/H vs. $H^2$ behaviour (Fig.23) and the fact that though above $T_{cr}$, $\chi$ (i.e. $\chi_{dc}$ (= M/H)) follows the Curie-Weiss behaviour, $\chi$= C/(T + θ), both C and θ are found to be H dependent (Fig.23(b)). The true Curie-Weiss $\chi$-behaviour (C, θ independent of H) is found in this system only above 650K [33]. Thus $T_{CF}$ = 650K. Fig.24 shows the IRM vs. T and M vs. H (T = 77K (a), 290K (b) and 365K (c)) curves for $Co_xZn_{1-x}FeCrO_4$, x ≅ 0.5, ($T_{M1}$ ~ 150K, $T_C$ ~ 230K (Fig.11(B))) [18]. These curves too confirm the presence of clusters (short range order) above $T_C$ (nonlinear M-H curve at 365K). Fig. 25 shows Mössbauer spectra recorded for this system [18] and as seen there considerable spectral broadening exists above $T_C$ (enhanced Mössbauer linewidth). Spectra show normal paramagnetic behaviour only for T ≳ 400K. Thus Figs. 24, 25 indicate the presence of clusters above $T_C$ with $T_{CF}$ ~ 400K (Sec. II.2). Mössbauer spectra recorded in presence of H confirm this conclusion [44]. Thus according to our model, the SG transition is a phase transition, though a cluster phase transition, and since magnetic viscosity (time effects etc.) exists this transition can be called as a phase transition controlled by dynamics. The presence of several ground states below $T_{SG}$ explains the absence of any specific heat anomaly at $T_{SG}$. Actually, the nonlinear susceptibility, $\chi_{nl} = \chi - \chi(0)$, is expected to diverge at $T_{SG}$ but experimentally it is difficult to measure this quantity [34,

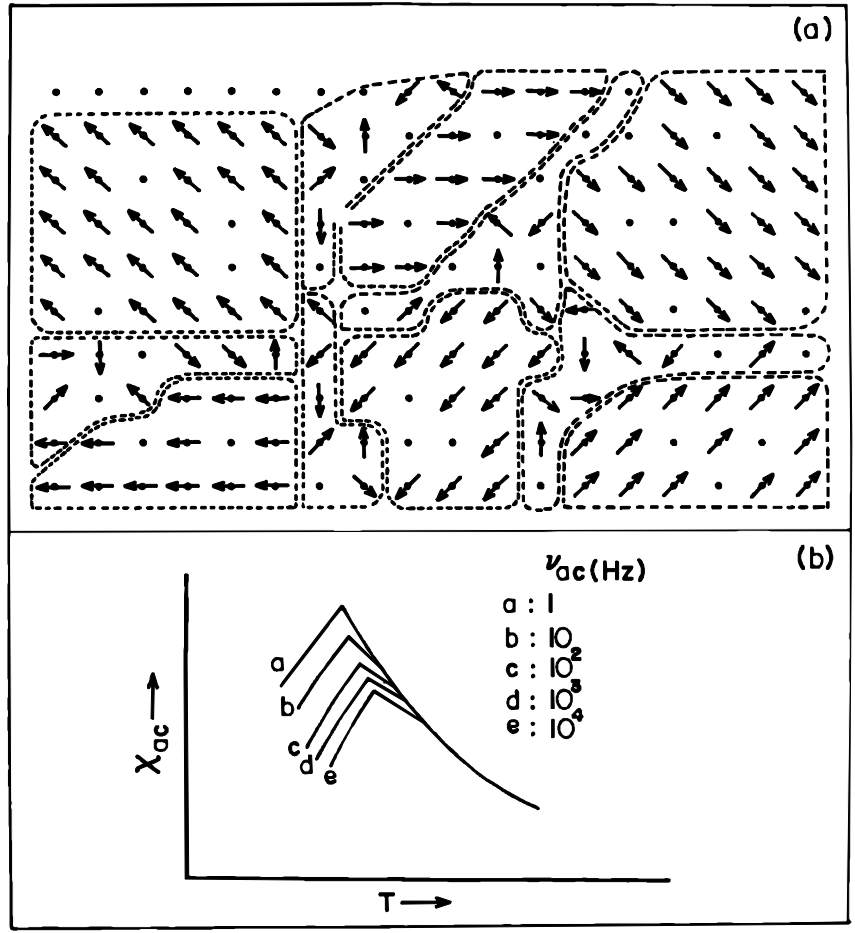

Fig.20. Schematic representation of randomly frozen clusters in a spin glass lattice (a) and applied a.c. frequency, $v_{ac}$, dependence of the cusp (peak) temperature observed in a $\chi_{ac}$ vs. T measurement (b).

39]. In the literature systems with various names have been reported like SG, cluster SG, cluster glass, reentrant SG, SG like system, frustrated system, 'good' SG, 'controversial' SG, semi disordered/ disordered system (SG), etc [43, 49]. *However a closer analysis of their data indicates that they all follow the cluster phase transition (CPT) model (Fig.21)*, though the details of their behaviour somewhat may differ owing to the difference in their cluster size distribution. For example in recent Au-Fe (undoped and Sn doped) Mössbauer experiments [49], the transferred hyperfine field [50] (THF), at the $^{197}$Au and $^{119}$Sn nuclei, vs. T curves have been found to show a kink at the same temperature, $T_{M1}$, at which the $^{57}$Fe hyperfine field [50] (HF) vs. T curve shows. This has been interpreted [49] as an evidence against the GT model since if the neighbouring (surrounding) Fe spins' T′-components froze randomly in a SG configuration at $T_{M1}$, then the $^{197}$Au, $^{119}$Sn - transverse THFs would have averaged out to zero and no kink would have been observed in THF vs. T curves at $T_{M1}$. This result strongly supports the CPT model according to which it is not the individual spin ($\vec{S}$) but the effective cluster spin ($\vec{S}'$) whose T′-components randomly freeze at $T_{M1}$. Thus inside the clusters the frozen T′-components of all $\vec{S}$ are parallel at $T_{M1}$ (and below) giving rise to a nonzero transverse THF. The fact that the external field (H ≠ 0) Mössbauer experiments rule out any global ordering (correlation), like global parallel alignment, for Fe spin's ($\vec{S}$'s) T′-components for T ≤ $T_{M1}$ [3, 49] is also consistent with the CPT model (Fig.21) since the random freezing of ($\vec{S}'_x, \vec{S}'_y$) for T ≤ $T_{M1}$ means that no global ordering for $\vec{S}_x, \vec{S}_y$ could exist. Thus according to our model, the three transitions at ~ 250K, 123K and 42K in Au- 19% Fe-2% Sn system [49] respectively represent the $T_{CF}$ -, $T_C$ - and $T_{M1}$ - transitions of Fig.21. From the above arguments it can also be easily seen that below $T_{M2}$ ( ~ 15K [2, 3]) whereas according to the GT model THFs (total (transverse + longitudinal)) should be zero, giving paramagnetic $^{197}$Au, $^{119}$Sn - Mössbauer spectra, as per CPT model THFs will be nonzero. Experimental results [49] can be clearly seen to favour the CPT model. Similar conclusion is also obtained from an analysis of the $^{119}$Sn - THF *vs*. T behaviour of CuCrSnS$_4$ spinel system [48].

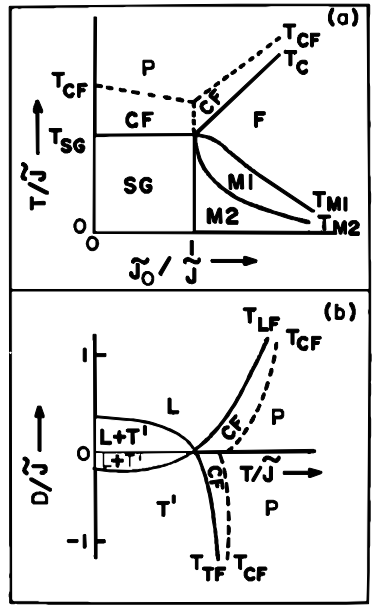

Fig.21. Modified spin glass phase diagrams. Symbols are described in the text. The diagrams are valid for both the metallic as well as nonmetallic systems.

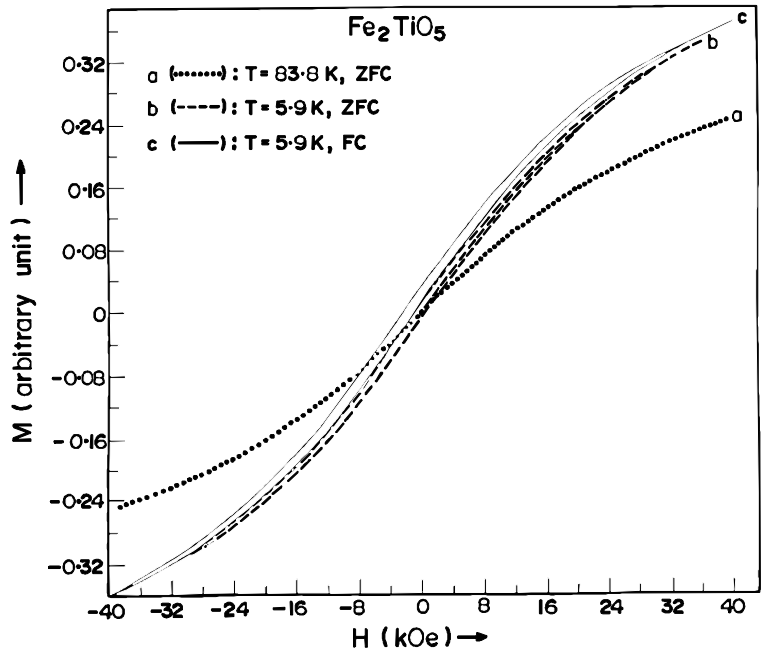

Fig.22. Fe$_2$TiO$_5$ magnetic hysteresis loop for the ZFC system ((a) and (b)) and FC system (c) (FC from 85K in +40 kOe) at few typical temperatures.

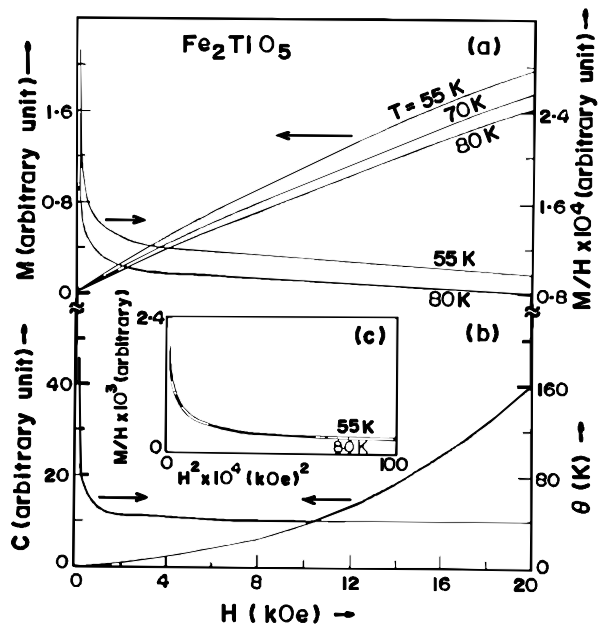

Fig.23. M, M/H, C and θ *vs.* H and M/H *vs.* $H^2$ for $Fe_2TiO_5$ at few typical temperatures. For C and θ, T > $T_{cr}$.

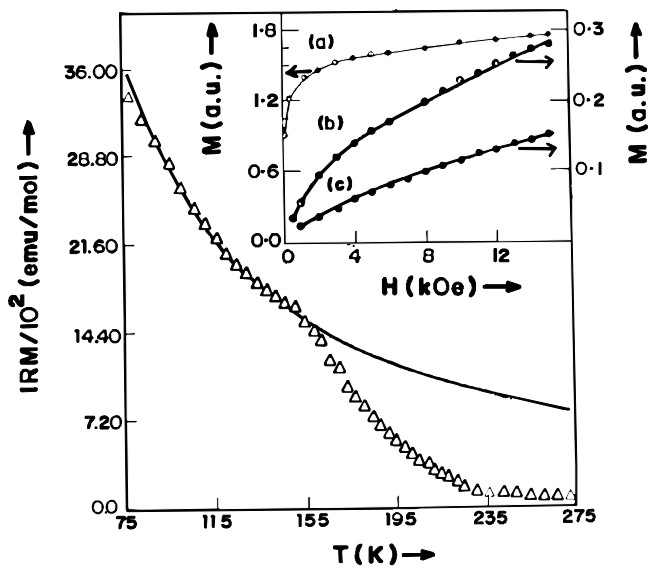

Fig.24. IRM *vs.* T and M *vs.* H (T = 77K (a), 290K (b) and 365K (c)) in $Co_xZn_{1-x}FeCrO_4$ (x ≅ 0.5). For IRM measurement, H = 15 kOe has been applied to the ZFC sample at 77K and then removed at the same temperature.

Due to space limitation, the phenomenon of entropic disordering [8,14,19] has not been described here. Preliminary attempts exist to study the domain pattern of double reentrant systems for $T_{MI} \leq T \leq T_C$ and see the change in pattern as the lattice is cooled [51]. However it may be noted that the clusters, being probably ~ few hundred Å or less in diameter [47], can not be directly mapped. Finally, the situation regarding the possible SG behaviour of high - $T_c$ (critical temperature) superconductors is discussed in the next part [52].

## II. SPIN GLASS FREEZING AND MECHANISM OF HIGH-$T_c$ SUPERCONDUCTIVITY

### II.1. INTRODUCTION

As discussed in I.1, according to the mean field theory of spin glasses [53], the spin glass (SG) transition is a phase transition occurring at well defined temperatures $T_{SG}$ or $T_{M1}$ or $T_{M2}$, when individual spins, or their components, freeze in random directions. Several people have proposed a 'cluster by cluster freezing', or 'continuous freezing', model according to which the SG freezing ( random freezing of moments) occurs continuously, and not at any defined temperature, as the lattice is cooled and instead of individual spins, it is the spin clusters which freeze. This happens because a SG system has different size clusters present and as the temperature decreases these clusters freeze one by one depending on their size, the smallest cluster freezing in the last [53]. As explained before, on the basis of our group work, and work of several other people, we have arrived at an intermediate model of SG freezing [53, 54], called cluster phase transition (CPT) model, according to which though the clusters are present in SG systems due to the magnetic frustration existing in the lattice [53], they are formed in the material's otherwise paramagnetic state itself at a certain temperature $T_{CF}$, above which clusters do not exist, and the SG freezing does not occur cluster by cluster but all size clusters freeze together (simultaneous freezing) at well defined temperatures, $T_{SG}$ or $T_{M1}$ or $T_{M2}$, due to cluster-cluster interaction. This model is found to be able to explain the experimentally observed behaviour of SG systems quite well [53,54]. In this part we present more details of this model and then show that the SG interactions, as envisaged in CPT model, are responsible for the high $T_c$ (critical temperature) and other behaviours of the newly discovered copper-oxide superconductors.

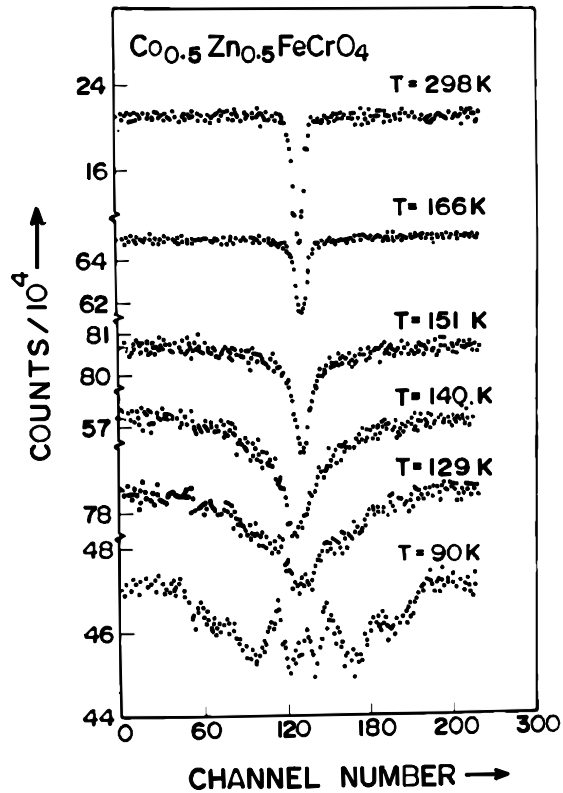

Fig.25. Typical Mössbauer spectra recorded for $Co_xZn_{1-x}FeCrO_4 (x \cong 0.5)$.

## II.2. SPIN GLASS CLUSTER PHASE TRANSITION

As mentioned in I.1, the infinite range Sherrington-Kirkpatrick mean field theory of Ising SG systems, which is based on Edwards-Anderson magnetic frustration model, has been extended by Gabay and Toulouse for vector SG systems [53]. Fig.26 (a) represents their phase diagram describing various symbols which are used in this article; $T_{SG}$ = SG temperature, $T_C$ = Curie temperature, $T_{M1}$=mixed phase 1 (M1) transition temperature, $T_{M2}$=mixed phase 2 (M2) transition temperature, P=paramagnetic phase, SG = SG phase, F = ferromagnetic phase, T=lattice temperature and $\tilde{J}_0, \tilde{J}$ = frustration parameters explained below. This phase diagram, shown for positive x-axis, is symmetrical on negative x-axis and thus F can also be taken as representing ferri- /antiferro-magnetic phases. Writing S for the individual spin value, the magnetisation $M = \frac{1}{N} <\Sigma_i <S_i>_T>_d$ and the Edwards-Anderson parameter $q = \frac{1}{N} <\Sigma_i <S_i>_T^2>_d$ where N=the number of the spins in the lattice, $<--->_T$ means thermal average and $<--->_d$ an average over the frustration disorder i.e. over, below explained, $p(J_{ij})$.

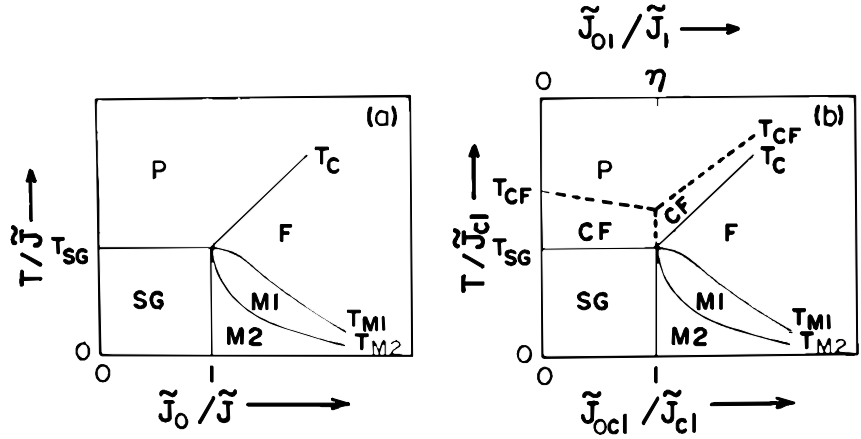

Fig.26. The spin glass phase diagram: (a) Gabay-Toulouse model, (b) CPT model. Details are described in the text; $\eta<<1$, H (external magnetic field)=0.

Taking the longitudinal (z-) direcon as the direction of Weiss field ($\vec{H}_W$) or external magnetic field, $\vec{H}$ (when $H > H_W$), $M_L$, $q_L$ = longitudinal (z-) component and $M_T$, $q_T$ = transverse (x-, y-) components of M, q respectively. For Fig.26 (a) [H = 0], in P, $M_L = M_T = q_L = q_T = 0$; in SG, $M_L = M_T = 0$, $q_L \neq 0$, $q_T \neq 0$; in F, $M_L \neq 0$, $M_T = 0$, $q_L \neq 0$, $q_T = 0$; in M1, $M_L \neq 0$, $M_T = 0$, $q_L \neq 0$, $q_T \neq 0$ and in M2, $M_L = 0$, $M_T = 0$, $q_L \neq 0$, $q_T \neq 0$. Thus whereas in SG and M2 phases the spin components are randomly frozen isotropically, in M1 phase only transverse spin components are randomly frozen. In the mean field SG theory [53], this random direction freezing of spins (SG freezing) is explained by assuming that due to the magnetic frustration in the lattice the exchange integral $J_{ij}$, for interaction between two spins $\vec{S}_i$ and $\vec{S}_j$, does not have an unique value but has a Gaussian distribution shown in Fig.27A (a) and the probability $p(J_{ij})$, for any $J_{ij}$ value, is given by ( Eq.2 ),

$$p(J_{ij}) = \frac{1}{J\sqrt{2\pi}} \exp\left[-\frac{(J_{ij} - J_0)^2}{2J^2}\right], \tag{8}$$

where $J_0$ = mean of the distribution and $J \propto \Delta$, the width of the distribution; $J_0 = \tilde{J}_0/N$ and $J = \tilde{J}/\sqrt{N}$.

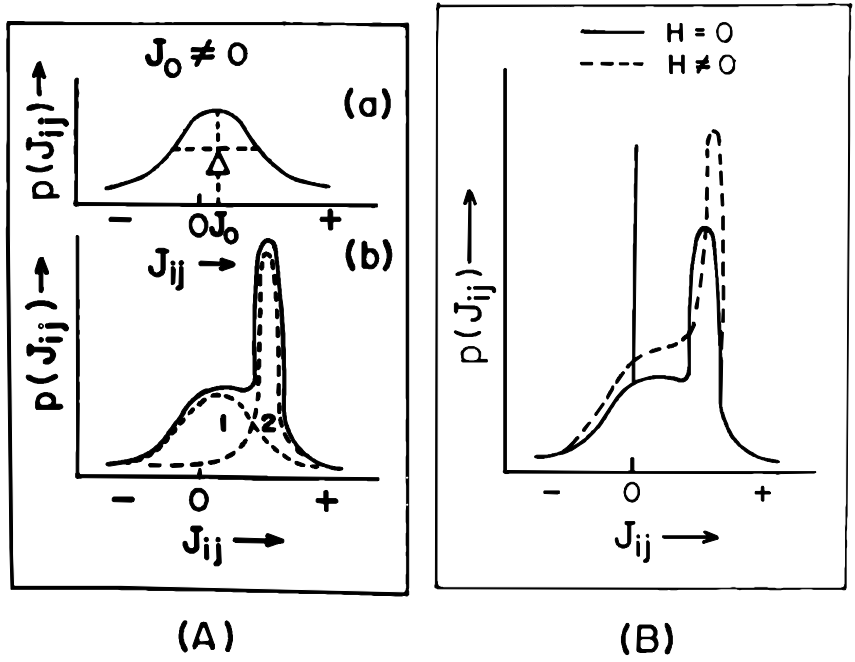

Fig.27. The exchange integral, $J_{ij}$, distribution in a spin glass lattice as per (i) Gabay-Tolouse (GT) model (Fig.A(a)), (ii) CPT model (Fig.A(b),B). Details are described in the text.

In the CPT model, the Gabay-Toulouse phase diagram of Fig.26(a) gets modified to that of Fig.26(b), where CF = cluster formed phase, $T_{CF}$ = cluster formation temperature and $T_C$, $T_{M1}$, $T_{M2}$, $T_{SG}$ represent the various transition temperatures of spin clusters (freezing simultaneously) and not of the individual spins as was the case in Fig.26(a); other symbols are described below. This phase diagram of Fig.26(b) can be obtained by using the $p(J_{ij})$ vs. $J_{ij}$ curve of Fig.27A(b), rather than that of Fig.27A(a). The CPT model thus assumes this modification.

We now have two regions in the $p(J_{ij})$ vs. $J_{ij}$ distribution (marked in Fig. 27A(b)). Using suffices 1 and 2 for various quantities belonging to the two regions respectively, for region 1, $\tilde{J}_{01}/\tilde{J}_1 \lesssim 1$ and for region 2, $\tilde{J}_{02}/\tilde{J}_2 \gg 1$. Thus, as can be seen from Fig.26(a), the spins of region 2 get magnetically ordered at a certain temperature $T_{C2}$. However at $T_{C2}$, these spins order in clusters owing to the presence of region 1 spins which, as seen from Fig. 26 (a), are either paramagnetic or frozen in SG configuration and thus form the cluster boundaries. $T_{C2}$ is thus $T_{CF}$. Due to magnetic frustration, the cluster-cluster exchange interaction too has a probability distribution of the form of Eq.(8). Writing the exchange integral, for interaction between the two cluster spins $\vec{S}_{icl}$ and $\vec{S}_{jcl}$, as $J_{ijcl}$, we have,

$$p(J_{ijcl}) = \frac{1}{J_{cl}\sqrt{2\pi}} \exp\left[-\frac{(J_{ijcl} - J_{ocl})^2}{2J_{cl}^2}\right]. \qquad (9)$$

Thus the SG, F, M1, M2 phases in Fig. 26(b) belong to the clusters and not to the individual spins as was the case in Fig.26(a). Consequently all the discussions given before concerning M, q are still valid but with $S_i$ (individual spin), N replaced by $S_{icl}$ (cluster spin), $N_{cl}$ where $N_{cl}$ = the number of the clusters in the lattice. In other words the Gabay-Toulouse calculation results [53] can still be used provided the individual spin $\vec{S}$ there is replaced by the cluster spin $\vec{S}_{cl}$. The $T_{CF}$, using physical arguments, can be written as follows.

(a) For $\widetilde{J}_{ocl}/\widetilde{J}_{cl} \lesssim 1$,

$$T_{CF} = T_{SG} + \frac{\alpha(\widetilde{J}_{av})^m}{\beta + \gamma(\widetilde{J}_{01}/\widetilde{J}_1)^n} \quad . \tag{10}$$

(b) For $\widetilde{J}_{ocl}/\widetilde{J}_{cl} \gtrsim 1$,

$$T_{CF} = T_C + \frac{\alpha(\widetilde{J}_{av})^m}{\beta + \gamma(\widetilde{J}_{01}/\widetilde{J}_1)^n} \quad . \tag{11}$$

Here α, β, γ are constants (system dependent), m, n~1, $\widetilde{J}_{av} = (\widetilde{J}_1 + \widetilde{J}_2)/2$, $\widetilde{J}_{oav} = (\widetilde{J}_{01} + \widetilde{J}_{02})/2$, $T_{SG} = \widetilde{J}_{cl} = \widetilde{J}_{av}$, $T_C = \widetilde{J}_{ocl} = \widetilde{J}_{oav}$ and $T_{CF} = \widetilde{J}_{02}$. Thus for clusters to exist $\widetilde{J}_{02} \neq 0$ and the conditions $\widetilde{J}_{01} = \widetilde{J}_{02}$ and $\widetilde{J}_1 = \widetilde{J}_2$ should not be satisfied simultaneously i.e. two regions should exist in Fig. 27A(b). In addition, $\widetilde{J}_{02} \gtrsim \widetilde{J}_1$. The Fig. 26(b) plot can now be understood more clearly. Whereas the upper x-axis scale there refers to the individual spin frustration parameters, the lower x-axis scale refers to the cluster spin frustration parameters and the two x-axes are coupled by a common y-axis. This coupling has been explicitly brought out in Eqs. (10), (11) where $T_{CF}$ is written in terms of $T_{SG}$ and $T_C$ though $T_{CF}$ could be written, and y-axis scale in Fig.26 (b) could be plotted, by replacing $T_{SG}$ and $\widetilde{J}_{cl}$ by $\widetilde{J}_{av}$ and $T_C$ by $\widetilde{J}_{oav}$ also which expresses the $T_{CF}$ plot in terms of individual spin frustration parameters only.

The Fig.27A (b) gives only part of the CPT model $p(J_{ij})$ vs. $J_{ij}$ curve. The complete curve is given in Fig. 27B where we see a peak at $J_{ij} = 0$ when H = 0. This is region 3 and it arises due to the presence of entropic spins [53,55] in the lattice. In a SG, due to magnetic frustration, a $H_W$ distribution exists owing to which for some spins $H_W = 0$ even at T = 0K. These are entropic spins (entropy nonzero even at T = 0K due to a temperature independent up, down spin direction fluctuation) and for them $\widetilde{J}_o = \widetilde{J} = 0$ i.e. in Fig.27B, $\widetilde{J}_{03} = \widetilde{J}_3 = 0$. Thus as per CPT model (Fig.27B), $p(J_{ij}) = \sum_{k=1}^{3} p_k(J_{ij})$ where $p_1(J_{ij})$ is the region 1 p (J$_{ij}$), etc. It is possible to obtain the Fig.27 (B) curve using a plaquette picture [53,55]. An H (Fig.27B dashed curve) removes, or reduces the intensity of, the region 3 peak by decreasing the entropic spin number to zero, or to a small value, as it diminishes the lattice frustration by decreasing the width (i.e. $\widetilde{J}_1$, $\widetilde{J}_2$, $\widetilde{J}_{cl}$) and leaving either undisturbed or somewhat enhanced (effectively) the mean (i.e. $\widetilde{J}_{01}$, $\widetilde{J}_{02}$ and $\widetilde{J}_{ocl}$) of frustration, the effect on width being more pronounced. Therefore $T_{CF}$ vs. H is complex ($\overline{H}$-, system-dependent) (Sec. II.6); a large H can change the nature of spin ordering inside clusters or can align the cluster and cluster boundary spins along its direction etc. and all such effects are also to be considered. Experimental results support the above described CPT model and confirm its validity for metallic, nonmetallic and amorphous systems [53, 54, 56, 57]; for example Au-Fe is discussed in Sec. I.5 [53] (for Au-19% Fe-2% Sn, $T_{CF}$ ~ 250K, $T_C$ ~ 123K, $T_{M1}$ ~ 42 K and $T_{M2}$ ~ 15K) and in $Zn_xCo_{1-x}FeCrO_4$ (x ~ 0.5), $T_{CF}$ ~ 400K, $T_C$ ~ 230 K, $T_{M1}$ ~ 150K and $T_{M2}$ ~ 20 K [53,56, Sec. I.5].

In principle, it is possible to have more regions in Fig.27B, and conditions somewhat different from what is mentioned before for various frustration parameter ratios, but experiments do not support that. Also in CPT model, it is possible to have magnetic domains for $T_{M1} \leq T \leq T_C$, where $M_L \neq 0$, but these domains, as well as the domain walls, will have clusters and cluster boundaries inside them. It is the cluster spins, thus, rather than individual spins, which form the domains and the domain walls.

## II.3. HIGH $T_c$ SUPERCONDUCTIVITY MECHANISM

### II.3.1. Conductivity Mechanism

The system $Fe_3O_4$ (ferrimagnetic below $T_C$ = 858 K, cation distribution $(Fe^{3+})_A [Fe^{2+}Fe^{3+}]_B O_4^{2-}$, $Fe^{2+} \xleftrightarrow{e'} Fe^{3+}$ B-site electron (e′) hopping present) is a typical example where several conductivity types, which could occur in an oxide material, exist in different temperature ranges (Fig.28(a), $\rho$ = resistivity). The conductivity regions in Fig.28(a) are: (i) for $T \leq T_V$ (Verwey temperature, 119K), $\rho$ is very high and experiments show distinct $Fe^{2+}$, $Fe^{3+}$ charge states for B-site Fe ions, (ii) for $T_V \leq T \leq T_m$ ($T_m \sim 350K$), $\rho$ is quite small, experiments show an average charge state of $\sim +2.5$ for either of the two B site ions indicating a fast $Fe^{2+} \Leftrightarrow Fe^{3+}$ fluctuation (hopping conductivity ($\sigma$) region ), (iii) $T_m \leq T \leq T_C$, $\rho$ increases with T now (metallic $\sigma$ region), experiments show a +2.5 limiting average charge state for both the B site Fe ions, (iv) $T_C \leq T \leq T_{wm}$ ($T_{wm} \sim 1100K$), $\rho$ decreases with T again due to cationic rearrangement and (v) $T \geq T_{wm}$, $\rho$ shows a weak metallic behaviour.

In a simple minded picture, oxide material's $\sigma$ can be written as [58] $\sigma = \dfrac{\sigma_0}{T} \exp\left[-\Delta_B / k_B T\right]$ where the energy barrier $\Delta_B$, which can be related to Hubbard on-site repulsion U and band width W [59,60], is the energy needed by an electron to get detached from the shell and make itself conduct in the lattice. In oxides, the presence of negative ions increases $\Delta_B$ (enhanced nearest neighbour-site repulsion V [59,60]). In Fig.28(a) region (i), $\Delta_B > k_B T$ and the system is insulating, in region (ii), $\Delta_B \sim k_B T$ and the system shows hopping $\sigma$, and in region (iii), $\Delta_B < k_B T$ where the system shows metallic $\sigma$; in this latter

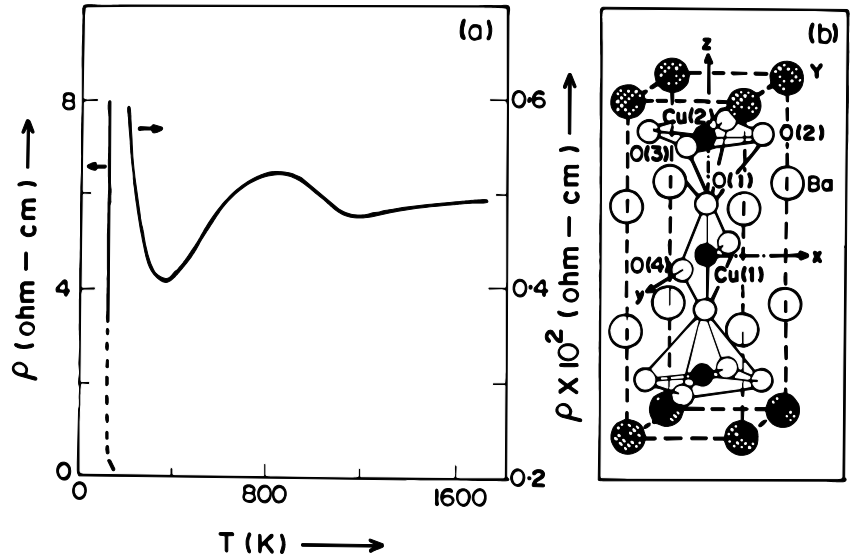

Fig.28. (a) Resistivity ($\rho$) vs. temperature curve of $Fe_3O_4$ system. (b) Unit cell of $YBa_2Cu_3O_7$ system.

($\Delta_B < k_B T$) case, the $\dfrac{\sigma_0}{T}$ term predominates in $\sigma$ expression.

In the high $T_c$ systems, $\Delta_B$ is very small due to strong hybridisation of Cu $3d_{x^2-y^2}$, O 2p orbitals and equal energies of these bands. Consequently even at much smaller temperatures than, for example, $Fe_3O_4$ $T_V$ and $T_m$, high $T_c$ systems acquire metallic nature. In this article we take the system $YBa_2Cu_3O_{7-\delta}$, $\delta = 0$ ($T_c \sim 93K$), as a typical example for discussion (Fig.28(b), unit cell dimensions are: a = 3.8231Å, b = 3.8863Å, c = 11.6809Å [60-63]). Band structure calculations [60-63] show strong Cu(2)-O hybridisation in the Cu-O plane, Cu(1)-O hybridisation along Cu-O chain and Cu(1)-O-Cu(2) hybridisation along the z direction (c axis). The Cu(1), Cu(2) 3d bands and O 2p bands have equal energy and all the three Cu ions in the unit cell have same effective charge state $\dfrac{7^+}{3}$ [60-63]. For charge neutrality the two Cu(2) ions should have charge 2+ and Cu(1) ion 3+ if all the unit cell oxygen ions are in 2- state. Thus a fractional charge state indicates fast $Cu^{2+} \xleftarrow{e'} Cu^{3+}$ fluctuation (or equivalently $Cu^{3+} \xleftarrow{h'} Cu^{2+}$ fluctuation, h′=hole, since e′, h′ symmetry exists in high $T_c$ systems [60, 64]). The above process can also be described as $O^{2-} \xleftarrow{e'} O^{1-}$ (or equivalently $O^{1-} \xleftarrow{h'} O^{2-}$) if one assumes that all the unit cell Cu ions are in 2+ state; in that case the effective charge state of each unit cell O ion is $\frac{13^-}{7}$ [60-63]. The two conductivity descriptions are equivalent, and the fractional charge is 'effective', because, due to hybridisation, it is actually the Cu-O bond charge that fluctuates; the actual process thus is $(Cu-O)^0 \xleftarrow{e'} (Cu-O)^{1+}$ [or $(Cu-O)^{1+} \xleftarrow{h'} (Cu-O)^0$] [60-63], which means that both the above mentioned conductivty descriptions are occurring with equal probability. Thus the conductivity occurs in the Cu(2)-O plane, along the Cu(1)-O chain and along the z-axis, though $\Delta_B$ is larger along the z-axis due to the missing O ions in the Y plane (break in Cu-O-Cu overlap). Just above $T_c$, both hopping σ and metallic σ have been observed depending on the sample treatment [60] which affects $\Delta_B$ by affecting δ (hybridisation, $Cu^{3+}(O^{1-})$ concentration) and sample microstructure. However the superconducting properties are similar [60] indicating the same superconductivity mechanism in both the cases. Higher temperature σ is always metallic [60].

### II.3.2. BCS - Migdal - Eliashberg Prediction

Experiments, especially those involving Y, Ba or Cu sites microscopically, indicate that the superconductivity occurs in the Cu-O plane, along the Cu-O chain and, most probably, along the c-axis by tunneling through the Y plane (Appendix 1). This is consistent with the σ behaviour described above (missing Y-plane oxygen effect). Most of the experiments further indicate the superconductivity to be of BCS [65] type [60,66-69] with singlet s-wave pairing, though mixed s-, d- wave or d- wave superconductivity has also been postulated (Appendix 1); however these latter postulate measurements are mostly magnetic in nature and the need of including the below discussed cluster effects in their analysis is there (see Sec. II. 5). The BCS theory, as modified by Migdal-Eliashberg approach [70-72], has been applied to the high $T_c$ system by Shiina and Nakamura [73]. We rewrite their procedure and result in a way suitable for the present purpose and also describe our result relevant for the present work. They [73] have calculated $T_c$, for $YBa_2Cu_3O_7$ system, vs. λ or S′ for $\mu^* = 0.1$, where $\lambda = 2\int_0^{\omega ph}[\alpha^2 F(\omega)/\omega]d\omega$, $S' = \int_0^{\omega ph} \alpha^2 F(\omega)d\omega$, $\mu^* = \mu/[1+\mu\ln(E_{el}/E_{ph})]$, $\mu = \alpha'\ln[(1+\alpha')/\alpha']$, $\alpha' = (m/2E_{el})^{1/2}(e^2/h)$, h = Planck's constant, e, m = electron charge and mass, the phonon energy $E_{ph} = \hbar\omega_{ph}$, $\omega_{ph} \sim \omega_D$ (Debye frequency), the electron energy $E_{el} \sim E_F$ (Fermi energy), spectral function $\alpha^2 F(\omega) = \alpha^2(\omega)G(\omega)$, $G(\omega)$ is obtained from experimental neutron diffraction data and $\alpha^2(\omega) = C_F$ (F-approximation) or $C_L\omega$ (L-approximation), $C_F$, $C_L$ = constant (adjustable). $T_c$ shows a monotonic increase with λ or S′ without any saturation and the L-approximation gives comparatively larger $T_c$. Our calculations show a pronounced effect of $\mu^*$ on $T_c$, $T_c$ increasing with decreasing $\mu^*$ [71,73]. This will be discussed in a later section.

Theoretically [66,70-72], $\alpha^2 F(\omega) = [N(E_F)\hbar\Omega_{BZ}/2M\omega]<I^2>$ where $<I^2> \sim CE^2_{el}/R^2_c$, $N(E_F)$=electronic density of states at $E_F$, $\Omega_{BZ}$=Brillouin zone volume, M, $R_c$=lattice ion's mass and core radius, and C=constant. Thus an increase in $E_{el}$ enhances $T_c$, via λ or S′ increase, by increasing $\alpha^2 F(\omega)$. Similarly an increase in $\omega_D$ (i.e. in Debye temperature $\theta_D$) increases λ or S′, and so can enhance $T_c$, by increasing the upper integration limit. An $E_{el}$ increase can also increase $T_c$ through a $\mu^*$ decrease. This will be discussed in more detail in the, above referred, later section.

### II.3.3. Cuprate Magnetic Frustration

The origin of oxide magnetic frustration is discussed in Part I [53] and the high $T_c$ system's frustration [74-77] can be understood on that basis, the random distribution of diamagnetic $Cu^{3+}$ ions (or equivalently paramagnetic $O^{1-}$ (S = 1/2) ions in the superexchange path) in the Cu-O plane and chain being the main cause of $Cu^{2+}$ (S = 1/2 [60,76]) moments' frustration. Assuming the $Cu^{2+}$, $Cu^{3+}$, $O^{2-}$ picture of the unit cell, Fig.29 shows nearest neighbour (nn) and next nn competing exchange interactions present at a central $Cu^{2+}$ ion, B in Cu-O plane (Fig. 29a) and C′ in Cu-O chain (Fig.29b). The nn exchange interaction $J_{BF}$ ($J_{C'A'}$) and the next nn $J_{BD}$ ($J_{C'G'}$), both antiferromagnetic through the superexchange paths, become competing when some of the nn ions like A or C or E′ are $Cu^{3+}$. A plane-chain coupling and weak plane-plane coupling through Y-plane, if present below certain temperature, can make the frustration three dimensional in nature as happens for the magnetic ordering in $YBa_2Cu_3O_6$ and related systems [77]. Though the $Cu^{3+}$ location could change due to charge fluctuation (short or long hop case), at any instant always same number of Cu ions are $Cu^{3+}$, randomly located, and so the same average frustration is always present. Similar discussion applies for the $O^{1-}$ induced frustration also.

### II.3.4. Singlet Paired Magnetic Clusters

As discussed in the CPT model description section, a consequence of cuprate magnetic frustration is the presence of magnetic clusters in these systems which are formed in the material's otherwise paramagnetic state ($T_{CF} > T_C$, $T_{SG}$). However these clusters differ from normal SG systems' clusters as they always occur in pairs (Fig.30(a), pair partner 1 has spins up and pair partner 2 has spins down). The two pair partners overlap each other and are coupled by the resonating valence bond (RVB) interaction present in these systems [74,78-80] due to which a spin of cluster 1 forms a singlet pair with a neighbouring spin belonging to the cluster 2. Two cluster pairs are separated by a cluster boundary (Fig.30b) and could have quantisation (spin alignment) axes along different directions. This description implies that $J_{02}S^2 \sim E_{RVB}$, the RVB interaction energy between a spin pair; some theoretical attempts too point in this direction [74,78-80] (Appendix 1). The effect of singlet pairing is that the coupled ion pair cannot cause any Cooper pair breaking via spin spin ( dipolar or exchange coupling induced [81-83]) interaction as the total pair spin, $\vec{S}_{pair}$ = 0. As discussed in earlier section, the picture (Fig.30) remains similar even if $Cu^{3+}$ location could shift due to charge fluctuation or the singlet valence bond resonate among different neighbours of a central ion or $O^{1-}$ presence is assumed in place of $Cu^{3+}$.

There is no evidence for a $T_C$ or $T_{SG}$ or $T_{M1}$ or $T_{M2}$ in high $T_c$ systems. It is difficult to ascertain the presence or absence of these temperatures below $T_c$ due to the diamagnetism of the superconducting state. However experiments do reveal the presence of magnetic clusters in these systems [84-87]. A careful analysis of Mössbauer data shows an average cluster volume of $\sim 5 \times 10^{-20}$ $cm^3$ for $YBa_2Cu_3O_7$ system [86, 87] implying a cluster diameter of ~ 50Å for a spherical cluster or a frustrated area diameter of ~100 Å in Cu-O plane and chain if one assumes the frustration to be confined within a unit cell along the c-axis.

### II.3.5. Cluster Pairing Consequences

For the $Cu^{3+}$, $Cu^{2+}$, $O^{2-}$ unit cell picture, as per earlier given σ description, at any instant on the average out of the three Cu ions one is in $Cu^{3+}$ state, the other is in $Cu^{2+}$ state and the third is the $Cu^{2+}$ ion which has lost its one electron in hopping or itinerant conduction and has become $Cu^{3+}$. The conducting electron (CE) produces a magnetic field, $H_{CE}$, due to its motion, at the ions of the singlet coupled ion pair (Fig.30) which have no spin spin interaction with CE as $\vec{S}_{pair} = 0$; $H_{CE} = \dfrac{ev_{el}}{r^2}$, $v_{el}$ =CE velocity and r = CE-ion distance. Calling an ion of cluster 1 (Fig.30a) as ion 1 and its singlet coupled partner of cluster 2 as ion 2, in the CE absence the magnetic field at ions 1, 2 is, $H_1$, $H_2 = H_w + H_{dip}$, where $\vec{H}_1$ and $\vec{H}_2$ have opposite directions and $H_{dip}$ is the dipolar field produced by the neighbouring spins. In CE presence, $H_1$, $H_2$ change to $H'_1$, $H'_2$ where $H'_1 = H_1 + H_{CE}$ and $H'_2 = H_2 - H_{CE}$ for $\vec{H}_1 \parallel \vec{H}_{CE}$ case. The crystal field ground state of $Cu^{2+}$ ion in high $T_c$ systems is orbital singlet and it is split by an amount $\Delta_1$, $\Delta_2$ by $H_1$, $H_2$ and $\Delta'_1$, $\Delta'_2$ by $H'_1$, $H'_2$; $\Delta_1 = g\mu_B H_1$ etc. (g=Landé's g-factor and $\mu_B$ = Bohr magneton). Thus when CE approaches ion 1, energy emission occurs due to upper level population decrease as $\Delta_1$ increases to $\Delta'_1$. Reverse happens when CE moves away from ion 1. Opposite is the situation for ion 2 since $\Delta'_2 < \Delta_2$. Consequently ions 1, 2 either both absorb energy or both emit energy when CE moves from ion 1 towards ion 2 or vice versa. Similar result is obtained for $\vec{H}_2 \parallel \vec{H}_{CE}$ and other cases. As shown in Appendix 3, the emitted energy is larger than the absorbed energy. The latter has to come from the lattice since moving nonrelativisitic CE does not emit energy quanta. However the emitted energy can go to both, the lattice and the CE. But due to orbital singlet ground state, $Cu^{2+}$ has weak spin lattice coupling. Consequently most of the emitted energy is taken by CE which gives part of this taken energy to the lattice through collisions (e′-lattice ion collisions, e′-e′-lattice ion collisions) and retains part of the energy as $\Delta E_{el}$, increase in $E_{el}$. As the energy absorbed from the lattice by ions 1, 2 is more than the energy given to it by them and CE, there is a net energy absorption from the lattice which reduces lattice ions' vibrational amplitude and enhances $\theta_D$ by $\Delta \theta_D$. An equilibrium $\Delta E_{el}$, and so $\Delta \theta_D$, occurs when any further energy received by CE is given completely to the lattice owing to enhanced collisions at large $E_{el}$. The CE can receive the ions 1, 2 emitted energy through $H_{\mu_{eff,CE}-\mu_{Cu^{2+}}} = 2\, \mu_{eff,CE}\, \mu_{Cu^{2+}}/r^3$ coupling, where $\mu_{Cu^{2+}}$ is the $Cu^{2+}$ ion magnetic moment and $\mu_{eff,CE} = r^3 H_{CE}/2$ is an effective magnetic moment associated with CE; this coupling is an order of magnitude larger than $Cu^{2+}$ spin lattice coupling (Appendix 3) and is similar to the 'spin current mechanism' of conduction electron spin relaxation [88].

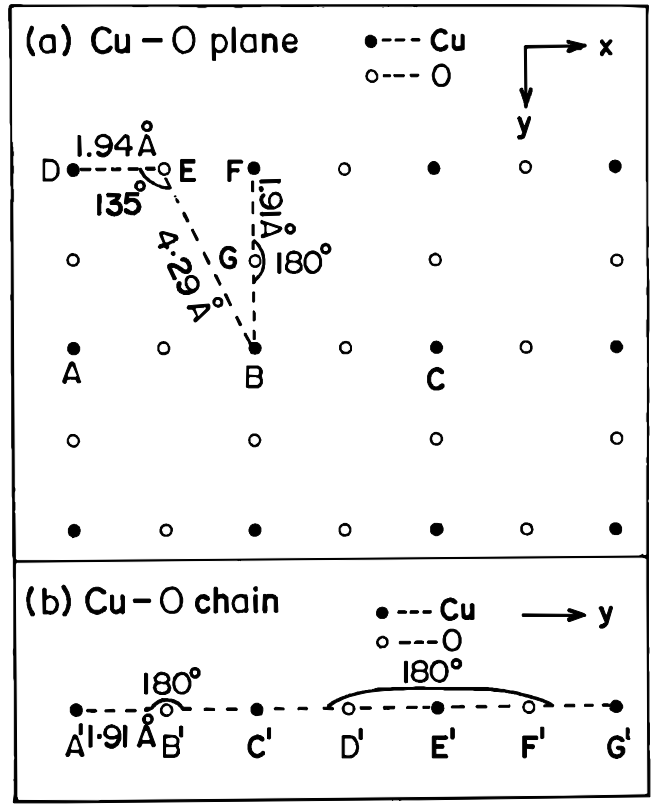

Fig. 29. Origin of magnetic frustration in (a) Cu-O plane and (b) Cu-O chain. A plane-chain coupling causes additional frustration.

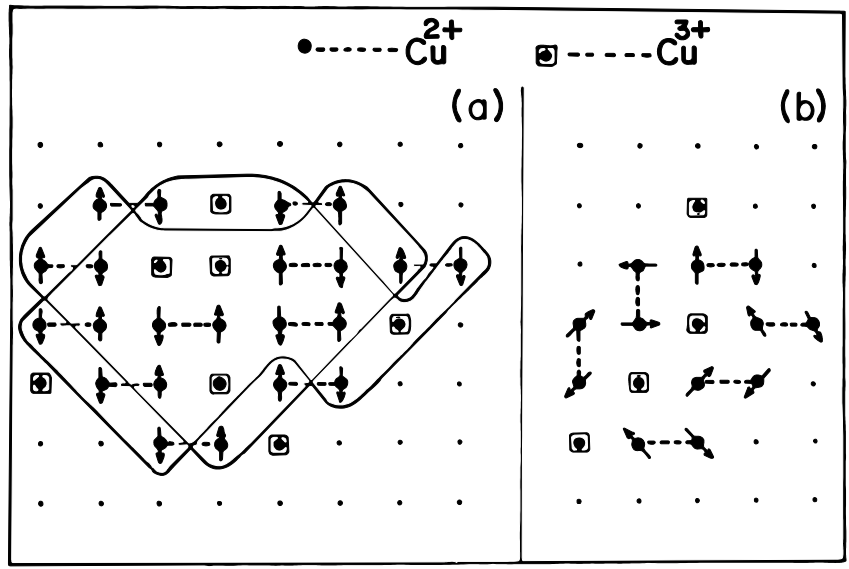

Fig.30. Schematic representation of (a) magnetic cluster and (b) cluster boundary for a high $T_c$ system. The dashed lines represent the RVB coupling between the spins (arrows) and as explained in the text, clusters exist in pairs due to the combined effect of the frustration and RVB interactions; cluster boundaries, compared to clusters, are rich in holes ($Cu^{3+}$ (or equivalently $O^{1-}$); $\tilde{J}_1/\tilde{J}_{01}>1$ (Sec. II.2, II.3.1). In (a), spins of one cluster point in one direction (which is opposite to the partner cluster's spin direction) and in (b), ions (cluster boundary ions) could also be in paramagnetic state (at higher T) (Sec. II.2, II.5).

A careful analysis of several experiments supports the above picture. For instance, anomaly observed at ~ 220 K in $YBa_2Cu_3O_{7-\delta}$ system in several studies like channeling, Mössbauer f-factor, NMR relaxation rate, transport, magnetic and elasticity etc. [60,89-93] is consistent with the above picture (Appendix 3). As discussed earlier, $E_{el}$ increase enhances $T_c$ via $\lambda$ or $S'$ increase and also $\mu^*$ decrease which occurs if $\Delta E_{el}/E_{el} \gtrsim \Delta \theta_D/\theta_D$ as is actually the case (Appendix 3). In this situation, $\theta_D$ increase too enhances $T_c$ via $\lambda$ or $S'$ increase. Appendix 3 gives quantitative details.

Enhanced anomaly at $T_c$ in the above mentioned studies [60,89-93] gives further support to the picture described above. At $T_c$, a fraction of CEs form Cooper pairs (CPs) which have no lattice or e′ collisions (coherence effect) and hence give much less energy to the lattice than CEs. This enhances $\Delta \theta_D$ and the related anomalies. An equilibrium $\Delta E_{CP}$, increase in CP energy, occurs when a CP starts losing coherence due to its enhanced velocity as compared to other CPs.

For cluster boundary ions (paramagnetic state(Sec. II.5)) $H_W = 0$ and $H_{dip}$ is produced by their pair partners. In such a case, as shown in Appendix 3, $\Delta \theta_D = 0$. Thus clusters' presence is necessary for $\Delta \theta_D$ and related anomalies to occur. The above description essentially assumes itinerant σ case. Same result is obtained for hopping σ case or for $O^{1-}$ unit cell picture, but $\Delta E_{el}$ is much larger there.

## II.4. COMPARISON WITH OTHER THEORIES

For a proper comparison of our model (called paired cluster (PC) model (Sec. II.3, [94])), which assumes BCS phonon coupling mechanism for CPs, with other theories, it is essential to understand the BCS theory physical picture. A CE moving in the lattice ($v_{el} \sim 10^8$ cm/s) distorts it, by pulling or pushing the lattice ions, at the sound velocity $v_s \sim 10^5$ cm/s and experiences a frictional motion, causing its own and neighbouring channels' narrowing. Since $v_{el} > v_s$, the distorting CE is far away from the distorted side. However if two other CEs, coming with almost equal velocity from opposite directions (which minimises their Coulomb repulsion and also helps in ~zero momentum pair formation needed for coherence (pair stability) discussed below), reach the distorted positive ion site simultaneously, they can get coupled through the distorted ion's Coulomb field, as in Fig. 31a (A, B, ..... ionic rows, 1,2,..... horizontal channels), provided $E_F - \hbar \omega_{ph} \lesssim E_{el1}, E_{el2} \lesssim E_F + \hbar \omega_{ph}$. Such a coupling removes lattice distortion, minimises the system's thermodynamic free energy and provides a frictionless motion to the coupled e′-e′ pair (CP) as that can drift without distorting the ionic row C with ~thermal velocity. However some friction still remains as the B, D rows could still distort. That too can get removed if several CPs move together (Fig.31b, various CPs are: aa′, bb′, cc′ and dd′). This actually happens in BCS theory as the CPs overlap each other (CP diameter >CPs' separation) and drift together with a common drift velocity (~ thermal velocity). It is this coherence (phase coherence) which is responsible for CPs' frictionless motion and inhibits any CP-e′,

CP-lattice ion collision. If the CPs are in singlet s state, as in original BCS theory, they are diamagnetic and avoid any magnetic scattering.

Fig.31a shows the shortest CP diameter case. In general a CP can have several rows between its partners and it is CPs' overlap and lattice coupling which prevent the lattice distortion. The two CEs can also get coupled, without any earlier distorting CE, by interacting through the distortion sites created by each other when felt by them at a retarded stage (Fig.31c).

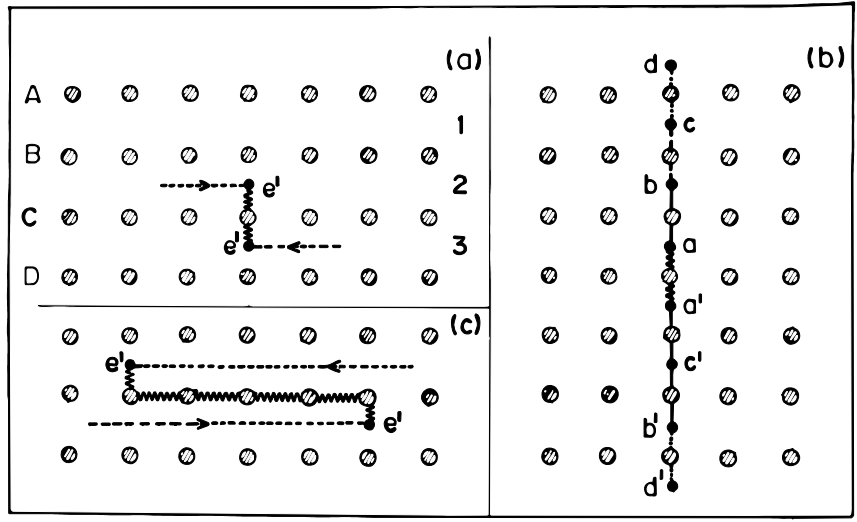

Fig. 31. Schematic representation of BCS model. Details are described in the text. The picture is valid for both, the itinerant σ case as well as the hopping σ case.

The above physical picture and the experiments mentioned before [60, 66-69] give a natural support to our model. Such a support is missing for other theories. For instance several problems exist with the d - density wave model, van Hove singularity based and antiferromagnetic spin fluctuation based mechanisms (Appendix 1) and theoretical difficulties exist [95-97] for the electronic and magnetic pairing theories and the spin bag theory [64, 98] (Appendix 1). The latter has physical problems too. Firstly, whereas our model is specific to high $T_c$ systems due to the RVB coupling presence [74], the spin bag formation should occur in other systems also where charge fluctuation and ionic moment exist like $Fe_3O_4$ discussed before. Secondly, as mentioned earlier, magnetic clusters are present in cuprates and if one tries to form spin bag with clusters, quasiparticle pair gets scattered due to cluster boundaries. The need for conducting pairs' phase coherence (overlap and its translational invariance) is discussed before. Such a coherence is absent in Hund's coupling mechanism theory [99] (Appendix 1) and also in RVB produced superconductivity mechanism [74, 78-80] though some attempts exist in this direction [100]. However, the RVB superconductivity mechanism assumes a spin and charge separation of conducting species and associated holon condensation which lacks evidence [101] (Appendix 1). In our model RVB coupling exists but by itself is not the cause of superconductivity. It is the interaction of the RVB coupled ions, in the cluster and at cluster boundaries, with CEs and phonon coupled CPs which gives rise to high $T_c$ and other properties of cuprates. Observations like almost absence of magnetic pair breaking, $T_c$ dependence on carrier concentration, measuring current, etc. [60, 66-69, 102-104] can be understood in a natural way in our model. The bipolaron superconductivity mechanism [105] too lacks phase coherence. If the coherence is brought by assuming large bipolarons, then the coupling becomes weak. Also physically it is difficult to visualise the polaron formation in metallic σ region, though its formation in insulating σ region is easily understood [58].In addition, the experimental result constraints are against the bipolaron model (Appendix 1). The existing 'gap' approach has its own problem [106] (Appendix 1, Sec. II.5, II.6). Further it basically assumes the neutron diffraction result, but the failure of neutron diffraction in detecting clusters in frustrated lattices is known [56, 107]. The channeling data, discussed before and in Appendix 3, too can not be understood on the basis of the present gap approach. For instance if any quasiparticle pair (phonon or nonphonon coupled) formed gap was responsible for the break in the ionic vibrational amplitude vs. T (i.e. $\theta_D$ vs. T) curve at $T_c$ and at some higher temperature, like T ~ 220K in $YBa_2Cu_3O_7$ case, then such a break should have been seen even in conventional (elemental, A15) superconductors at $T_c$. However no such lattice mode behaviour (hardening or softening ) has been observed in them. In our model it is the interaction of the CEs, CPs with the singlet coupled ion pairs which is responsible for the above mentioned breaks. Thus it is the ignorance of this interaction which makes people feel that the high $T_c$ superconductivity is either nonphononic or non-BCS type or completely new in character and the quasiparticle pairs are distinctly different from the CPs. The nature of superconductivity, s-wave or mixed s-, d- wave or d- wave, can be handled within the BCS model due to the singlet pairing of magnetic clusters and the retarded nature of the interaction between the CP forming CEs. At this stage it is proper to point out the limitations of our model. For any quantitative comparison between theory and experiment, the approximations used in the BCS theory calculations obviously provide the limitations of our model. These approximations are discussed in [73]. Further, doubts have been raised about the correctness of the BCS theory for underdoped superconductors and the importance of non-BCS pairing gap approach stressed for them [108]. However we show in the following section that even their results can be understood easily on the basis of our model. But before that, we would like to point out that the interactions described in our model are important even for a non-BCS gap case as thay affect any gap. The model is thus basically general in nature and can be extended to include additional interactions too if present in some specific cases. We now discuss below the pseudogap originating due to the interactions described in our model.

### II.5. PSEUDOGAP ORIGIN AND CONSEQUENCES

A large number of experimental and theoretical works have appeared in the literature concerning pseudogap presence in cuprates' normal state [Sec. II.4, 106, 109-121] and it has been suggested that a competent high $T_c$ superconductivity theory should be able to explain the pseudogap's origin and nature [115, 117, 118, 120]. The pseudogap has been observed in overdoped, underdoped and optimally doped superconductors, its presence is independent of the number of Cu-O layers in the cuprate unit cell and similar gap behaviour has been found for the electron doped and the hole doped superconductors [109, 110, 112, 113, 116, 118]. It has d-wave symmetry, develops, as one cools the lattice, in the superconductor's normal state at a temperature much above $T_c$ and remains present upto $T_c$ [106, 109-115, 118, 120]. Below $T_c$ the observed superconducting state (SS) energy gap properties are abnormal like almost temperature (T) independent gap width, presence of states in the gap, gap's nondisappearance at $T_c$, unconventional gap symmetry, which could be a mixed s-, d- wave or anisotropic s-wave or d- wave symmetry, etc. [60, 111, 112, 114, 118, 120, 122-141] (Appendix 1). Several theories exist for pseudogap's origin [Sec. II.4, 106, 115, 117, 119, 121, 142-145 ], but they are found wanting in some respect or other [Sec. II.4, 106, 111, 112, 117, 118, 120, 146] (Appendix 1). In this section we show that our (PC) model of high $T_c$ superconductivity is able to explain the pseudogap's origin and nature, the above described abnormal SS energy gap properties and also several other experimental results, like the uncommon temperature variation of NMR spin relaxation rate [147], which are related to the pseudogap and SS energy gap behaviours.

According to our model (Sec. II.3), as the magnetically frustrated cuprate lattice cools magnetic clusters are formed at a temperature $T_{CF}$ much above $T_c$. Different experimental techniques may sense $T_{CF}$, i.e. the clusters' presence, at somewhat different temperatures depending on their characteristic measuring times and the nature of their interaction with the clusters. Thus in practice $T_{CF}$ is either the temperature at which the clusters are formed, if the technique can sense the clusters immediately on formation, or the temperature at which the clusters are felt by the measuring technique. The clusters formed at $T_{CF}$ exist as singlet coupled pairs. As discussed before, for $T_c \leq T \leq T_{CF}$ the conducting electrons (CEs) interact with the singlet coupled ion pairs in the cluster and the cluster boundaries by a process described in Sec. II.3. This interaction enhances the CE energy, $E_{el}$, to $E_{el} + \Delta E_{el}$ and the lattice Debye temperature, $\theta_D$, to $\theta_D + \Delta\theta_D$. However for a nonzero $\Delta E_{el}$ (or $\Delta\theta_D$) a Weiss field, $H_W$, is required to exist at the ion pair site. Such a Weiss field is present for the cluster ions (CIs) since $T_{CF}$ = cluster $T_C$ (Curie temperature ) but is absent for the cluster boundary ions (CBIs) which are in the paramagnetic state at higher temperatures, $T_c \lesssim T \leq T_{CF}$.

Thus for higher T only CIs contribute to $\Delta E_{el}$ and $\Delta\theta_D$ enhancements. At lower temperatures, $T \lesssim T_c$, when the CBIs freeze, which happens to be the case as is discussed later on, in spin glass (SG) state (T < $T_{SG}$, their SG temperature) [Sec. II.2, II.3] (Fig. 30b), a nonzero $H_W$ develops at their site which enables them to contribute to $\Delta E_{el}$ and $\Delta \theta_D$. However at any T, unlike the CIs' $H_W$, which has an unique value given by the cluster magnetisation Brillouin function temperature dependence, the CBIs' $H_W$ has a distribution, ranging from zero (for certain ions) to a maximum value (for certain other ions), since in the SG state such a $H_W$ distribution is known to exist [53, 148]. We will mention more about this at an appropriate place. As has been discussed in Sec. II.3, $\Delta E_{el}$ and $\Delta \theta_D$ enhancements cause $T_c$ increase by affecting several $T_c$ influencing parameters.

For $0 \leq T \leq T_c$, Cooper pairs (CPs) exist and also those CEs which have not formed CPs. In this T range therefore both the CEs and CPs interact with the singlet coupled cluster and cluster boundary ion pairs due to which $E_{el}$ is increased to $E_{el} + \Delta E_{el}$, the CP energy, $E_{CP}$, to $E_{CP} + \Delta E_{CP}$ and the lattice $\theta_D$ to $\theta_D + \Delta \theta_D$. However the $\Delta \theta_D$ for $T \leq T_c$ is much larger than the $\Delta \theta_D$ for $T_c \leq T \leq T_{CF}$ (Sec. II.3). We will concentrate below on $\Delta E_{el}$ and $\Delta E_{CP}$ enhancement effects.

The effect of above mentioned $\Delta E_{el}$ and $\Delta E_{CP}$ enhancements, which may also be called as $\Delta E_{el}$ and $\Delta E_{CP}$ scatterings, is to cause a redistribution of the filled electronic density of states (DOS), $D_F(E_{el})$, at any temperature below $T_{CF}$. This gives rise to a pseudogap in the electronic DOS distribution (redistributed $D_f$ ($E_{el}$) vs. $E_{el}$ curve) at $T_{CF}$ which persists at lower temperatures and gets superimposed over the SS energy gap (BCS energy gap) below $T_c$. The superimposition is responsible for the abnormal SS energy gap properties. Also since $T_{CF} = T^*$ (pseudogap temperature), different experimental techniques may sense $T^*$ (i.e. pseudogap formation), like $T_{CF}$, at somewhat different temperatures (Appendix 1). We describe the details below for different temperature ranges.

### (i) $T_c \leq T \leq T_{CF}$

In this temperature range only $\Delta E_{el}$ scattering is present as CPs do not exist. For obtaining the redistributed $D_f(E_{el})$ vs. $E_{el}$ curve the quantities needed are, the electronic DOS vs. $E_{el}$ curve which would have existed if no $\Delta E_{el}$ enhancement effect was present, T, $E_F$ (Fermi energy), $\Delta E_{el}$ and $N_P$, the percentage of $CE_S$ for which $\Delta E_{el}$ enhancement occurs.

The calculation of $\Delta E_{el}$ is given in Sec. II.3, Appendix 3 and it ($\Delta E_{el}$) depends on T and $E_{el}$ through the magnetic fields $H_W$, $H_{dip}$, $H_{CE}$ (and also $H_{CP}$ for $T < T_c$) and relaxation times $\tau'$ and $\tau$. The $H_W$, to a good approximation, can be obtained by assuming a Brillouin function temperature dependence for the cluster magnetisation; this is since $H_W$ does not exist for CBIs at high T. However the calculation of $\tau'$, $\tau$ is generally not possible (Appendix 3) and only an order of magnitude estimate can be made for them. Using such estimates, $\Delta E_{el}$ is found to have an oscillatory dependence on T. For a given $E_{el}$, as T decreases below $T_{CF}$ $\Delta E_{el}$ first increases upto a certain temperature, then becomes almost constant over a temperature range below which it decreases. Such a behaviour arises owing to the exponential dependence of the ionic level Boltzmann population on $H_W/T$ and the $H_W$'s Brillouin function T dependence due to which $\Delta H_W/\Delta T$ decreases as T decreases, tending to zero as $T \to 0$. Approximately $T/T_C (\equiv T/T_{CF}) \sim 0.5$ could be taken as the temperature at which the $\Delta E_{el}$ is maximum.

At a given T, $\Delta E_{el}$ depends on $E_{el}$. The nature of this dependence is governed by the $E_{el}$ dependence of $\tau$ which can not be exactly calculated (Appendix 3) and therefore various possibilities are to be examined. Thus $\Delta E_{el}$ vs. $E_{el}$ could be flat ($\Delta E_{el}$ independent of $E_{el}$) or linear or exponential depending on the magnitude and $E_{el}$ dependence nature (flat, linear or exponential) of $\tau$. Physically, an exponential dependence of $\tau$ on $E_{el}$ is more probable as is seen for the paramagnetic ion concentration dependence of the ionic spin-spin relaxation time [81, 149].

The $N_P$ is another needed quantity. At high T when CBIs do not contribute to $\Delta E_{el}$ enhancement, $N_P \sim 50\%$ if one assumes almost equal number for CIs and CBIs i.e. almost equal total occupied volumes for the clusters and the cluster boundaries in the lattice. At low T, CIs' contribution to $\Delta E_{el}$, and $\Delta E_{CP}$, diminishes and finally vanishes. However CBIs contribute to $\Delta E_{el}$, and $\Delta E_{CP}$, at these temperatures. But owing to $H_W$ distribution only few (a fraction of) CBIs, which have right magnitude $H_W$ at the working T, contribute. Thus $N_P$ is small and has no systematic T variation. $N_P \sim 50\%$ is also an approximation since the cluster boundary volume is not known. It is therefore better to use the experimental results as guidelines for estimating $N_P$ when comparing experiment and theory. This is true for other parameters also.

Thus for getting the theoretical results, the parameter values used have been obtained by using both the theoretical considerations (Appendix 3) and the experimental results [112, 126-128, 132, 150] as guidelines. The two guidelines have been found to give consistent estimates. Further though the results discussed below are for typical parameter values, calculations have been done for other parameter values also and the results obtained are similar in nature. This is mentioned at appropriate places describing the parameters used and the results obtained.

Fig. 32 shows a typical result where the electronic DOS, D ($E_{el}$), is plotted against $E_{el}$. The dotted curve is the total (filled plus empty) DOS, $D_t$ ($E_{el}$), (quadratic, free electron approximation [66], Sec. IV.1), the dashed curve is the density of filled states, $D_f$ ($E_{el}$), which would have existed at temperature T if there was no $\Delta E_{el}$ enhancement present, the full line curve is the density of filled states redistributed, i.e. redistributed $D_f(E_{el})$ or $D_{fr}(E_{el})$, due to $\Delta E_{el}$ enhancement effect and in Fig. 32(a) the dash-dot curves a, b are the same as the full line curve there but have been obtained for different $N_P$ values; N($E_{el}$), the number of CEs at energy $E_{el}$, = 2 D($E_{el}$). Fig. 32(a) shows the results for T ~ $T_{CF}$ case. For YBa$_2$Cu$_3$O$_7$, $T_{CF}$ ~ 220K and $E_F$ ~ 310 meV (Sec. II.3, Appendix 3). For other cuprates also $T_{CF}$ and $E_F$ are of similar order. We therefore assume these values for the present calculations. However we have done calculations for the higher and lower values of $T_{CF}$ and $E_F$ also and the results obtained are similar to the results given here. For Fig. 32(a), the various parameters used in the calculation are, T = 200K (T ~ $T_{CF}$), $E_F$ = 310 meV, $\Delta E_{el}$ ($E_F$), the value of $\Delta E_{el}$ at $E_{el} = E_F$, = 300 meV, an exponential dependence, of the form given below, is assumed for $\Delta E_{el}$ on $E_{el}$ and $N_P$ = 50% (full line curve), 40% (curve a) and 60% (curve b). The $\Delta E_{el}$ vs. $E_{el}$ form is: $\Delta E_{el} = (\Delta E_{el})_0$ [1-exp (-$\alpha E_{el}$)], where $(\Delta E_{el})_0$, ~ $E_{el}$ ($E_F$), = 300 meV, $\alpha$ = 0.7 (meV)$^{-1}$ i.e. a quick rising, fast saturating $\Delta E_{el}$ variation with $E_{el}$ is assumed. The $\Delta E_{el}$ ($E_F$) given above matches with the theoretical estimate (Appendix 3) and with that deduced from the experimental results [112]. Owing to Pauli principle, the CE energy can be enhanced from $E_{el}$ to $E_{el} + \Delta E_{el}$ only if the energy state at $E_{el} + \Delta E_{el}$ is either partially or completely empty. Due to this the full line curve (Fig. 32a) does not start from the origin but merges with the dashed curve at $E_{el}$ ~ 3 meV; this effect can be seen more clearly in the below discussed Fig. 32 (b). Similar behaviour exists for the curves a, b (Fig. 32a) also but is not shown in the figure for clarity near $E_{el}$ ~ 0 region. If instead of an exponential $E_{el}$ dependence, an $E_{el}$ independence is assumed for $\Delta E_{el}$, then the full line, a, b curves start from the origin but the other results, like their nature etc., are similar. Similarly, for a linear $E_{el}$ dependence of $\Delta E_{el}$, though the full line, a, b, curves merge with the dashed curve at $E_{el} \gg$ 3meV, the other results obtained are similar to those shown in the figure (Fig. 32a) (Appendix 4). The full line, a, b, curves are calculated by adding the scattered CEs, at any energy state, to the CEs which had remained there after the scattering from that state had taken place, if the state was not completely empty (Appendix 5, Part IV). A comparison of the full line, a, b curves shows the $N_p$ value's effect.

As mentioned above, for calculating the Fig. 32(a) $D_{fr}$ ($E_{el}$) curves we have used the dotted $D_t$ ($E_{el}$) curve. However we have done calculations for other types of $D_t$ ($E_{el}$) curve also which instead of increasing with $E_{el}$ in the $E_{el}$ ~ $E_F$ region, as in Fig. 32(a), either remain flat ($E_{el}$ independent) or decrease with increasing $E_{el}$. The results obtained in all the cases are similar to what is shown in Fig. 32(a). A pseudo-energy gap (pseudogap) is clearly seen in the $D_{fr}$ ($E_{el}$) vs. $E_{el}$ distribution in Fig. 32(a) at the Fermi surface. Similar pseudogap is seen in Fig. 32(b) also where all descriptions, and parameter values, are same as those of Fig. 32(a) except T = 100K, $\Delta E_{el}$($E_F$) = 150 meV and $N_P$ = 40%. Thus Fig. 32(b) describes a situation where $T_c$ < T < $T_{CF}$. The full line curve in Fig.32(b) meets the dashed curve at much higher $E_{el}$ than the curves of Fig. 32(a) due to a smaller $\Delta E_{el}$ owing to which the Pauli principle does not allow CEs from the states below the energy state marked E to be scattered since the empty (partially or fully) states occur from the energy state marked F onwards in the $D_t$ ($E_{el}$) vs. $E_{el}$ distribution.

Thus we see that a pseudogap appears in the $D_{fr}(E_{el})$ vs. $E_{el}$ distribution (Fig. 32) for $T \leq T_{CF}$. This pseudogap persists for $T \leq T_c$ also since $\Delta E_{el}$ scattering is present below $T_c$ too where, in addition, $\Delta E_{CP}$ scattering also occurs whose effect we will discuss in a later section (Sec. II.5(ii)). The $D_{fr}(E_{el})$ vs. $E_{el}$ distribution of Fig. 32 when translated into the tunneling conductance curves [151, 152] shows agreement with the experimental results [112] like the shape, size, location and T dependence of the pseudogap. Since the pseudogap arises due to the CE ($\Delta E_{el}$) scattering, it has d-symmetry (actually mixed d-, p- ($d_{x2-y2}$-, $p_x$-, $p_y$-) symmetry which experimentally can not be distinguished from a d- symmetry) in the normal state. This is shown by the experiments [111, 112, 114, 115, 118, 120, 143]. Below $T_c$, $\Delta E_{CP}$ scattering and SS energy gap are also present and this situation gap symmetry we will describe in the later section (Sec. II.5(ii)). Our results are also consistent with various other experiments which conclude that the pseudogap arises in the charge excitation spectrum [113, 117, 118, 120].

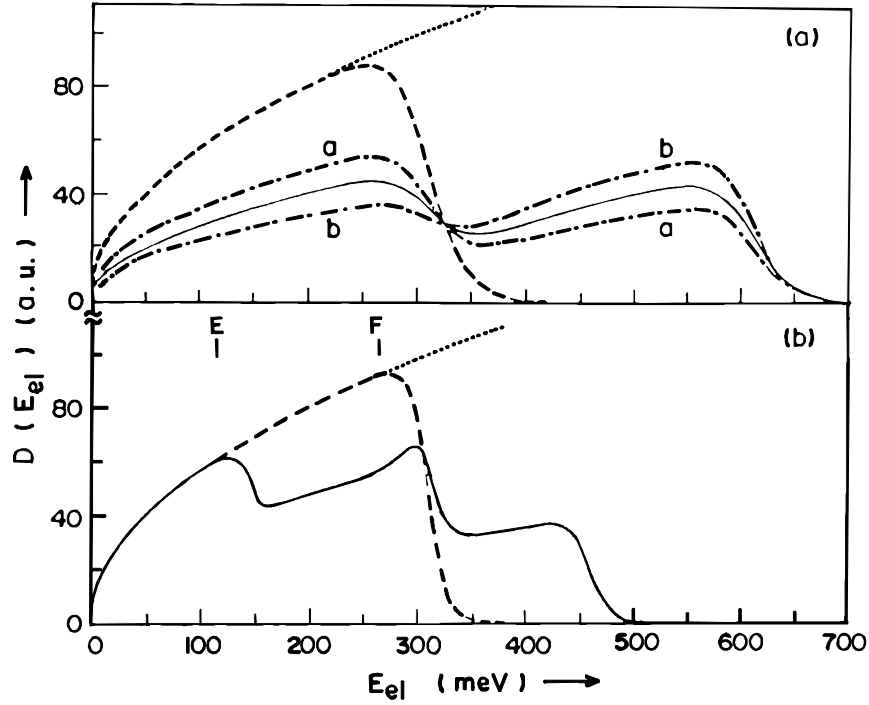

Fig. 32. Dependence of the electronic density of states, $D(E_{el})$, on electrons' energy, $E_{el}$, for $T > T_c$; $T_c$ = critical temperature, a.u. = arbitrary unit. Details are described in the text.

**(ii) $0 \leq T \leq T_c$**

As mentioned above, below $T_c$ the SS energy gap (BCS energy gap [66]) appears which is present alongwith the $\Delta E_{el}$ scattering and additional $\Delta E_{CP}$ scattering (enhancement) near $T_c$; this is since, as will be discussed later, $\Delta E_{CP} \sim 0$ for T away from $T_c$. This ($0 \leq T \leq T_c$) case is more complex and the results obtained are summarised in Figs. 33, 34. We first discuss the Fig. 33 results where $T \sim T_c$ (Fig. 33a) and $T < T_c$ (Fig. 33b). In Fig. 33(a), 33(b), $E_{el}$ variation is shown for $D_t(E_{el})$ (dotted curve), $D_f(E_{el})$ (dashed curve), $D_{fr}(E_{el})$ (full line curve) and redistributed $D_e(E_{el})$, the redistributed density of empty states (shown only for above the BCS energy gap region), (dash-dot curve). The dash-double dot curve is an extension of the $D_t(E_{el})$ curve which would have existed if there were no SS peaks and energy gap present. In a tunneling conductance measurement, the SS energy gap at T, $2\Delta(T)$, is measured either by the SS peaks' (P, Q in Fig. 33a) separation or by A, A′ (Fig. 33a), the points at which the dash-double dot curve intersects the SS energy gap edges, separation [112, 150]. The curve a′(b′) is obtained by subtracting the unoccupied (empty) side ($E_{el} > E_F + \Delta(T)$, $\Delta(T)$ = SS energy gap width/2) portion of the curve a(b) from the $D_t(E_{el})$ curve. Writing $D_t(E_{el}) = D_f(E_{el}) + D_e(E_{el}) = D_{fr}(E_{el}) + D_{er}(E_{el})$, curve a′(b′) is actually the empty side portion of the $D_{er}(E_{el})$ (i.e. redistributed $D_e(E_{el})$) curve which gets merged with the dotted curve.

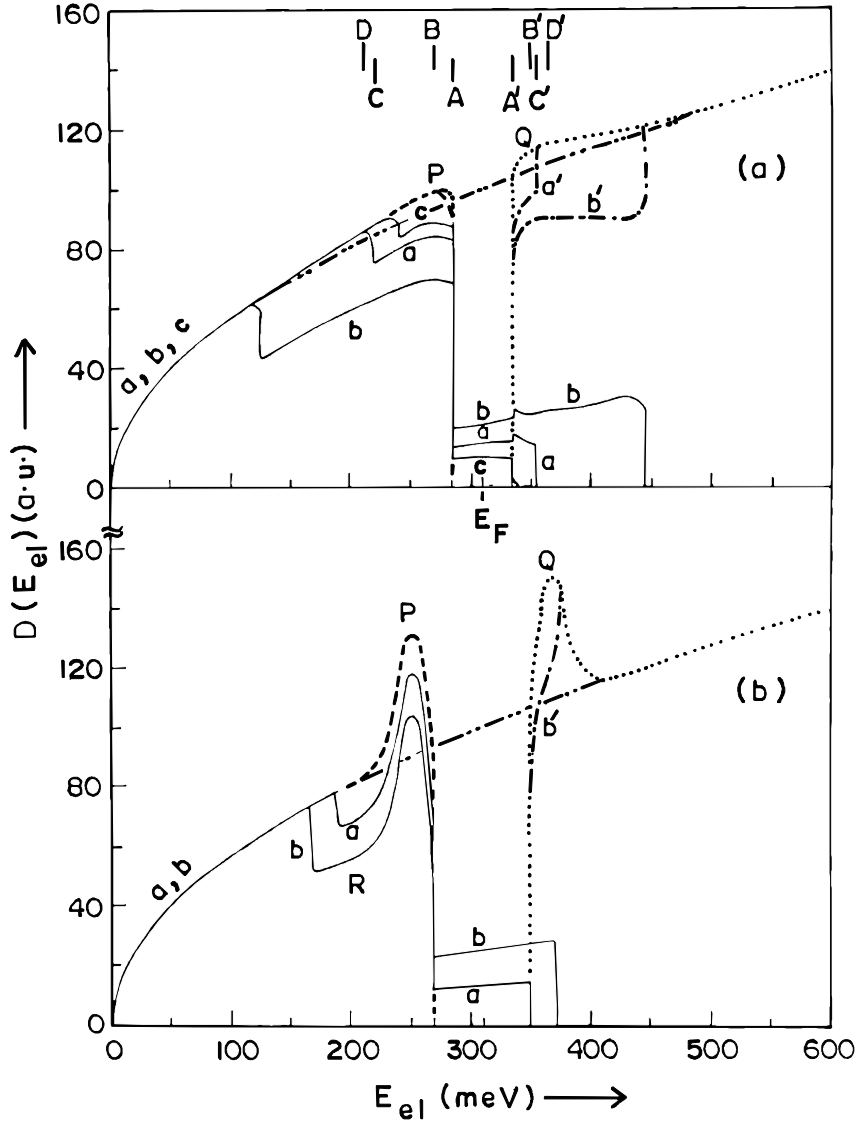

Fig. 33. Dependence of the electronic density of states, $D(E_{el})$, on electrons' energy, $E_{el}$, for $T \lesssim T_c$; $T_c$ = critical temperature, a.u. = arbitrary unit. Details are described in the text.

As before, various Fig. 33 parameters have been chosen using theoretical and experimental (Appendix 3, [112]) considerations. In Fig. 33(a), T = 75K, $E_F$ = 310 meV, $2\Delta(T)$ = 50 meV (i. e. T spread of $D_f(E_{el})$ curve's tail portion ~ $2\Delta(T)$) and, as before, an exponential dependence is assumed for $\Delta E_{el}$ on $E_{el}$. This (exponential dependence) assumption has been used in Fig. 33(b), 34 calculations also. For the curve a, $\Delta E_{el}(E_F)$ = 70 meV and $N_P$ = 15%, for curve b, $\Delta E_{el}(E_F)$ = 165 meV and $N_P$ = 30%, and for the curve c, $\Delta E_{el}(E_F)$ = 50 meV and $N_P$ = 10%. As mentioned before, CPs also exist at this T and calculations show that for them $\Delta E_{CP} \sim \Delta E_{el}(E_F)$ for all $E_{el}$ and, for obvious reasons, they have the same percentage scattering as the CEs (i.e. $(N_P)_{CP} = (N_P)_{CE}$). We have done Fig. 33(a) calculations assuming these but even for somewhat different values of $\Delta E_{CP}$ and $(N_P)_{CP}$, the results obtained are similar mainly because the CPs are much smaller in number than the CEs (Appendix 3) and so have only small influence on the results. Also in these calculations, as well as in Fig. 33(b), 34 calculations, it has been assumed that CEs can have scattered states in the SS energy gap as happens in the case of the inelastic electron scattering and the proximity effect [151, 153-156]. However even without this assumption similar results are obtained when the redistributed DOS curves are translated into the tunneling conductance curves [112]. This happens due to the existence of CEs' scattered states outside the SS energy gap. The assumption of nonzero CE scattering in the SS energy gap, however, looks justified as the experimental results too support the presence of states in the SS energy gap [112, 126, 127, 132]. It has also been assumed that CPs do not get scattered in the SS gap. Though even without this assumption the results similar to those of Figs. 33, 34 are obtained, mainly owing to the much smaller percentage of CPs, the assumption is physically justified since the interactions causing $\Delta E_{el}$, $\Delta E_{CP}$ scatterings are not pair breaking; they rather enhance $T_c$, $\Delta(0)$, the T =0K SS energy gap width/2. As has been explained before at these temperatures (T ~ 75K and below, when $T_{CF}$ ~ 220K) CIs' contribution to $\Delta E_{el}$, $\Delta E_{CP}$ decreases and CBIs' contribution becomes significant due to their SG freezing. However CBIs have $H_W$ distribution and a relatively smaller maximum $H_W$. The $N_P$ and $\Delta E_{el}$ are therefore smaller at these temperatures.

Fig. 33(a) results can now be understood. The curves a, a′ show that in this case due to the filled and empty states' redistribution owing to $\Delta E_{el}$, $\Delta E_{CP}$ scatterings, both, the filled side and the empty side, SS peaks, P and Q, have got modified in location, shape and size. Their peak positions have shifted from B, B′ to D, D′ showing P, Q separation enhancement. Similarly A, A′ separation has increased to C, C′ separation. Since, as mentioned before, $\Delta(T)$ is measured by these separations [112, 150], the effect of these separation enhancements, which are relatively larger at higher T due to large $\Delta E_{el}$, $\Delta E_{CP}$, is to cause an increase in the value of $2\Delta(T)$. This, to a good extent, nullifies the $\Delta(T)$'s BCS increasing - T decrease. As a result the observed $\Delta(T)$ shows weak T dependence [112]. The $\Delta(T)$'s increasing- T decrease is also partially compensated by the following effect, more effective at higher T, of the $\Delta E_{el}$, $\Delta E_{CP}$ scattering induced DOS redistribution. If 2f(T) is the fraction of broken CP CEs (quasiparticles), i.e. broken CPs' fraction, near the SS energy gap edge at T (2f(0) = 0, 2f($T_c$) = 1), then [108, 152] approximately $\Delta(T) \propto$ [1-2f(T)]. Since due to the DOS redistribution, f(T) is reduced near the gap edge, as $D_{fr}(E_{el}) < D_f(E_{el})$, near the filled side gap edge, shows (Figs. 33, 34), $\Delta(T)$ gets enhanced. The f(T) decrease near the gap edge is a consequence of the fact that alongwith the normal CEs, broken CP CEs also get scattered due to the $\Delta E_{el}$ scattering. Experimentally [112] the SS energy gap observed below $T_c$ does not vanish at $T_c$ whereas BCS $\Delta(T)$ =0 at $T_c$. This happens because at $T_c$ the earlier described $\Delta E_{el}$ scattering induced normal state pseudogap is present. Since like the pseudogap just above $T_c$ the SS energy gap just below $T_c$ has $\Delta E_{el}$ scattering induced states in the gap, there is no conspicuous change in the gap's nature (width, shape, size, location) at $T_c$. The gap vanishes only at $T_{CF}$. These results agree with the experiments [112].

The SS energy gap symmetry can also be understood now. The observed SS gap is the BCS energy gap modified by the $\Delta E_{el}$, $\Delta E_{CP}$ scattering induced effects (pseudogap effect). As discussed before, the $\Delta E_{el}$ scattering induced pseudogap has a d-wave symmetry. The BCS gap in principle can have any symmetry [Sec. II.4, 60, 124, 139, 141]. However physically for the cuprates (high $T_c$, anisotropic crystal structure), an anisotropic s-wave symmetry is more likely for the BCS energy gap since non-s wave superconductors are known to have low $T_c$ [157] and s-wave CP coupling is more stable against magnetic perturbations (paramagnetic ions, electric current, magnetic field). Thus the observed SS energy gap below $T_c$ is expected to have a mixed anisotropic s-, d- wave symmetry. Experimentally it is not possible to distinguish easily between an anisotropic s-wave or d-wave or mixed s-, d- wave symmetry and therefore experimentalists favour one symmetry or the other [Sec. II.3, 60, 111, 112, 114, 118, 120, 122-141] (Appendix 1).

The curves b, b′ (Fig. 33(a)) show that in this case both, the P and Q, peaks have disappeared as a result of the DOS redistribution. Thus even though we are below $T_c$, experiments, like the tunneling experiments, may indicate a normal state, with a pseudogap, for the system suggesting a lower $T_c$ value. Also such a DOS redistribution may be one of the reasons for the NMR spin relaxation rate coherence peak's absence in cuprates since P peak's presence is necessary for the coherence peak's existence [147]. The curve c (Fig. 33(a)) shows that for this case only the peak P has got modified (height, width decreased, position shifted) and the peak Q has remained undisturbed. In this case also the measured $\Delta(T)$ will be larger than the $\Delta(T)$ which would have been obtained if no DOS redistribution existed. Such a $\Delta(T)$ enhancement, as mentioned before, partially compensates for any $\Delta(T)$'s T decrease. All these results, and the nature (height, width, shape, location) of the modified P, Q peaks and of the observed SS energy gap, agree with the experiments [112].

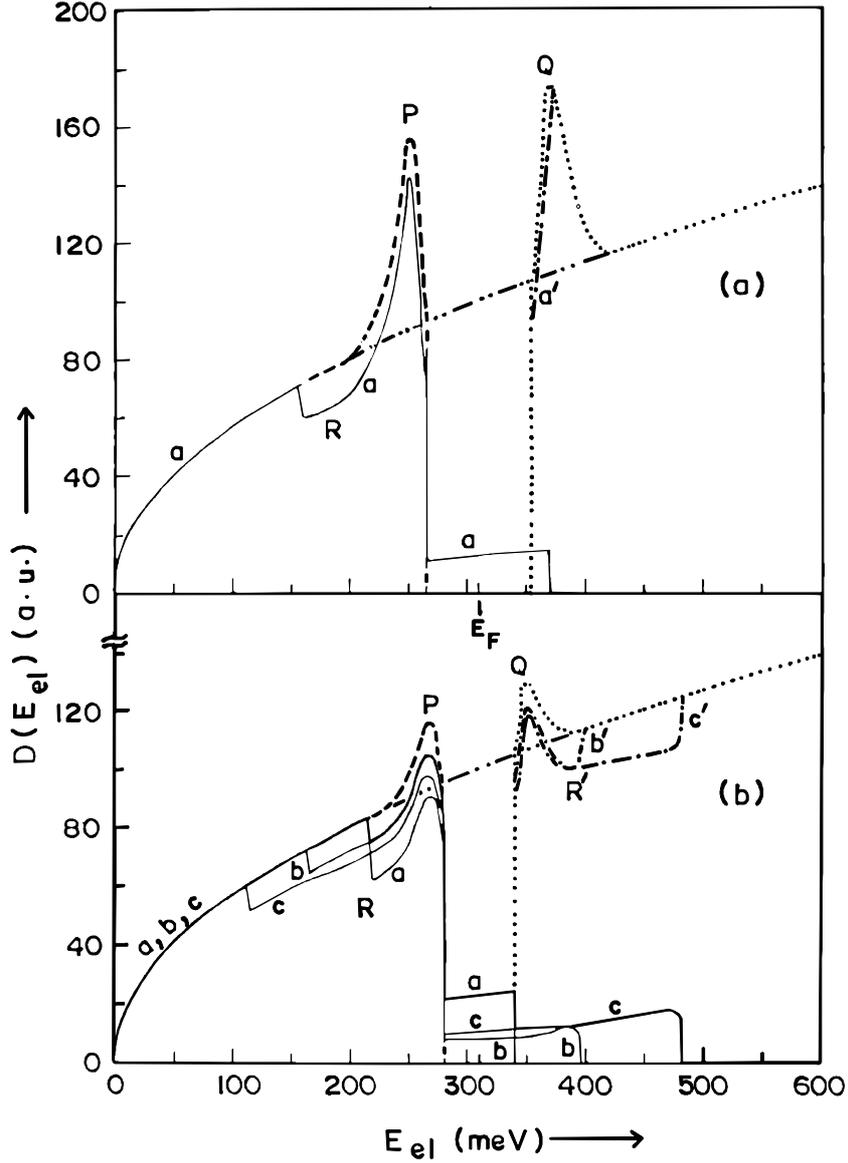

Fig. 34. Dependence of the electronic density of states, $D(E_{el})$, on electrons' energy, $E_{el}$, for $T < T_c$; $T_c$ = critical temperature, a.u. = arbitrary unit. Details are described in the text.

In Fig. 33(b), T = 50K, $E_F$ = 310meV and $2\Delta(T)$ =80 meV (i.e. $D_f(E_{el})$ curve's tail portion's T spread < $2\Delta$). For the curve a, $\Delta E_{el}(E_F)$ = 80 meV, $N_P$ = 15% and for the curve b, $\Delta E_{el}(E_F)$ = 100 meV, $N_P$ = 30%. For both the cases, $\Delta E_{CP} \sim 0$ ( < 1 meV), both for CIs and CBIs, and therefore the $\Delta E_{CP}$ scattering effect is negligible. For the case of curve a, due to the DOS redistribution, owing primarily to the $\Delta E_{el}$ scattering, peak P height, width decrease, peak Q remains undisturbed and dip R (a part of the curve a) appears on the P peak side. This dip is more pronounced for the curve b where both P and Q have got modified. These results and the relative modified peaks' and dips' heights, widths, locations, shapes etc. agree with the experimental results [112]. Also for both the cases, the measured $\Delta(T)$ is greater than the non-DOS redistributed case's $\Delta(T)$. This, as before, makes $\Delta(T)$'s T dependence very weak.

We can now discuss the results of Fig. 34 where the full line curve, dotted line curve etc. have the same meaning, and the curves a′, b′, c′ are obtained in the same way, as in Fig. 33 and $T << T_c$ (Fig. 34(a)), T near $T_c$ (Fig. 34(b)). We first discuss Fig. 34(a) results where the parameters used are, T = 4.2K, $E_F$ = 310 meV, $2\Delta(T)$ = 90meV (i.e. $D_f(E_{el})$ curve's tail portion's T spread ( $\sim 0$) << $2\Delta$), $\Delta E_{CP} \sim 0$ (< 1 meV), $\Delta E_{el}$ ($E_F$) = 105 meV and $N_P$ = 15%. As before, the DOS redistribution's effect is to enhance $2\Delta(T)$. However since $\Delta(4.2K) \sim \Delta(0)$, this enhancement means enhanced $2\Delta(0)/k_BT_c$ ratio ($k_B$ = Boltzmann's constant). Thus to some extent the large $2\Delta(0)/k_BT_c$ ratio (> BCS weak coupling limit) for cuprates can be understood from the DOS redistribution. However for accounting it fully, the effect of somewhat large $\lambda$, $S'$ parameters, defined in Sec. II.3, is to be taken into account. The relative heights, widths, locations, shapes etc. of the dip and the peaks obtained here (Fig. 34(a)) agree with the experimental results [112].

For Fig. 34(b), T = 70K, $E_F$ = 310 meV, $2\Delta(T)$ = 60 meV (i. e. $D_F(E_{el})$ curve's tail portion's T spread $\sim 2\Delta(T)$), $\Delta E_{CP} \sim \Delta E_{el}(E_F)$ for all $E_{el}$ and $(N_P)_{CP} = (N_P)_{CE}$. This figure (34(b)) shows some specific cases. For the curve a, $\Delta E_{el}(E_F)$ = 60 meV, $N_P (\equiv (N_P)_{CE})$ = 25%, for the curve b $\Delta E_{el}(E_F)$ = 115 meV, $N_P$= 10% and for the curve c, $\Delta E_{el}(E_F)$ = 165 meV, $N_P$ = 15%. Whereas for the curve a's case, peak P disappears and peak Q remains undisturbed, for the case of curves b, b' both P, Q almost symmetrically decrease and a dip R' appears on the peak Q side. This dip is broadened for the case of curves c, c' where the peak P has very small amplitude. These results agree with the experimental results [112]. As before these DOS redistributions too enhance $\Delta(T)$, partially compensating for $\Delta(T)$'s T decrease and making $\Delta(T)$'s T dependence weak. R, R' having locations dependent on P, Q (gap edges) positions are experimentally seen [112, 127].

## II.6. CONCLUSION

In conclusion, the cuprate properties can be understood on the basis of what is described above. This is true even for those properties which have not been specifically discussed here. As an example we explain below the anomalous T behaviour of the NMR spin relaxation rate ($1/T_1$) [147]. The coherence peak's absence in $1/T_1$ vs. T, near $T_c$, has already been explained earlier. The other $1/T_1$ vs. T anomalies can be understood in a similar way. For instance, $1/T_1$ decreases with decreasing T due to lattice cooling and shows drops at $T^*$ ($\gg T_c$, $\sim$ 220K for $YBa_2Cu_3O_7$), $T_c$ [147]. The drop at $T_c$ is seen in conventional (elemental, A15) superconductors also and results from the CPs' presence. Assuming same explanation for cuprates' $T_c$ drop, the drop at $T^*$ is not yet properly understood. For example one of the explanations assumes CPs' formation at $T^*$ itself [144]; CPs grow in size with decreasing T, developing overlap and phase coherence below $T_c$ to give superconductivity. Experiments do not support this explanation [Sec. II.4, 118, 120] (Appendix 1). Also CPs' stability is difficult to understand without their phase coherence (Sec. II.4). If one assumes too tightly bound a CP at $T^*$, then the problems of lattice distortion, lattice stability and polaron formation in cuprates come into the picture (Sec. II.4). In our model, $T^* = T_{CF}$ and the lattice excessively cools at this temperature due to the interactions described in our model (Sec. II.3, Appendix 3). This excessive cooling is responsible for the drop in $1/T_1$ at $T^*$. The drop at $T_c$ results from both the CPs' presence and the excessive lattice cooling at $T_c$. Thus the anomalies of the $1/T_1$ vs. T behaviour, like the Mössbauer f-factor vs. T anomalies (Sec. II.3), can be understood on the basis of the channeling r.m.s. lattice ions' vibration amplitude vs. T data [89]. Similarly $T^*$ (pseudogap temperature) vs. H can also be understood on the basis of our model. Apart from influencing the frustration parameters ($\widetilde{J}_0, \widetilde{J}$), H can also affect (Sec. II.5, Appendix 3) $\alpha$, $\tau$, $\Delta E_{el}$, $\Delta E_{CP}$, $N_P$, electron motion (via magnetoresistance), RVB singlet coupling of Fig. 30 by tilting (aligning) the moments towards (along) $\vec{H}$, etc. Therefore $T^*$ vs. H behaviour is complex ($\vec{H}$-, system- dependent). However, a large enough $\vec{H}$ can appreciably weaken the RVB singlet coupling by, as mentioned before, tilting (aligning) the spins towards (along) it and thus making it necessary to go to a lower temperature to get the RVB singlet coupling back. Thus the pseudogap temperature ($T^*$) gets decreased with H for large H. This has been experimentally observed (Appendix 1 (v)). In the same way, the Zeeman scaling relation observed recently between the T$\sim$ 0K pseudogap closing field, $H_{pg}(0)$, and the pseudogap temperature $T^*$ (Appendix 1(v)) too can be understood easily on the basis of our (PC) model according to which the RVB singlet coupling energy (Fig. 30) $\sim k_BT^* \sim \mu H_{pg}(0)$. Thus $\mu H_{pg}(0) \sim k_BT^*$, where $\mu = g\mu_BS$ and S=1/2, g=2 for $Cu^{2+}$ ions which have orbital singlet ground state in the cuprate crystal field (Appendix 3, Sec. II.3.5). This is what has been observed experimentally as Zeeman energy scaling relation. (Physically $H_{pg}(0)$ is large enough to align all the cluster and cluster boundary ions along $\vec{H}_{pg}(0)$ and break the RVB singlet coupling at T=0K (which closes the pseudogap). Thus $\mu H_{pg}(0)$ gives the RVB singlet coupling energy. Similarly H=0 $T^*$ (i.e. $T^*(H=0) \equiv T_{CF}$) is the large enough temperature above which RVB singlet coupling is destroyed and system becomes paramagnetic. Thus $k_BT^*$ also give the RVB singlet coupling energy. Therefore $\mu H_{pg}(0) \sim k_BT^*$. This is similar to the relation $\mu H_W(0) \sim k_BT_N(T_C)$ of magnetically ordered systems.)

## II.7. SUMMARY

We have described here a new mechanism of high $T_c$ (critical temperature) superconductivity. According to our cluster phase transition (CPT) model of spin glass systems, magnetic clusters are present in the frustrated magnetic lattices in the material's otherwise paramagnetic state below a temperature $T_{CF}$ (cluster formation temperature). After giving some details of this model, it has been envisaged that in high $T_c$ superconducting cuprate systems, which are magnetically frustrated, these magnetic clusters exist in pairs. The two pair partners are interpenetrating and an ionic spin of one cluster forms a singlet pair with a corresponding ionic spin of the partner cluster. This singlet pairing could occur due to a resonating valence bond (RVB) interaction. The conducting electrons (CEs) and the Cooper pairs (CPs), formed by BCS-Migdal-Eliasberg phonon coupling, interact with the singlet coupled ion pairs and this interaction is responsible for the $T_c$ enhancement and other properties of cuprate superconductors. Consequences of this model and its comparison with other existing theories have been discussed.

Further the above model, called paired cluster (PC) model of high $T_c$ superconductivity, is able to explain the pseudogap origin and other gap related properties of high $T_c$ superconductors (cuprates). The interaction of CEs for temperatures $T \geq T_c$ and of both the CEs and the CPs for $T < T_c$ with the singlet coupled ion pairs enhances the CE energy, $E_{el}$, by $\Delta E_{el}$ and the CP energy, $E_{CP}$, by $\Delta E_{CP}$ causing a redistribution of the filled electronic density of states (DOS). Due to this a pseudogap appears in the electronic DOS at the Fermi surface, for $T_{CF} \geq T \geq T_c$, with d-wave symmetry which, slightly modified by $\Delta E_{CP}$ enhancement, superimposes over the BCS superconducting state (SS) energy gap for $T < T_c$ resulting in (i) a mixed s-, d- wave symmetry for the observed below $T_c$ energy gap if one assumes the BCS energy gap to have anisotropic s-wave symmetry for cuprate crystal lattice (high $T_c$, anisotropic, almost no magnetic pair breaking), (ii) nondisappearance of the gap at $T_c$ on heating and almost temperature independence of the gap width, (iii) presence of states in the gap and (iv) several other gap behaviour related properties, like the absence of NMR spin relaxation rate coherence peak, which give impression of a non-BCS, nonphononic cuprate superconductivity with conducting pairs distinctly different from BCS CPs.

## III. PSEUDOGAP CRITICAL CONCENTRATION, VORTEX CORE PSEUDOGAP AND STRIPE PHASE IN HIGH -$T_c$ SUPERCONDUCTORS

### III.1. INTRODUCTION

We have presented in earlier parts [159-161] a new mechanism of high-$T_c$ (critical temperature) superconductivity (HTSC), called paired cluster (PC) model, and explained the high $T_c$, normal state pseudogap (NSPG), superconducting state (SS) energy gap and several other cuprate properties on the basis of the proposed new mechanism. In this part we show that our new mechanism, the PC model, can explain the pseudogap critical dopant concentration [162], vortex core pseudogap [163] and possible stripe phase [164] in high- $T_c$ superconductors. However before that, we give below a brief outline of the PC model and its consequences in a way suitable for the present part.

The PC model is an extension of the earlier proposed cluster phase transition (CPT) model of spin glass (SG) systems which uses the concept of clusters' presence in magnetically frustrated SG systems [159,165]. According to the PC model since high $T_c$ superconductors (cuprates) are also magnetically frustrated, magnetic clusters are present in their lattices too. However, unlike CPT model clusters, these (cuprate) clusters occur in pairs. The two pair partners are interpenetrating (Fig.30) and a spin of one cluster (spin 1) forms a singlet pair, due to the resonating valence bond (RVB) interaction, with a corresponding spin of the partner cluster (spin 2). According to the PC model the conducting electrons (CEs) for $T \geq T_c$, and both the CEs and the drifting Cooper pairs (CPs) for $T < T_c$, interact with the singlet coupled ion pairs in the cluster and cluster boundaries and this interaction is responsible for the high $T_c$ and other properties of cuprates. A consequence of the CE-, CP- singlet coupled ionic spin pair interaction is to enhance the CE energy, $E_{el}$, by $\Delta E_{el}$, CP energy, $E_{CP}$, by $\Delta E_{CP}$ and the lattice Debye temperature, $\theta_D$, by $\Delta\theta_D$ (as one cools the lattice through $T_{CF}$, $T_c$ where $T_{CF}$ is the temperature at which the PC model clusters are formed). All the cuprate properties can be understood on the basis of these enhancements (scatterings), considered either individually or collectively or in some combination, and the clusters' and cluster boundaries' presence in the lattice. We show this below starting from the critical dopant concentration origin. Though the results given here are for typical parameter values, as discussed in Part II [159] they have been checked to be general in nature. The cluster boundaries and the clusters (Fig. 30) have different frustration parameters ($\tilde{J}$, $\tilde{J}_0$) since cluster boundaries, compared to clusters, are rich in holes ($Cu^{3+}$ (or equivalently $O^{1-}$) [159]; $\tilde{J}_1/\tilde{J}_{01} > 1$ (Fig. 27)(Sec. III.4) [159]).

### III.2. CRITICAL DOPANT CONCENTRATION

As discussed in Part II [159] the NSPG has been observed in underdoped, optimally doped and overdoped samples, where by optimally doped sample we mean a sample where dopant concentration ($\delta$(as say in $YBa_2Cu_3O_{7-\delta}$) or x (as in $La_{2-x}Sr_xCuO_4$ or $Y_{1-y}Ca_yBa_2Cu_3O_{6+x}$)) is such that the $T_c$ is maximum (i. e. $x(\delta) = x_{opt}(\delta_{opt})$). However in the overdoped region there is a critical dopant concentration, $x_{crit}$ or $\delta_{crit}$,(system dependent) above which NSPG is not observed [162]. We explain below the existence of such a $x_{crit}$ on the basis of the PC model.

As shown in Part II [159], in the normal (nonsuperconducting) state ($T_c \leq T \leq T_{CF}$) due to $\Delta E_{el}$ scattering (enhancement) the electronic density of states (DOS), $D(E_{el})$, gets disturbed and there is redistribution of the filled DOS, $D_f(E_{el})$, due to which NSPG (i.e. pseudogap in the redistributed $D_f(E_{el})$, or $D_{fr}(E_{el})$, vs. $E_{el}$ curve ) gets created at the Fermi surface; $N(E_{el})$, the number of CEs at energy $E_{el}$, = $2 \times D(E_{el})$. The magnitude (size (amplitude, width)), shape of NSPG depends on $\Delta E_{el}$, T (temperature), nature of total DOS, $D_t(E_{el})$, $E_F$ (Fermi energy) and $N_P$, the percentage of CEs for which $\Delta E_{el}$ enhancement occurs [159]. In the overdoped region as the dopant concentration increases, more and more $Cu^{2+}$ ions get converted to $Cu^{3+}$ [162] (assuming $Cu^{2+}$, $Cu^{3+}$, $O^{2-}$ picture of the unit cell [159]; the description with $Cu^{2+}$, $O^{1-}$, $O^{2-}$ unit cell picture is equivalent). Due to this $N_P$ gets decreased since $N_P$ is larger if more paired $Cu^{2+}$ ions, with $H_W$ (Weiss field) at their site, are present in the lattice. The $N_P$ decrease is more than just suggested by the $Cu^{2+}/Cu^{3+}$ ratio decrease since as $Cu^{3+}$ increases, $\tilde{J}_{02}$, defined in Part II [159], gets decreased which enhances the CBIs' (cluster boundary ions') number and diminishes the cluster size i. e. the CIs' (cluster ions') number. Since in the $T_c \leq T \leq T_{CF}$ range CBIs do not contribute to $\Delta E_{el}$ scattering, decrease in CIs' number causes considerable $N_P$ decrease. The $\Delta E_{el}$ is another quantity which gets significantly affected by the $Cu^{3+}$ increase. Our calculations (Part II) [159] show for the overdoped region a decrease in $\Delta E_{el}$ with increasing x. This happens due to decrease in $H_W$, caused by the $\tilde{J}_{02}$ decrease, and decrease in $N'_{CE}/N_{CE}$ ratio, defined in Part II [159], occurring since $N'_{CE}$ decreases faster than any $N_{CE}$ decrease owing to the above discussed $N_P$ decrease reason; these findings are consistent with the experimental results [166,167]. A consequence of $N_P$, $\Delta E_{el}$ decrease with concentration x is that for x larger than a certain concentration $x_{crit}$ (critical concentration), $N_P$, $\Delta E_{el}$ become small enough to make the normal state pseudogap disappear. This is typically shown in Fig. 35 where all descriptions, and parameter values, are same as those of Fig. 32 (b) of Part II [159], to facilitate comparison, except $\Delta E_{el}(E_F) = 100$ meV, $N_P = 10\%$ (curve a) and $\Delta E_{el}(E_F) = 50$ meV, $N_P = 5\%$ (curve b); (the dotted curve is the total (filled + empty) DOS, $D_t(E_{el})$, [159], the dashed curve is $D_f(E_{el})$ and curves a, b are $D_{fr}(E_{el})$ [159]; $\Delta E_{el}(E_F)$ = the value of $\Delta E_{el}$ at $E_{el} = E_F$ [159]). The pseudogap absence is clearly seen there. This absence gets further confirmed when these curves (a, b) are translated into the tunneling conductance curves [159, 166]. Even for somewhat higher values of $N_P$ than the curve a, b $N_P$ values pseudogap is not seen when $\Delta E_{el}$ is small, $\lesssim 5k_BT$; $k_B$ = Boltzmann's constant. The pseudogap effect absence, i. e. appreciable $\Delta E_{el}$ scattering effect absence, for $x > x_{crit}$ makes these samples behave more like conventional BCS superconductors; this is experimentally seen [162].

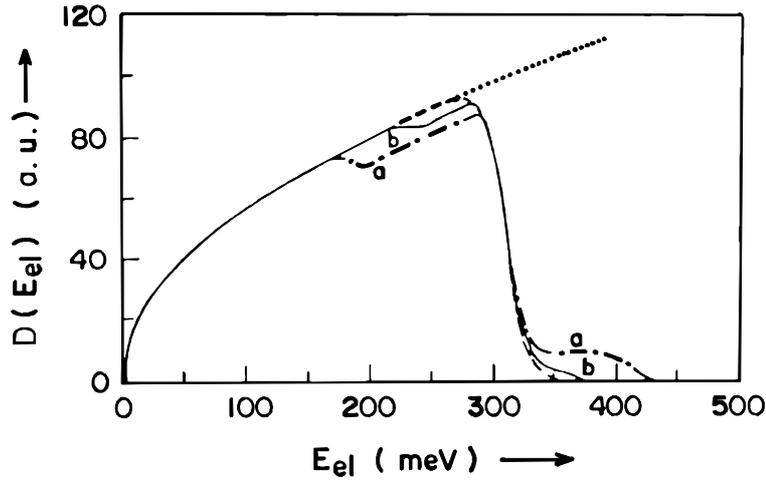

Fig. 35. Dependence of the electronic density of states, D($E_{el}$), on electrons' energy, $E_{el}$, for T > $T_c$ and x, δ (dopant concentration) > $x_{crit}$, $δ_{crit}$ (critical dopant concentration); $T_c$ = critical temperature, a. u. = arbitrary unit. Details are described in the text.

      In the above discussion, as mentioned in the beginning, it has been assumed that the CPs do not exist in the $T_c ≤ T ≤ T_{CF}$ range. This is because our (PC) model can explain the cuprate experimental results without assuming the CPs' presence above $T_c$. However, there are theories [168] which assume the CPs' existence, without any phase coherence, in the range $T_c ≤ T ≤ T_{CF}$. We want to point out here that even if such CPs are assumed to be present above $T_c$, our calculations show that the results given in Fig. 35 (and also those given in Fig. 32 of Part II [159]) remain almost same. This is since the CPs' number is much smaller than the CEs' number, their (CPs') number varying from ~ 10% of the CEs' number at T ~ $T_c$ to zero at T ~ $T_{CF}$, assuming the CPs' formation at $T_{CF}$, as theories assume [159, 168], and BCS temperature variation for the CPs' number [159]. Thus in T ~ $T_{CF}$ region, CPs' number is very small, ~ 1-2% of CEs' number, and therefore they do not affect the Fig. 35 results any significantly. In the T ~ $T_c$ region, where their number is slightly more, ~ 9-10% of CEs' number, our calculations (Part II) [159] show that $ΔE_{CP}$ ~ $ΔE_{el}(E_F)$ and thus these CPs get scattered in the same way as the CEs. Consequently in the T ~ $T_c$ region too CPs' presence shows no appreciable effect on results. Thus the presence of pseudogap does not prove or disprove the CPs' existence above $T_c$. Similarly in Part II [159] we have explained the NMR relaxation rate (1/$T_1$) vs. T anomaly at $T_{CF}$ (i.e. $T^*$ drop; $T_{CF}$ = $T^*$, the pseudogap formation temperature as one cools the lattice from above [159]) without assuming CPs' presence above $T_c$. However, this on its own does not rule out such a presence. For instance, if CPs are present above $T_c$, they will only act in cooperation with the effects of our model and further enhance the $T^*$ drop. But we do not favour this explanation since experimental, and also physical, doubts exist on the CPs' presence above $T_c$ [159, 167, 169] (Sec. II.4, II.5, Appendix 1).

      The pseudogap effect absence above $T_c$ does not necessarily mean $ΔE_{el}$, $ΔE_{CP}$ scattering effect absence below $T_c$ since, as explained in Part II [159], at lower temperatures CBIs can have SG freezing which enhances $N_P$, $ΔE_{el}$, $ΔE_{CP}$. Thus three different situations arise depending on whether the CBIs' $T_{SG}$ ~ $T_c$ or < $T_c$ or 0; $T_{SG}$ = SG temperature [159]. For instance, for the $T_{SG}$ = 0 case the system behaves like the conventional BCS superconductor right from T = 0K. Similarly for $T_{SG}$ ~ $T_c$, the system will have non-BCS like behaviour only below $T_c$. Some evidence exists in this direction [162]. For more detailed discussion, x vs. $T_{CF}$ and x, T behaviour of experimental data, like say Mössbauer effect, channeling and tunneling conductance data, are needed at least for few cuprates.

### III.3. VORTEX CORE PSEUDOGAP

The tunneling experiment [163] has shown the presence of pseudogap in vortex core in high-$T_c$ Bi-compound at $T \ll T_c$ (T=4.2K, H (external field) = 60kOe) which is very similar to the NSPG and changes slowly to the SS gap as one moves away from the vortex core and reaches to a sufficiently separated distance, away from all the other neighboring vortices. The observation of such a vortex core pseudogap has a natural explanation in our (PC) model according to which, as discussed in Part II [159], the pseudogap (NSPG) persists below $T_c$ and gets superimposed over the BCS SS gap for $T < T_c$. This happens because $\Delta E_{el}$ scattering is present below $T_c$ too where $\Delta E_{CP}$ enhancement also occurs. Thus according to the PC model, since inside the vortex core system is in the normal (nonsuperconducting) state BCS SS gap is absent and so only NSPG, with parameters (shape, size etc.) modified by low T, H, is observed. Since the magnetic field, which is constant inside the vortex core, decreases, exponentially, as one moves away from the vortex core [170], the pseudogap observed in the vortex core slowly changes from NSPG shape to SS gap shape with increasing distance from the vortex core. This is similar to the NSPG changing slowly to SS gap as T decreases from $T_{CF}$ (=T*) to below $T_c$ where system changes from normal (nonsuperconducting) state to SS. Thus the PC model explanation is consistent with the experimental results [163, 166]. In the normal state ($T_c \leq T \leq T_{CF}$ (=T*)), NSPG changes with T [163, 166]. Similar changes can be observed for the pseudogap inside the vortex core with H.

The observation of NSPG at very low temperatures ($T \ll T_c$) (in the vortex core) puts doubt on the theories which assume CPs' formation, without phase coherence, above $T_c$ owing to high temperature ($T > T_c$) superconducting fluctuations and associate the NSPG formation to such fluctuations [168]. Since for $T < T_c$ superconducting fluctuations are absent, NSPG at low T ($T \ll T_c$) should have not been seen if the above mentioned theories were correct. Also these theories predict the NSPG presence only for underdoped cuprates whereas NSPG has now been observed in optimally doped and overdoped systems also [159, 163, 166]. The recent observation of zero bias conductance peak (ZBCP) in the tunneling conductance spectra of high-$T_c$ Bi-compound and other measurements [171] also argue against the preexisting CPs as the possible cause of vortex core pseudogap [163]. Some connection in the x dependence of NSPG and SS ($T < T_c$) gap is possible since $\Delta E_{el}$ affects both the NSPG and $T_c$, and also since the NSPG superimposes over the BCS SS gap below $T_c$ (i. e. $\Delta E_{el}$ effect is present below $T_c$ too) [159].

### III.4. STRIPE PHASE

A large number of theoretical [172-177] and experimental [178-189] works exist concerning the possibility of a stripe phase in cuprates. Theories predict either a macroscopic phase separation [172, 173, 178, 186-188] or a stripe formation [172, 173, 179, 180, 184, 186-188] in cuprates and similar systems. In a macroscopic phase separation the lattice is divided on macroscopic scale into hole full (e.g. $Cu^{3+}$ ($O^{1-}$) [159] full in cuprates) and holeless (e. g. $Cu^{2+}$ ($O^{2-}$) full in cuprates) regions. Such a phase separation has been observed in insulating $La_2 CuO_{4+\delta}$ [172, 173, 178, 186-188]. In a stripe phase we have alternate hole full and holeless regions. For cuprate and similar lattices, which have two dimensional planar structure, theories predict, in the Cu-O or similar planes, $Cu^{2+}$ (or similar spin) full antiferromagnetically ordered regions (domains) separated by one dimensional (or slightly wider) hole ($Cu^{3+}$ or similar ion) full stripes (domain walls). Such a separation occurs as antiferromagnetically ordered regions expel the holes to minimize region's thermodynamic free energy. Experimentally such a stripe phase has been observed in insulating nickelates, manganites [172, 173, 179, 180, 186-188] and in $La_{2-x}Sr_xCuO_4$, $La_{2-x}Ba_xCuO_4$, $La_{1.6-x}Nd_{0.4}Sr_xCuO_4$ and similar La-Sr-Cu-O based systems for x~1/8 where the system becomes insulating [172, 173, 182, 184, 186-188]. So far such a regularly arranged stripe phase has not been observed in any superconducting cuprate sample. Some stripe phase is indicated by the experiments in superconducting La-Sr-Cu-O based samples (when x is away from 1/8 value) but the stripes there are dynamically fluctuating in space (location) and are not static [172, 173, 181, 182, 185-188]. In other superconducting cuprates indications of stripe phase are not conclusive [164, 172, 173, 183, 186-188]. Physically if all the holes (say $Cu^{3+}$ [159]) stayed in one region and all the spins (say $Cu^{2+}$) in the other region with holes not allowed to jump to the spin region, metallicity would not occur and superconductivity will not be possible. On the other hand if the hole full region is assumed to have some spins in it, i.e. it is not a hole full region but a hole rich region, then conductivity (metallicity) will occur only along the one dimensional hole rich stripe and thus the position (location) of stripes will remain fixed in the space. This is true even if one assumes CPs' formation above $T_c$ in the stripe region and assumes tunneling of CPs, CEs between the stripes to give two-dimensional conductivity. Thus experimentally what we have is, for superconducting cuprates, hole rich and hole poor regions (rather than hole full and holeless regions) with hole rich region stripes either fluctuating in space ( as in La-Sr-Cu-O based systems for x ≠ 1/8) or being not observed at all (as in other cuprates). Experiments do not say anything conclusively about size (width), shape or spatial arrangement of fluctuating stripes. We show below that these results can be understood on the basis of our (PC) model.

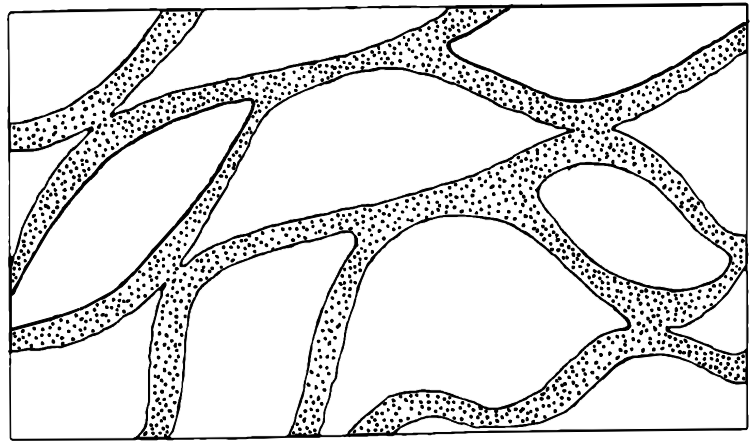

Fig. 36. Schematic representation of possible stripe phase in superconducting cuprates. The cluster boundaries (dotted area) form stripes in Cu-O plane. Details are described in the text.

According to the PC model, we have cluster and cluster boundaries in the superconducting cuprate lattice (Fig. 30). The cluster boundaries are hole ($Cu^{3+}$ ($O^{1-}$)) rich and the clusters are hole poor ($Cu^{2+}$ ($O^{2-}$) rich). This happens because, like the frustrating ions in a SG lattice [165], the doped holes are distributed randomly in the cuprate lattice (at any given instant). Due to this, in the regions where the holes are in excess frustration is large ($\widetilde{J}_1 / \widetilde{J}_{01} > 1$ [159]) resulting in the formation of cluster boundaries, and the spin ($Cu^{2+}$) rich regions (hole poor regions) form clusters ($\widetilde{J}_{o2} / \widetilde{J}_2 > 1$ [159]). Any preferential segregation of holes, as proposed for insulating lattices [172-177], is not possible for superconducting cuprates since the random $Cu^{2+} \Leftrightarrow Cu^{3+}$ ( or equivalently $O^{1-} \Leftrightarrow O^{2-}$ [159]) charge fluctuation present in their lattice will destroy preferential hole segregation in any region. The cluster boundaries can form stripes in Cu-O plane (Fig. 36) if, as discussed in Part II [159], the frustration is confined within a unit cell along the c-axis and the interplanar coupling is weak, otherwise three dimensional canal (tubular) structure could exist with two dimensional stripe projection on the Cu-O plane. However these stripes (Fig. 36) will have a width, will run randomly in the lattice and will randomly fluctuate in position with time (due to random $Cu^{2+} \Leftrightarrow Cu^{3+}$ ( or $O^{1-} \Leftrightarrow O^{2-}$ ) charge fluctuation). At different instants different types of stripe pattern (arrangement) will be formed in the Cu-O plane and Fig. 36 shows one of the possible stripe patterns which may exist at any particular instant. The details (size, shape, arrangement of stripes) will depend on x, T and sample microstructure [159, 190]. The width of the stripes will vary from place to place, owing to cluster size distribution in the lattice [159], and may even be just one lattice spacing wide at some places [159] or may even not exist at some places (instantaneously broken stripes). For an average 100 Å diameter frustrated area in the Cu-O plane [159], the average stripe width will be ~25 Å if the cluster boundaries and clusters have equal area in the lattice. Thinner stripes will occur if the average cluster boundary area is comparatively smaller (owing to say smaller $Cu^{3+}$ concentration or better lattice microstructure etc.). When a stripe is shared by several clusters, it will be wider. Whereas in most superconductiong cuprates these stripes will be completely randomly located at any instant (complete disorder), in some systems they may have some kind of partial order [174] owing to some specific lattice property like the characteristic buckling of Cu-O plane in La-Sr-Cu-O based superconducting systems [172-174, 182, 186-188]. For partially ordered stripes, some indications may be seen experimentally as fluctuating (dynamic) incommensurate (superlattice) reflections. This seems to be the case in La-Sr-Cu-O based superconducting systems [172-174, 182, 186-188]. Completely disordered stripes will not be detected experimentally. Thus the observed results [172, 173, 178-189] are consistent with the physical picture provided by our (PC) model. Some theories [172-177] too feel the need for bending of stripes, random running of stripes in the lattice, larger width of the stripes, existence of hole rich and hole poor regions (rather than hole full and hole free regions), instantaneously broken stripes, etc. for understanding the experimental results. The PC model provides a physical basis for the occurrence of the above desired picture. It is also consistent with a recent theoretical work (Appendix 1 (xvii)) which prohibits the static, charge ordered, stripe formation in superconducting cuprates owing to the presence of next nearest neighbour hopping effect in their lattice. Such a hopping is required for metallicity, and so superconductivity, but it suppresses the static stripe formation.

### III.5. CONCLUSION

Thus we see that the cuprate properties can be understood on the basis of the PC model. This is true even for those properties which are not specifically discussed here or in Part II [159]. For example, Fig. 34(b) shows the $D(E_{el})$ vs. $E_{el}$ distribution obtained on the basis of the PC model for $T < T_c$ case [159]. The parameters used for calculating various curves (a; b, b′; c, c′) in Fig. 34(b) are given in Part II [159] and a dotted curve is $D_t(E_{el})$, a dashed curve is $D_f(E_{el})$, a full line curve is $D_{fr}(E_{el})$ and a dash-dot curve is the redistributed density of empty states, $D_{er}(E_{el})$. The curve b′ (c′) is obtained by subtracting the unoccupied (empty) side ($E_{el} > E_F + \Delta(T)$, $\Delta(T)$ = SS energy gap width/2) portion of the curve b(c) from the $D_t(E_{el})$ curve. A dip R is seen for curve a (Fig. 34(b)) on P peak side; for the curve b, b′ R appears on P peak side and another dip R′ appears on the Q peak side, and for the curve c, c′ these dips are broad. The location of R, R′ depends on P, Q (gap edges) positions. These results have been experimentally seen [163, 166, 191]. However a variety of explanations have been given for R, R′ existence such as unusually strong pair coupling, d-wave superconductivity, band structure effect, thermal broadening (smearing) effect, mismatch of the pseudogap width and SS gap width at ~ $T_c$, ($\pi$, $\pi$) scattering of electrons in the Brillouin zone, etc. [163, 166, 167, 191, 192]. We see here (Fig. 34(b)) that these dips naturally occur in the PC model in certain circumstances (for certain parameter values) owing to the presence of $\Delta E_{el}$ and $\Delta E_{CP}$ scatterings which, as described before, also explain several other cuprate properties. Thus earlier different explanations have been given for different phenomena, e.g. some earlier explanations for R, R′ are mentioned above, the absence of NMR coherence peak has been associated to enhanced electron-phonon scattering, near $T_c$, owing to the high value of $T_c$ [193], SS stripe phase to the presence of antiferromagnetic correlations [172-177], pseudogap to existence of preformed CPs [159, 168] or spin gap presence etc. [159], high value of $T_c$ to nonphononic Cooper pairing etc. [159], and several phenomena remained unexplained like the T dependence of the SS gap width [166], increase in $\theta_D$ at $T^*$ ($\equiv T_{CF}$) on cooling [159], etc. *We see here that all these phenomena can be explained by a common picture, namely $\Delta E_{el}$, $\Delta E_{CP}$, $\Delta \theta_D$, clusters', cluster boundaries' presence, in our (PC) model.* Thus the PC model seems to provide a correct explanation for the HTSC nature and mechanism. The model is flexible enough to incorporate other interactions if needed in some special cases [159].

### IV. ORIGIN OF PSEUDOGAP AND STRIPE PHASE IN HIGH-$T_C$ SUPERCONDUCTORS IN TWO DIMENSIONAL PICTURE

### IV.1. INTRODUCTION

In earlier parts [194-196] we have explained the origin of pseudogap and stripe phase in magnetically frustrated cuprate superconductors assuming a three dimensional (3D) electronic density of states (DOS), $D(E_{el})$, for conducting electrons (CEs) and using a paired cluster (PC) model developed by us to describe the physics of high-$t_c$ superconductivity; $E_{el}$ = CE energy, $T_c$ = critical temperature, $D_t(E_{el})$, total electronic DOS (at $E_{el}$), = $D_f(E_{el})$ ( filled electronic DOS) + $D_e(E_{el})$ ( empty electronic DOS ), $D_t(E_{el})$ = $AE_{el}^{1/2}$ for $T \geq T_c$ and for $T < T_c$,

$$D_t(E_{el}) = A \operatorname{Re}[( E_F \pm \sqrt{[(E_{el} + i\Gamma) - E_F]^2 - \Delta^2} \ )^{1/2} \{\frac{|(E_{el} + i\Gamma) - E_F|}{\sqrt{[(E_{el} + i\Gamma) - E_F]^2 - \Delta^2}}\}] \quad ,$$

where - (minus) sign before the square root, in the first round bracket, applies for $E_{el} \leq E_F$ and + sign for $E_{el} > E_F$, Re means real part, A= proportionality constant, T = working temperature, $i = (-1)^{1/2}$, $\Delta$ = BCS ( superconducting state (SS) ) energy gap/2, $E_F$ = Fermi energy and $\Gamma$, explained later, = broadening due to Cooper pair (CP) lifetime decay. The assumption of 3D $D(E_{el})$ looks justified due to the following reasons. The conductivity ($\sigma$), and the superconductivity, are 3D in nature in cuprates. Eventhough the c-axis resistivity ($\rho_c$) is higher than a-,b-axis resistivity ($\rho_a$, $\rho_b$), it is still in the m$\Omega$-cm range [197, 198], which is similar to $\rho$ (resistivity) observed in some metals like Au-Fe spin glass[199]. $\rho_c$ is only about an order of magnitude higher than $\rho_a$, $\rho_b$ (due to reasons described in Sec. II.3.1 [194-196]) and thus the system (cuprate) is not an insulator along the c-axis; $\rho$ for an insulator is in M$\Omega$-cm range [199, 200]. Also careful measurements show same $T_c$ along the a-,b-,c-axes [197, 198]. A m$\Omega$-cm range $\rho$ is observed in powdered samples (powdered cuprates) also, where a-,b-,c-axes are random[197, 198]. If $\rho$ was insulating along any direction, powder sample would have shown insulating $\rho$ (M$\Omega$-cm range value). Also experiments show same $\rho$ mechanism for $\rho_a$, $\rho_b$, $\rho_c$ below, and above, $T_c$ [194, 197, 198]. Thus the use of 3D $D(E_{el})$ looks justified. However at the same time it is also true that the cuprates have layered (two dimensional (2D)) structure where Cu-O planes are a-,b- planes, stacked one over the other along the c-axis in the crystal unit cell [194, 197, 198]. It is the coupling between the Cu-O planes which gives three dimensionality, but this coupling has different strength in different cuprates. Thus some 2D nature (two dimensionality) is always present in cuprates [197, 198]. In some systems, like Bi-cuprates [198, 201], this two dimensionality may be more pronounced. It is thus necessary to ascertain that the results obtained in earlier parts [194-196] using 3D $D(E_{el})$ are valid even for 2D cuprate $D(E_{el})$. In this part we examine this point and find that the 2D $D(E_{el})$ based results are same as those obtained in earlier parts [194-196] using 3D $D(E_{el})$. In the following sections we give the details.

### IV.2. METHODOLOGY AND RESULTS

Since band structure calculation conduction bands for 3D lattices can be approximated to a parabolic shape, quadratic approximation DOS can be used there reasonably well [194, 197, 198, 202, 203]. However this is not the case for a 2D cuprate lattice (Cu-O planar structure with small plane-plane coupling) owing to the presence of van Hove singularity in their electronic DOS (conduction band) at $E_{el}=E_s$ where $E_s$ (the singularity point energy) is close to $E_F$ [197, 198, 201, 204-209]. Several people have tried to associate the cuprate high-$T_c$ with the presence of this singularity [197, 198, 205]. However since experimentally such a singularity has not been observed, it is believed that any small coupling between the Cu-O planes, which can always be present, turns this singularity into a broad peak at $E_s$ and therefore the broadening, $\Lambda$, should be taken into account in any calculation [206, 207]. Thus for the 2D Cu-O planes [197, 198, 201, 205-212],

$$D_t(E_{el}) = C[1 + \frac{1}{2}\ln\frac{E_s}{\sqrt{(E_{el}-E_s)^2 + \Lambda^2}}] \quad (12)$$

for $T \geq T_c$, and for $T < T_c$,

$$D_t(E_{el}) = C \operatorname{Re}[\{1 + \frac{1}{2}\ln\frac{E_s}{\sqrt{([\{(E_{el}+i\Gamma)^2 - E_F\}^2 - \Delta^2]^{1/2} + E_F - E_S)^2 + \Lambda^2}}\}\{\frac{|(E_{el}+i\Gamma) - E_F|}{\sqrt{[(E_{el}+i\Gamma) - E_F]^2 - \Delta^2}}\}], \quad (13)$$

where C = proportionality constant and $\Gamma$, as mentioned before, is the broadening present owing to the CP lifetime decay arising due to the inelastic electron- electron and electron- phonon scatterings [198, 210-212]. We examine below the effect of this (Eq. 12, 13) 2D $D(E_{el})$ on the pseudogap and stripe phase in cuprates. For the pseudogap calculation, we discuss the $T \geq T_c$ and $T < T_c$ ranges separately and assume a small plane- plane coupling ($\Lambda/E_s = 5\%$) which is sufficient to give a broad singularity peak.

**(i) $T_c \leq T \leq T_{CF}$**

In earlier parts [194-196], for the 3D $D(E_{el})$ case, we have presented $D(E_{el})$ vs. $E_{el}$ variation for $D_t(E_{el})$, $D_f(E_{el})$ and $D_{fr}(E_{el})$, the density of filled states redistributed. $D_f(E_{el}) = D_t(E_{el}) \times f(E_{el})$, where $f(E_{el})$ is the Fermi function, and $D_{fr}(E_{el})$ is the modified $D_f(E_{el})$ i.e. the $D_f(E_{el})$ which has got modified owing to the $\Delta E_{el}$ scattering which occurs for $T \geq T_c$ in the PC model [194-196]. According to the PC model (Part II) [194-196], paired magnetic clusters are present in the cuprate lattice below a temperature $T_{CF}$ (> $T_c$); $T_{CF}$ = cluster formation temperature (pseudogap appearance temperature) [194-196]. The CEs for $T \geq T_c$, and both the CEs and the Cooper pairs, CPs, for $T < T_c$, interact with these clusters, by an interaction described in Part II [194-196], and this interaction enhances the CEs' energy, $E_{el}$, by $\Delta E_{el}$ and CPs' energy, $E_{CP}$, by $\Delta E_{CP}$. The modification in $D_f(E_{el})$ due to the $\Delta E_{el}$ enhancement for $T \geq T_c$, and due to both the $\Delta E_{el}$ and $\Delta E_{CP}$ enhancements below $T_c$, gives rise to $D_{fr}(E_{el})$. The earlier parts' calculations [194, 196] have been presented for some typical values of the parameters used (i.e. T, $E_F$, $\Delta E_{el}$, $N_P$ where $N_P$, $\equiv (N_P)_{CE}$, is the percentage of CEs for which $\Delta E_{el}$ enhancement occurs and its value depends on the relative space occupied by the clusters and the cluster boundaries in the cuprate lattice [194-196]; below $T_c$ we have $\Delta E_{CP}$ and $(N_P)_{CP}$ (percentage of CPs for which $\Delta E_{CP}$ enhancement occurs) also present). As has been explained there [194, 196], though the results are presented for certain typical parameter values they have been checked to be general in nature. The typical parameter values have been chosen for obtaining results similar to those of the tunneling conductance experiments mentioned in Part II [194, 196], or to bring out any specific nature of the $D_{fr}(E_{el})$, or $D_{er}(E_{el})$, vs. $E_{el}$ curve, and are consistent with the theoretical estimates (Part II) [194-196].

For the present 2D lattice, $D_t(E_{el})$ is given by Eq. (12) for $T \geq T_c$ and the other quantities are as follows.

$$D_f(E_{el}) = D_t(E_{el}) \times f(E_{el}), \tag{14}$$

where $f(E_{el}) = 1/\{\exp[(E_{el} - E_F)/k_B T]+1\}$ and $k_B$ = Boltzmann's constant. Assuming that the Pauli principle permits the above mentioned $\Delta E_{el}$ scattering, i.e. empty states are available for such a scattering, we have [194, 196],

$$D_{fr}(E_{el}) = D_f(E_{el}) - N_P D_f(E_{el}) + N_P D_f(E_{el} - \Delta E_{el}). \tag{15}$$

Details for the occurrence of this scattering for any $E_{el}$ are discussed in Part II [194, 196]. Fig. 37 shows the results obtained, where $E_{el}$ dependence of $D_t(E_{el})$ (dotted curve), $D_f(E_{el})$ (dashed curve) and $D_{fr}(E_{el})$ (full line and dash-dot (a, b) curves) are plotted. The parameter values chosen are the same as used in Parts II, III [194, 196] to facilitate comparison except $E_s$, $\Lambda$ which are new and, as mentioned before, have been fixed to $\Lambda/E_s = 5\%$; T=200K, $E_F = E_s = 310$ meV, $\Lambda = 15$ meV, $\Delta E_{el}(E_F) = 300$ meV, $N_P = 50\%$ (full line curve), 40% (curve a) and 60% (curve b), and $\alpha = 0.7$ (meV)$^{-1}$, where $\Delta E_{el}(E_F)$ is the value of $\Delta E_{el}$ at $E_F$ and as discussed in Part II [194, 196] $\alpha$ describes the dependence of $\Delta E_{el}$ on $E_{el}$. The results obtained in Fig. 37 are same as those obtained in Part II [194, 196]. The Fig. 37 results have been checked to be valid for $E_s$ close to $E_F$, either > $E_F$ or < $E_F$, cases also. Thus the results which have been obtained for a 3D $D(E_{el})$ in Part II [194, 196] for $T \sim T_{CF}$ are valid for 2D cuprate $D(E_{el})$ also. Similarly the other results obtained in

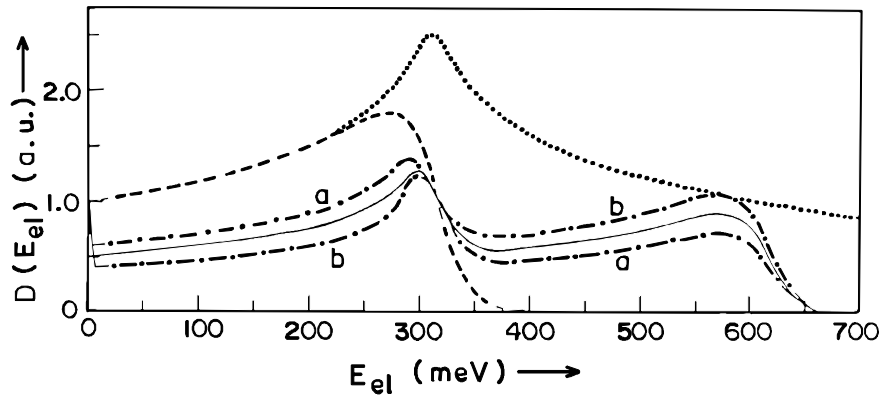

Fig.37. Dependence of the electronic density of states, $D(E_{el})$, on electrons' energy, $E_{el}$, for $T > T_c$; $T_c$= critical temperature, a.u.= arbitrary unit. Details are described in the text.

Part II [194, 196] for T ≥ $T_c$ have also been found to exist in the present 2D $D(E_{el})$ case. Further, as discussed in Part II [194, 196], even if CPs' presence is assumed in $T_c \leq T \leq T_{CF}$ range, results obtained here remain same.

### (ii) $0 \leq T < T_c$

Below $T_c$, BCS energy gap is also present alongwith the $\Delta E_{el}$ and $\Delta E_{CP}$ scatterings in the $D(E_{el})$ vs. $E_{el}$ distribution (Part II) [194-196]. In this case $D_t(E_{el})$ is given by Eq.(13), $D_f(E_{el})$ by Eq.(14) and $D_e(E_{el}) = D_t(E_{el}) - D_f(E_{el})$. To facilitate comparison we have used the parameters of 3D $D(E_{el})$ calculation case (Part

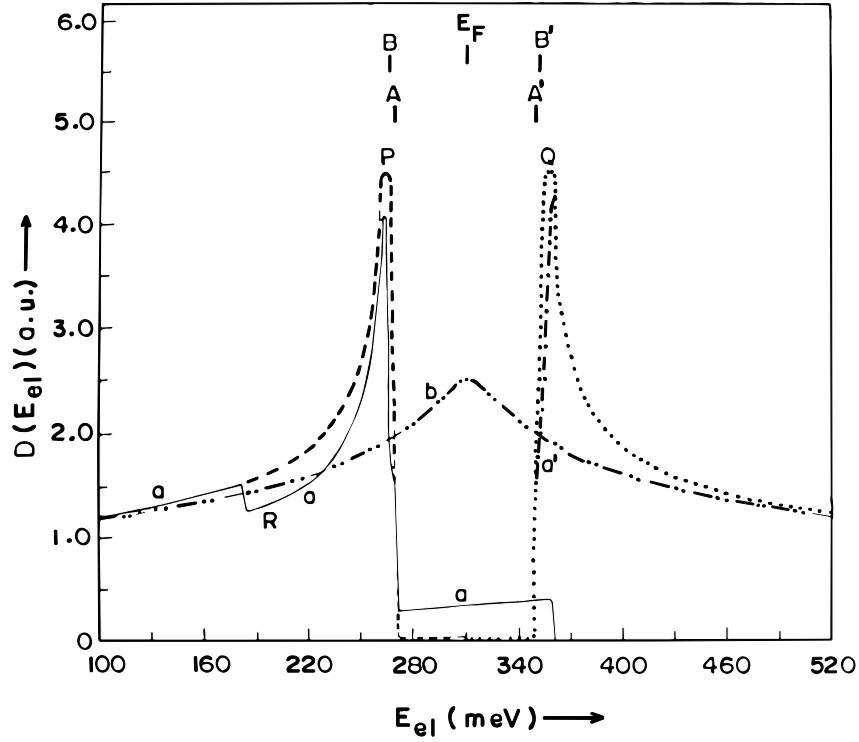

Fig.38. Dependence of the electronic density of states, $D(E_{el})$, on electrons' energy, $E_{el}$, for $T< T_c$; $T_c$= critical temperature, a.u.= arbitrary unit. Details are described in the text.

II) [194, 196] and done calculation for the present 2D $D(E_{el})$ case. Fig. 38 shows a typical result where $D_t(E_{el})$ (dotted curve), $D_f(E_{el})$ (dashed curve), $D_{fr}(E_{el})$ (full line curve) and $D_{er}(E_{el})$ (dash - dot curve) vs. $E_{el}$ is shown. The parameters used are T = 4.2K, $E_F = E_s$ = 310 meV, $\Lambda$ = 15 meV ($\Lambda/E_s$ = 5%), $\Delta$ = 45 meV, $\Gamma$ = 2.25 meV ($\Gamma/\Delta$ = 5%), $\Delta E_{CP} \sim 0$ (< 1 meV), $\Delta E_{el}(E_F)$ = 90 meV, $N_P$ = 20% and $\alpha$ = 0.7 $(meV)^{-1}$. The dash - double dot curve (curve b) is the total electronic DOS curve which would have existed at 4.2K if there was no BCS gap present (i.e. $D_t(E_{el})$ of Eq.(12)). To distinguish it from the $D_t(E_{el})$ of Eq.(13) (dotted curve in Fig. 38), we denote it (curve b) by $D_t'(E_{el})$. Since below the gap ($E_{el} < (E_F - \Delta)$), $f(E_{el})$ = 1 at 4.2K, curve b also represents the filled electronic DOS which would have existed at 4.2K, for $E_{el} < (E_F - \Delta)$, if no BCS gap was present, i.e. $D_f'(E_{el})$. This means below the gap $D_t'(E_{el}) = D_f'(E_{el})$. Thus $D_f(E_{el})$ gives the filled DOS at 4.2K for those CEs which have not formed CPs and $D_f(E_{el}) - D_f'(E_{el})$ for those which have formed CPs. Since, as mentioned above in the parameter values, $\Delta E_{CP} \sim 0$ at 4.2K, we have for the region below the gap ($E_{el} < (E_F - \Delta)$) (Part II) [194, 196],

$$D_{fr}(E_{el}) = D_f(E_{el}) - N_P D_f'(E_{el}), \tag{16}$$

and for the regions inside and above the gap,

$$D_{fr}(E_{el}) = N_P D_f'(E_{el} - \Delta E_{el}). \tag{17}$$

In Eq.(16), the term $N_P D_f'(E_{el} - \Delta E_{el})$ is absent on the right hand side (R.H.S.) since all the $D_t'(E_{el})$ states below the gap are completely filled at 4.2K ($D_t'(E_{el}) = D_f'(E_{el})$). In Eq.(17) only one term appears on R.H.S. since inside and above the gap there exist negligible, or no, filled states in the absence of $\Delta E_{el}$ scattering. However if any significant number of filled states are present in those regions in some case, then they will get added to the R.H.S. of Eq.(17) (Appendix 5).

The results obtained in Fig. 38, like change in the size, shape, location of BCS peaks (P, Q), appearance of dip R, enhancement of A, A′ separation to B, B′ separation, etc., due to the $\Delta E_{el}$ scattering presence are same as obtained for the 3D $D(E_{el})$ case in Part II [194, 196]; A, A′ are the points where the curve b intersects the gap edges when no $\Delta E_{el}$ scattering is present and B, B′ when $\Delta E_{el}$ scattering exists [194, 196].

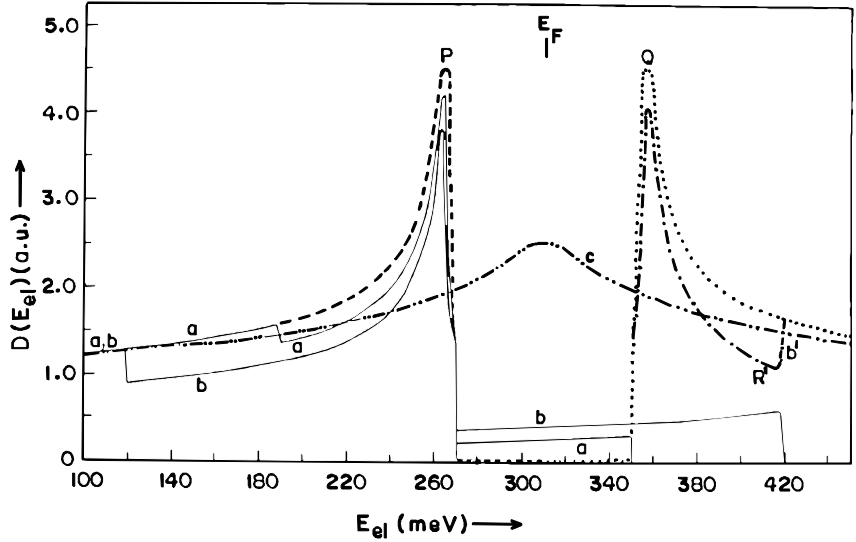

Fig.39. Dependence of the electronic density of states, $D(E_{el})$, on electrons' energy, $E_{el}$, for $T < T_c$ and some parameter values different from the Fig. 38 parameter values; $T_c$ = critical temperature, a.u.= arbitrary unit. Details are described in the text.

The Fig. 39 shows another typical result. Here the dotted, dashed, dash - dot, dash - double dot and full line curves have the same meanings as in Fig. 38 and all the parameter values are same as those of Fig. 38 except $N_P = 15\%$, $\Delta E_{el}(E_F) = 80$meV for the curve a and $N_P = 30\%$, $\Delta E_{el}(E_F) = 150$meV for the curve b. A comparison of Fig. 39 results with those obtained in Part II [194, 196] shows that they are the same. Similarly for the other situations discussed in Part II [194, 196], like $T \sim T_c$ where $\Delta E_{CP}$ scattering too is present (Appendix 5), also we get same results for the 3D $D(E_{el})$ and 2D $D(E_{el})$ cases. Thus the conclusions drawn on the basis of the 3D $D(E_{el})$ calculation results (Part II) [194, 196] are valid for the present 2D $D(E_{el})$ calculations also.

### (iii) Stripe phase

For the 3D case, the stripe phase is discussed in Part III [194, 196] and as mentioned there the 3D stripes will have tubular structure with 2D planar projection on Cu-O planes. However for the 2D case, the stripes will be located in Cu-O planes only. The other stripe properties, like bending of stripes, stripe fluctuation etc. (Sec. III.4) [194, 196], remain same for the 3D and 2D cases. Like 3D stripes, the 2D stripes will also have a width and run randomly in the Cu-O plane changing their location with time as the $Cu^{2+} \leftrightarrow Cu^{3+}$ fluctuation [194-196] occurs in superconducting cuprates.

## IV.3. CONCLUSION

In conclusion, the results obtained for the 3D $D(E_{el})$ case (Part II, III) [194, 196] are valid for the 2D $D(E_{el})$ case also. This is true even for those results which are not specifically described above. For example, as have been discussed in Sec. III.2 [194, 196] above a certain critical dopant concentration, in the overdoped region, pseudogap does not occur in cuprates. This absence has been explained there (Sec. III.2) [194, 196] on the basis of the decrease in $\Delta E_{el}$, $N_P$ values at higher dopant concentrations. In Fig. 40 we have repeated the 3D $D(E_{el})$ calculations for the 2D $D(E_{el})$ case using the same parameter values as those of 3D case (Sec. III.2) [194, 196]; T = 100K, $E_s = E_F = 310$ meV, $\Lambda = 15$ meV, $\alpha = 0.7$ (meV)$^{-1}$ and, $\Delta E_{el}(E_F) = 100$ meV, $N_P = 10\%$ for the curve a and $\Delta E_{el}(E_F) = 50$ meV, $N_P = 5\%$ for the curve b. A comparison of the Fig. 40 results with those of 3D $D(E_{el})$ results (Sec. III.2) [194, 196] shows that they are the same. Similar is the case with the other results also, obtained in Part II, III [194, 196] for the $T \geq T_c$ or $T < T_c$. Thus the results of the 3D case, and the conclusions drawn from them, are valid for the 2D case also. It may be mentioned here that the present results, and so also those of Part II, III [194, 196], are consistent with some recent experiments which link the pseudogap origin to the change in the electronic DOS near $E_F$ and indicate that the underdoped and overdoped cuprates have the same superconductivity origin [213]. Similarly the $\Delta\theta_D$ break at $T_{CF}$, mentioned in Part II [194-196] ($\theta_D$ = Debye temperature), can be attributed only to the presence of clusters, and cluster boundaries, interacting with CEs [194-196], since both the $\Delta\theta_D$ break and the CE- cluster, -cluster boundary, interaction are present in superconducting cuprates only whereas the regular charge ordered stripe phase (resulting from the absence of $Cu^{2+} \leftrightarrow Cu^{3+}$ type charge fluctuation) is present in nonsuperconducting cuprates where no $\Delta\theta_D$ break is observed [194] eventhough the experimental $T_{CF}$ (higher $\Delta\theta_D$ - break temperature) vs. doping concentration, monitored by superconducting samples' study, shows a slow variation (matching with the doping concentration dependence of $T^*$ (pseudogap temperature) seen by several other experimental techniques).Thus $T_{CF}$'s $\Delta\theta_D$ break can not be associated with stripe formation. It may also be mentioned here that the anomalies observed at $T_{CF}$ ($T^*$), $T_c$ are $\Delta\theta_D$ breaks only arising due to a change in lattice ion's r.m.s. vibrational amplitude, and its temperature dependence, and are not a result of any static lattice distortion at $T_{CF}$ or correlated vibration of lattice ions for $T \leq T_c$. This is because these anamolies are observed in Mössbauer f-factor vs. T measurements also as breaks, and change in curve's slope, at $T_{CF}$, $T_c$ (and Mössbauer f-factor senses only the r.m.s. vibrational amplitude of lattice ions and not their static distortion or correlated vibration). In addition, direct measurements of lattice parameters vs. T have also not shown any break at $T_{CF}$, $T_c$ in superconducting cuprates. Only techniques like Mössbauer effect (where Fe substitutes for Cu) or channeling (which can isolate the Cu-O row, or Cu row, selectively) can sense the $T_{CF}$-, $T_c$- $\Delta\theta_D$ break appreciably. Other techniques (bulk or otherwise) may not be able to get appreciable $\Delta\theta_D$- break effect in their studies. Even channeling measurements when carried out for the Y(Er)-Ba row, in Y(Er)Ba$_2$Cu$_3$O$_{7-\delta}$, do not find any appreciable $\Delta\theta_D$ break. This also rules out any lattice distortion, like the change in unit cell orthorhombicity, to be the cause of the $\Delta\theta_D$ break since any such distortion, if present, would have affected all the unit cell ions in which case $\Delta\theta_D$ break would have existed for the Y(Er)-Ba row also once it existed for the Cu-O row. It may additionally be mentioned here that whereas the stripe phase theory pedicts a stripe formation temperature very much higher than the pseudogap temperature, with $\Delta'$ (difference between the two temperatures) increasing with decreasing doping concentration, in the underdoped region [168], experimentally wherever stripes (dynamical) have been observed, by experiments like neutron diffraction and NMR, in the underdoped region, the pseudogap temperature and the stripe formation temperature have been found to exist simultaneously; any minute difference in their values, if at all present, may be arising due to the difference in the way the two temperatures are sensed by the experimental technique. This too speaks against the stripe phase theory prediction and so against the $T_{CF}$ - $\Delta\theta_D$ break's association with stripe formation. Similarly it may further be mentioned that the Mössbauer hyperfine field vs. T measurements too favour the earlier discussed low temperature SG freezing of the CBIs.

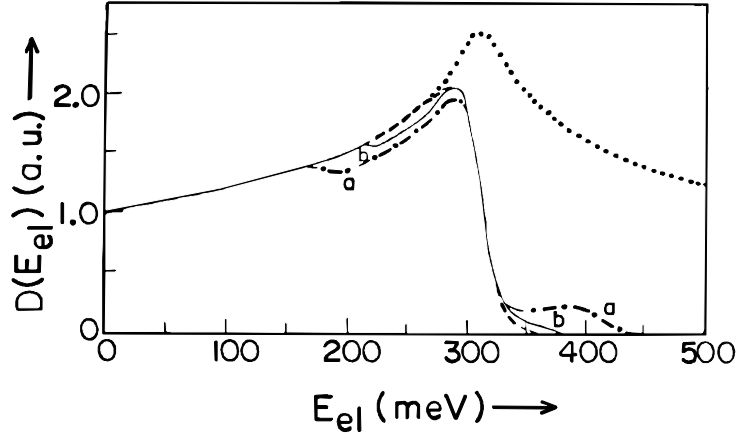

Fig.40 Dependence of the electronic density of states, $D(E_{el})$, on electrons' energy, $E_{el}$, for $T>T_c$ and dopant concentration > critical dopant concentration; $T_c$=critical temperature, a.u.= arbitrary unit. Details are described in the text.

### IV.4. SUMMARY

The details of the pseudogap origin and other gap related properties discussed in earlier parts for cuprates, in the framework of the paired cluster (PC) model, using three dimensional (3D) electronic density of states (DOS) [194-196], are valid even when a two dimensional (2D) cuprate electronic DOS is used. Similarly the stripe phase description is also similar for the 3D and 2D cases. It may be mentioned here that same results are obtained if extended van Hove singularity is used in the DOS calculation instead of logarithmic singularity.

### V. PAIRED CLUSTER MODEL EXPLANATION OF HIGH-$T_c$ MAGNETIC SUPERCONDUCTIVITY

Recently intense studies have been going on for high-$T_c$ ruthenocuprates as they show coexistance of magnetism and superconductivity, and also because people are trying to find out if high-$T_c$ superconductivity can be achieved without Cu-O planes i.e. only by Ru-O planes. We find that our paired cluster (PC) model is able to explain, in a very natural way, the properties of high-$T_c$ ruthenocuprates including its magnetic superconductivity. These details are described in Appendix 1 (xviii).

### C. CONCLUDING REMARKS

Thus as an overall conclusion we find that the PC model is capable of explaining the HTSC properties either in its present form or in a somewhat modified form. Predictions of the model [194-196] made some years ago, like the coexistence of pseudogap and superconducting state energy gap below $T_c$, are being found true by the experiments now confirming the correctness of the model. *Several other such results too point in the same direction.*

### ACKNOWLEDGEMENTS

The author is extremely grateful to all his collaborators in India and abroad for their very fruitful collaboration in his spin glass and HTSC research programmes. Thanks are also due to all those with whom the author has ever discussed these topics either orally or through correspondance. The author is indebted to those who have helped him at any time in his scientific pursuit. He dedicates this article to the departed millennium which has seen the birth & growth of his research activities culminating into what seems to be the unfolding of the HTSC mystery. It has been a hard work carried out under difficult circumstances and with time coming nearer for hanging off the shoes, the author would like to be permitted to quote, "*With them the Seed of Wisdom did I sow, And with my own hand laboured it to grow, And this was all the Harvest that I reaped, I came like Water and like Wind I go.*" (From Rubaiyat of Omar Khayyam). Finally, the author leaves it to the world to evaluate the efforts concerning putting the HTSC physics in right perspective. It was an Arabian sea (Institute (TIFR) sea shore) sunset with which the development of the model was started and when the work was completed to its present form, it was again a sunset at the same Institute site. Between the two sunsets, days flew and years passed, seasons changed, close ones left, new ones joined – all without getting noticed which shows a complete mental engagement and the people who suffered most due to this mental engrossment were the family members. The author acknowledges with deep gratitude their patience and understanding. To end, it is a pleasure to bring out this work to the attention of the people in this World Year of Physics (2005).

## APPENDIX 1

### (i) Works Concerning Problems with the Spin Bag Model

Some typical works discussing the limitations of spin bag model are, Solid St. Commun. **65**, 343 (1988); ibid **67**, 1059 (1988); Phys. Rev. Lett. **61**, 2813 (1988); Phys. Rev. B **40**, 10796 (1989); ibid **42**, 6408 (1990); ibid **62**, 9059 (2000); ibid **42**, 6222 (1990).

### (ii) Works Concerning Problems with the van Hove Singularity Based Mechanisms

Some typical works discussing the limitations of van Hove singularity based mechanisms are, Phys. Rev. B **48**, 15957 (1993); ibid **44**, 12565 (1991); ibid **48**, 16557 (1993); ibid **49**, 6864 (1994); ibid **50**, 9554 (1994); ibid **46**, 5466 (1992); ibid **44**, 12525 (1991); ibid **48**, 13127 (1993); ibid **53**, R 11972 (1996); Phys. Rev. Lett. **68**, 1089 (1992); ibid **68**, 1090 (1992); ibid **85**, 3692 (2000).

### (iii) Works Concerning Problems with the Antiferromagnetic Spin Fluctuation Based Mechanism

Some typical works discussing the limitations of antiferromagnetic spin fluctuation based mechanism are, Physica C **235-240**, 71 (1994); ibid **153-155**, 21 (1988); ibid **282-287**, 1551 (1997); J. Low Temp. Phys. **99**, 397 (1995); ibid **105**, 699 (1996); Phys. Rev. B **51**, 9253 (1995); ibid **52**, 13043 (1995); ibid **48**, 15957 (1993); ibid **42**, 6196 (1990); ibid **41**, 7264 (1990); ibid **53**, 6786 (1996); ibid **47**, 9124 (1993); ibid **46**, 11975 (1992); ibid **53**, 5137 (1996); ibid **43**, 10530 (1991); Phys. Rev. Lett. **74**, 2303 (1995); Sov. Phys. JETP **71**, 519 (1990); "Electron Correlations in Atoms and Solids", eds. A. N. Tripathi & Ishwar Singh (Phoenix Pub., New Delhi, 1998).

### (iv) Works Concerning Problems with the RVB Model

Some typical works discussing the limitations of RVB model are, Phys. Rev. Lett. **87**, 197002 (2001); J. Phys. Condens. Matter **10**, 11385 (1998); P. Brusov, "Mechanisms of High Temperature Superconductivity, Vol. II" (Rostov State Uni. Pub., Rostov on Don, 1999), pp. 14, 66, 176, 455.

### (v) Works Concerning Problems with the Preformed Cooper Pairs Based Mechanism

Some typical works discussing the limitations of preformed Cooper pairs based mechanism are, Physica C **282-287**, 1215 (1997); Phys. Rev. B **55**, 3173 (1997); ibid **61**, 9748 (2000); ibid **64**, 134504 (2001); ibid **54**, R6909 (1996); ibid **52**, R3884 (1995); ibid **60**, R9947 (1999); Phys. Rev. Lett. **82**, 3725 (1999); ibid **86**, 1614 (2001); ibid **85**, 4787 (2000); ibid **86**, 2657 (2001); ibid **82**, 4918 (1999); ibid **84**, 5860 (2000); ibid **86**, 5763 (2001); ibid **82**, 2784 (1999); ibid **80**, 377 (1998); ibid **67**, 3140 (1991); Physica C **341-348**, 831 (2000); cond. mat/0103507 (D. N. Basov et. al.); Phys. Rev. Lett. **85**, 2805 (2000); ibid **84**, 5848 (2000); ibid **82**, 177 (1998).

### (vi) Works Concerning Problems with the Hund's Coupling Based Mechanism

Some typical works discussing the limitations of Hund's coupling based mechanism are, Solid St. Commun. **74**, 371 (1990).

### (vii) Works Arguing for s- or d- or Mixed s, d- Wave Superconductivity

Some typical works discussing the s- or d- or mixed s, d- wave superconductivity are, Physica C **282-287**, 1649 (1997); ibid **282-287**, 1701 (1997); ibid **282-287**, 1487 (1997); ibid **282-287**, 1703 (1997); ibid **282-287**, 1841 (1997); ibid **282-287**, 1701 (1997); ibid **282-287**, 1843 (1997); ibid **282-287**, 1515 (1997); ibid **282-287**, 1509 (1997); ibid **282-287**, 4 (1997); ibid **235-240**, 1899 (1994); Phys. Rev. B **36**, 2386 (1987); ibid **51**, 8680 (1995); ibid **54**, 16216 (1996); ibid **54**, R9682 (1996); ibid **45**, 5585 (1992); ibid **53**, 5137 (1996); ibid **53**, R522 (1996); ibid **41**, 11128 (1990); Phys. Rev. Lett. **70**, 85 (1993); ibid **83**, 4168 (1999); ibid **85**, 182 (2000); ibid **82**, 4914 (1999); ibid **83**, 2644 (1999); ibid **78**, 721 (1997); ibid **74**, 797 (1995); ibid **72**, 1084 (1994); ibid **75**, 1626 (1995); ibid **83**, 4160 (1999); ibid **83**, 616 (1999); ibid **71**, 2134 (1993); ibid **74**, 4523 (1995); ibid **74**, 1008 (1995); ibid **71**, 4277 (1993); ibid **71**, 4278 (1993); ibid **88**, 107002 (2002); J. Low Temp. Phys. **105**, 527 (1996); ibid **105**, 733 (1996); ibid **105**, 545 (1996); ibid **105**, 539 (1996); Science **261**, 337 (1993); Nuovo Cimento D **19**, 1131 (1997).

### (viii) Works Concerning Problems with the Nonphononic, Electronic or Magnetic Mechanisms

Some typical works discussing the problems with the nonphononic, electronic or magnetic mechanisms are, Solid St. Commun. **67**, 1073 (1988); ibid **75**, 971 (1990); ibid **76**, 1019 (1990); Phys. Rev. B **44**, 4473 (1991); Sov. Phys. JETP **71**, 519 (1990); Phys. Rev. Lett. **82,** 2784 (1999).

### (ix) Works Concerning Problems with the Bipolaron Model

Some typical works discussing the problems with the bipolaron model are, Phys. Rev. Lett. **81**, 433 (1998); Phys. Rev. B **44**, 5148 (1991); ibid **40**, 11378 (1989).

### (x) Works Discussing Intrinsic Josephson Couplings Presence

Some typical works discussing intrinsic Josephson effects are, Physica C **282-287**, 2431 (1997); ibid **235-240**, 289 (1994); ibid **282-287**, 2293 (1997); Phys. Review B **50**, 3511 (1994); Science **261**, 337 (1993).

### (xi) Works Discussing Electron-Phonon Interaction Importance

Some typical works discussing electron phonon coupling are, Physica C **235-240**, 1325 (1994); Phys. Rev. Lett. **84**, 4192 (2000); ibid **86**, 4899 (2001); ibid **69**, 359 (1992); Phys. Rev. B **42**, 8702 (1990).

### (xii) Works Describing Antiferromagnetic Spin Fluctuation Based Mechanism

Some typical works describing antiferromagnetic spin fluctuation based mechanism are, Phys. Rev. B **50**, 16015 (1994); ibid **48**, 3896 (1993); ibid **42**, 6711 (1990); ibid **47**, 6069 (1993); ibid **43**, 275 (1991); ibid **49**, 426 (1994); ibid **46**, 14803 (1992); Phys. Rev. Lett. **69**, 961 (1992); ibid **71**, 208 (1993); ibid **67**, 3448 (1991); ibid **74**, 2303 (1995); J. Phys. Condens. Matter **8**, 10017 (1996); Z. Phys. B **103**, 129 (1997); Physica C **263**, 277 (1996).

### (xiii) Works Concerning Fermi Liquid Picture

Some typical works concerning Fermi liquid / near Fermi liquid picture are, Phys. Rev. B **45**, 4930 (1992); ibid **44**, 4727 (1991); ibid **42**, 1033 (1990); ibid **39**, 11633 (1989); ibid **48**, 3356 (1993); P. Brusov,

"Mechanisms of High Temperature Superconductivity, Vol. II" (Rostov State Uni. Pub., Rostov on Don, 1999), pp. 455, 456; Physica. C **161**, 415 (1989); ibid **153-155**, 21 (1988); ibid **162-164**, 300 (1989); ibid **263**, 277 (1996); J. Phys. Condens. Matter **8**, 10017 (1996); Phys. Rev. Lett. **63**, 1700 (1989); ibid **82**, 5345 (1999); ibid **82**, 5349 (1999); Science **245**, 731 (1989).

### (xiv) Recent Works Supporting the Predictions of Paired Cluster Model (made in 1997) Concerning the Coexistence of Pseudogap and Superconducting State Energy Gap Below $T_c$

Some recent experiments supporting the predictions of paired cluster model made in 1997 concerning the coexistence of pseudogap and superconducting state energy gap below $T_c$ are, Phys. Rev. Lett. **85**, 4787 (2000); ibid **86**, 4911 (2001); ibid **86**, 2657 (2001); ibid **82**, 4918 (1999); ibid **84**, 5860 (2000).

### (xv) Works Showing Different Techniques may Sense the Pseudogap Formation Temperature Somewhat Differently

Some typical works showing that the different techniques may sense the pseudogap temperature somewhat differently are, Phys. Rev. Lett. **84**, 5848 (2000).

### (xvi) Works Describing the d- Density Wave (DDW) Model and the Problems Associated with it

Some typical works describing the d- density wave (DDW) model and its limitations are, Phys. Rev. B **63**, 094503 (2001); Phys. Rev. Lett. **87**, 077004 (2001); ibid **88**, 057002 (2002); ibid **88**, 097002 (2002); Phys. Rev. B **71**, 134503 (2005).

### (xvii) Works Concerning Problems with the Stripe Phase Picture

Some typical works discussing the problems with the stripe phase picture are, Phys. Rev. B **65**, 020509(R) (2001).

### (xviii) Works Concerning Paired Cluster (PC) Model Explanation of High-$T_c$ Magnetic Superconductivity

Some typical works discussing the paired cluster model explanation of high-$T_c$ magnetic superconductivity are, Phys. Letters A **324**, 71 (2004).

## APPENDIX 2

The infinite range *GT* m-dimensional vector Hamiltonian for a two sublattice system can be written as

$$H = -\sum_{(ij)} J_{ij}^A \sum_\mu S_{i\mu}^A S_{j\mu}^A - \sum_{(ij)} J_{ij}^B \sum_\mu S_{i\mu}^B S_{j\mu}^B - \sum_{i,j} J_{ij}^{AB} \times \sum_\mu S_{i\mu}^A S_{j\mu}^B - H^A \sum_i S_{i1}^A - H^B \sum_i S_{i1}^B, \qquad (A2.1)$$

where the A and B sublattices refer to A and B site spins, being *N* in number for each site, and $\sum_{\mu=1}^m S_\mu^2 = m$; m=3 for Heisenberg lattice. Each $J_{ij}$ is independently distributed with the Gaussian probability distribution,

$$p(J_{ij}^{A,B,AB}) = \frac{1}{J^{A,B,AB}(2\pi)^{1/2}} \exp(\frac{-(J_{ij}^{A,B,AB} - J_0^{A,B,AB})^2}{2(J^{A,B,AB})^2}). \tag{A2.2}$$

Writing

$$J_0^A = \tilde{J}_0^A/N, \qquad J_0^B = \tilde{J}_0^B/N, \qquad J_0^{AB} = \tilde{J}_0^{AB}/2N,$$
$$J^A = \tilde{J}^A/\sqrt{N}, \qquad J^B = \tilde{J}^B/\sqrt{N}, \qquad J^{AB} = \tilde{J}^{AB}/(2N)^{1/2},$$

and using the standard replica symmetric approach( Phys. Rev. Lett. **49**, 158 (1982); ibid **47**, 201 (1981); ibid **35**, 1792 (1975); Phys. Rev. B **17**, 4384 (1978)), the free energy per spin is obtained as,

$$f = \frac{\tilde{J}_0^{AB}}{2} \sum_\mu \frac{1}{2}(2(M_\mu^{A,\alpha})_0 (M_\mu^{B,\alpha})_0 + \frac{3d_0^A - (d_0^A)^2}{d_0^A + 1}(M_\mu^{A,\alpha})_0^2$$

$$+ \frac{3d_0^B - (d_0^B)^2}{d_0^B + 1}(M_\mu^{B,\alpha})_0^2) + \frac{(\tilde{J}^{AB})^2}{4kT} \sum_\mu [\frac{1}{2}(2(q_\mu^{A,\alpha})_0 (q_\mu^{B,\alpha})_0$$

$$+ \frac{3(d^A)^2 - (d^A)^4}{(d^A)^2 + 1}(q_\mu^{A,\alpha})_0^2 + \frac{3(d^B)^2 - (d^B)^4}{(d^B)^2 + 1}(q_\mu^{B,\alpha})_0^2)$$

$$- (2(q_\mu^{A,(\alpha\beta)})_0 (q_\mu^{B,(\alpha\beta)})_0 + \frac{3(d^A)^2 - (d^A)^4}{(d^A)^2 + 1}$$

$$\times (q_\mu^{A,(\alpha\beta)})_0^2 + \frac{3(d^B)^2 - (d^B)^4}{(d^B)^2 + 1}(q_\mu^{B,(\alpha\beta)})_0^2)]$$

$$- \frac{kT}{2(2\pi)^m} \int_{-\infty}^{\infty} (\prod_\mu dt_\mu^A dt_\mu^B) \exp\{-\frac{1}{2} \sum_\mu [(t_\mu^A)^2 + (t_\mu^B)^2]\} \ln Q_0, \tag{A2.3}$$

where

$$Q_0 = Q_0^A Q_0^B = m(2\pi)^{m-1}(|a^A|_{m-1})^{(3-m)/2}(|a^B|_{m-1})^{(3-m)/2}$$

$$\times (\int_{-\sqrt{m}}^{\sqrt{m}} ds^A \exp\{a_1^A s^A + b_1^A (s^A)^2 + b_{\mu>1}^A [m - (s^A)^2]\}$$

$$\times [m - (s^A)^2]^{(m-3)/4} I_{(m-3)/2}\{|a^A|_{m-1}[m - (s^A)^2]^{1/2}\})$$

$$\times (\int_{-\sqrt{m}}^{\sqrt{m}} ds^B \exp\{a_1^B s^B + b_1^B (s^B)^2 + b_{\mu>1}^B [m - (s^B)^2]\}$$

$$\times [m - (s^B)^2]^{(m-3)/4} I_{(m-3)/2}\{|a^B|_{m-1}[m - (s^B)^2]^{1/2}\}),$$

$$a_\mu^A = (\tilde{J}^{AB}/kT)[(q_\mu^{B,(\alpha\beta)})_0 + (d^A)^2 (q_\mu^{A,(\alpha\beta)})_0]^{1/2} t_\mu^A$$
$$+ (\tilde{J}_0^{AB}/kT)[(M_\mu^{B,\alpha})_0 + d_0^A (M_\mu^{A,\alpha})_0] + (H^A/kT)\delta_{1,\mu},$$

$$a_\mu^B = (\tilde{J}^{AB}/kT)[(q_\mu^{A,(\alpha\beta)})_0 + (d^B)^2(q_\mu^{B,(\alpha\beta)})_0]^{1/2} t_\mu^B$$
$$+ (\tilde{J}_0^{AB}/kT)[(M_\mu^{A,\alpha})_0 + d_0^B(M_\mu^{B,\alpha})_0] + (H^B/kT)\delta_{1,\mu},$$

$$b_\mu^A = [(\tilde{J}^{AB})^2/2(kT)^2][(q_\mu^{B,\alpha})_0 + (d^A)^2(q_\mu^{A,\alpha})_0 - (q_\mu^{B,(\alpha\beta)})_0 - (d^A)^2(q_\mu^{A,(\alpha\beta)})_0],$$

$$b_\mu^B = [(\tilde{J}^{AB})^2/2(kT)^2][(q_\mu^{A,\alpha})_0 + (d^B)^2(q_\mu^{B,\alpha})_0 - (q_\mu^{A,(\alpha\beta)})_0 - (d^B)^2(q_\mu^{B,(\alpha\beta)})_0],$$

$$d_0^A = \tilde{J}_0^A/\tilde{J}_0^{AB}, \qquad d_0^B = \tilde{J}_0^B/\tilde{J}_0^{AB}, \qquad d^A = \tilde{J}^A/\tilde{J}^{AB}, \qquad d^B = \tilde{J}^B/\tilde{J}^{AB}.$$

$$b_{\mu>1}^{A,B} = b_2^{A,B} \text{ or } b_3^{A,B} \text{ or..... } b_m^{A,B}.$$

$I_\nu(z)$ is a modified Bessel function of the first kind,

$$\left|a^{A,B}\right|_{m-1} = [(a_2^{A,B})^2 + (a_3^{A,B})^2 + \ldots\ldots + (a_m^{A,B})^2]^{1/2},$$

$$(M_\mu^{A,\alpha})_0 = \left\langle\left\langle s_\mu^A\right\rangle_T\right\rangle_d, \qquad (M_\mu^{B,\alpha})_0 = \left\langle\left\langle s_\mu^B\right\rangle_T\right\rangle_d,$$

$$(q_\mu^{A,\alpha})_0 = \left\langle\left\langle (s_\mu^A)^2\right\rangle_T\right\rangle_d, \qquad (q_\mu^{B,\alpha})_0 = \left\langle\left\langle (s_\mu^B)^2\right\rangle_T\right\rangle_d,$$

$$(q_\mu^{A,(\alpha\beta)})_0 = \left\langle\left\langle s_\mu^A\right\rangle_T^2\right\rangle_d, \qquad (q_\mu^{B,(\alpha\beta)})_0 = \left\langle\left\langle s_\mu^B\right\rangle_T^2\right\rangle_d.$$

$\langle...\rangle_T$ refers to the thermal average and $\langle...\rangle_d$ refers to the average over the bond disorder i.e. over $p(J_{ij})$. By putting A = B in the above expression, the free energy per spin given by GT theory for one sublattice case (Section I.1, Phys. Rev. Lett. **49**, 158 (1982); ibid **47**, 201 (1981)) is easily obtained.

Assuming $H^A = H^B = H = 0$, $\tilde{J}_0^A = \tilde{J}_0^B = \tilde{J}_0^{AB} = 0$ and $\tilde{J}^A > \tilde{J}^B$, for the $\tilde{J}^{AB} < (\tilde{J}^A \tilde{J}^B)^{1/2}$ case,

$$T_{SG}^A = (1/\sqrt{2})\left[(d^A)^2 + (d^B)^2 + \left\{\left[(d^A)^2 - (d^B)^2\right]^2 + 4\right\}^{1/2}\right]^{1/2},$$

$$T_{SG}^B = (1/\sqrt{2})\left[(d^A)^2 + (d^B)^2 - \left\{\left[(d^A)^2 - (d^B)^2\right]^2 + 4\right\}^{1/2}\right]^{1/2}. \tag{A2.4}$$

## APPENDIX 3

### (a) Calculation for Cluster Ions

#### (i) T ~ 220 K

Though the calculations are given for typical parameter values, they are valid for other values too and thus the results mentioned are general in nature. For T ~220K anomaly, assuming T = 220K = $T_{CF}$ = cluster $T_C$ = $T_{C2}$ and Brillouin function $B_{1/2}(x)$ dependence for $Cu^{2+}$ (S = 1/2, g = 2) magnetisation, $H_W(0)$

= $H_W$ (T = 0K) = 3275.21kOe and $H_W(T)/H_W(0)$ = 0.05 for $T/T_C$ = 0.995, giving $H_W$ (219K) = 163.76kOe. For a typical r = 0.2Å, $H_{CE}$ = 132.5kOe for $v_{el} = v_F$ (Fermi Velocity) = $0.33 \times 10^8$ cm/s ($YBa_2Cu_3O_7$ system) [60]. $r_{Cu^{2+}-Cu^{2+}}$ = 3.88Å, 3.82Å along x-, y- axes in Cu-O plane and chain which gives $H_{dip}$ = 0.841kOe at a central $Cu^{2+}$ ion from the antiparallel nn and parallel next nn spins. When CE reaches near ion 1 (r = 0.2Å), the energy emitted by ion 1 per second, $E_{e1} = (n_1 - n'_1)[1-\exp(-t/\tau)]\Delta'_1 N_{CE}$, where $n_1 = (\frac{s_1}{1+s_1})$, $s_1 = \exp(-\frac{\Delta_1}{k_B T})$, $n'_1 = (\frac{s'_1}{1+s'_1})$, $s'_1 = \exp(-\frac{\Delta'_1}{k_B T})$, t = time difference between the successive arrival of two CEs at the ion 1 site, τ = ion 1 spin relaxation time for the process which takes away the emitted energy and $N_{CE}$ = number of CEs passing through the ion 1, or ion 2, per second; $N_{CE} = 7.57 \times 10^{14}$ and $t = \frac{1}{N_{CE}} = 1.321 \times 10^{-15}$s since $YBa_2Cu_3O_7$ has one CE per unit cell.

The emitted energy is taken by the lattice through the spin lattice coupling ($H_{SL}$) and by the CE through $H_{\mu_{eff,CE}-\mu_{Cu^{2+}}}$. For the orbital singlet ground state of $Cu^{2+}$, the spin lattice relaxation time $\tau_{SL} \sim 10^{-3}$s and the spin lattice coupling energy, responsible for $\tau_{SL}$, $E_{SL} \cong 8.75 \times 10^{-17}$ erg. This is an order of magnitude estimate for $YBa_2Cu_3O_7$ system for T ~220K [60,81-83,109]. $E_{\mu_{eff,CE}-\mu_{Cu^{2+}}} \cong 1.23 \times 10^{-15}$ erg ~14 $E_{SL}$ for r = 0.2Å. Thus $\tau_{\mu_{eff,CE}-\mu_{Cu^{2+}}} \sim \frac{1}{14}\tau_{SL} \sim 0.5 \times 10^{-4}$s. Since $\tau^{-1} = \tau_{SL}^{-1} + \tau^{-1}_{\mu_{eff,CE}-\mu_{Cu^{2+}}}$, $\tau \sim 0.5 \times 10^{-4}$s. These are order of magnitude estimates; exact relaxation time calculation is generally not possible [81-83,158].

When CE has moved away ($H_{CE}$ = 0), the energy absorbed by ion 1 per second, $E_{a1} = (n_1 - n'_1)[1-\exp(-t/\tau')]\Delta_1 N_{CE}$, where τ′ = ion 1 spin relaxation time for the process through which the ion 1 absorbs energy = $\tau_{SL} \sim 10^{-3}$s. When CE reaches near ion 2 (r = 0.2Å), the energy absorbed by ion 2 per second, $E_{a2} = (n'_2 - n_2)[1-\exp(-t/\tau')]\Delta'_2 N_{CE}$, where $n'_2 = (\frac{s'_2}{1+s'_2})$, $s'_2 = \exp(-\frac{\Delta'_2}{k_B T})$, $n_2 = n_1$ (∵ $\Delta_2 = \Delta_1$) and τ′ = $\tau_{SL} \sim 10^{-3}$s since ion 1, ion 2 relaxation times are same. When CE has moved away ($H_{CE}$ = 0), the energy emitted by ion 2 per second, $E_{e2} = (n'_2 - n_2)[1-\exp(-t/\tau)]\Delta_2 N_{CE}$, where τ ~ $0.5 \times 10^{-4}$s. Thus the energy emitted per ion per second by the ion pair 1, 2 (τ = $0.5 \times 10^{-4}$s, τ′ = $10^{-3}$s), $E_e = (E_{e1}+E_{e2})/2 = 17.26 \times 10^{-13}$ erg and the energy absorbed per ion per second by the ion pair 1, 2, $E_a = 3.674 \times 10^{-14}$ erg. Channeling experiments [89] show $\Delta\theta_D$ = 80K = $1.104 \times 10^{-14}$ erg when T decreases below ~220K. Thus out of $E_e$, the energy which goes to the lattice through collisions and spin lattice coupling = $(3.674-1.104) \times 10^{-14} = 2.57 \times 10^{-14}$ erg and the energy which goes to $N_{CE}$ CEs = $17.003 \times 10^{-13}$ erg giving $\Delta E_{el} = 17.003 \times 10^{-13} \times \frac{N'_{CE}}{N_{CE}}$ erg, where $N'_{CE}$ = number of $Cu^{2+}$ ions which CE meets per second = $2.88 \times 10^{14}$; this means an increase in $v_{el}$ from $v_F$ to $1.5135 v_F$ ($0.4994 \times 10^8$ cm/s). Thus, $(E_{el})_{T>220K} = E_F = 4.96 \times 10^{-13}$ erg [60] and $(E_{el})_{T<220K} = E_F + \Delta E_{el} = 11.36 \times 10^{-13}$ erg. Using channeling data [89], $(E_{ph})_{T>220K} = k_B(\theta_D)_{T>220K} = 4.694 \times 10^{-14}$ erg and $(E_{ph})_{T<220K} = 5.8 \times 10^{-14}$erg. This gives, $(\mu^*)_{T>220K} = 0.2774$ and $(\mu^*)_{T<220K} = 0.2179$. Our calculations show for this μ* decrease, a ~25% $T_c$ increase for λ = 2.5 and a ~30% $T_c$ increase for λ = 4.7 in F approximation, the effect being more pronounced in L approximation. Further a ~33% μ* decrease decreases λ by ~25% for obtaining the same value of $T_c$ (~96K) in $C_F$ case, the effect being larger in $C_L$ case. If we use τ = $10^{-4}$s in above calculation, $(\mu^*)_{T>220K} = 0.2774$ and $(\mu^*)_{T<220K} = 0.2394$ showing a 13.7% μ* decrease below 220 K. Also calculations done for $H_W(T) = 0.1 H_W(0)$ ($T/T_C \sim 0.99$) give similar result. For r > 1Å, $H_{SL}$ becomes comparable to, or larger than, $H_{\mu_{eff,CE}-\mu_{Cu^{2+}}}$ but in that case the energy emission and absorption are very small due to small $H_{CE}$.

(ii) **T ~ $T_c$**

The number of CEs which have formed CPs at $T_c$ can be estimated as follows. Since $YBa_2Cu_3O_7$ has one CE per unit cell, $n_{CE}$ = number of CEs per $cm^3$ = $5.762 \times 10^{21}/cm^3$; this agrees with experimental value [60]. If $n_{CPCE}(0)$ = number of CEs which form CPs at T = 0K, to a good approximation $n_{CPCE}(0)/n_{CE}$ = $1.5(\hbar\omega_D/E_F)$, where $\hbar\omega_D = k_B\theta_D$ and $\theta_D$ = 450K (average value) [60,66]. $\therefore n_{CPCE}(0) = 1.0824 \times 10^{21}/cm^3$. Since $2\hbar\omega_D = 2\Delta(0)$ = BCS T = 0K energy gap, $n_{CPCE}(T)/n_{CPCE}(0) = \Delta(T)/\Delta(0)$. BCS theory gives for $T/T_c \cong 0.99$, $\Delta(T)/\Delta(0) = 0.1$ [65]. For $YBa_2Cu_3O_7$ system, $T_c \sim 93K$. $\therefore$ For $T \sim 92K$, $n_{CPCE}(T) = 1.0824 \times 10^{20}/cm^3$. $\therefore n'_{CE}(T)$ = number of CEs which have not formed CPs = $n_{CE} - n_{CPCE}(T) = 5.65376 \times 10^{21}/cm^3$. $\therefore$ Fraction of CEs which do not form CPs, $f_{n'CE} = n'_{CE}(T)/n_{CE} = 0.98121$ and the fraction of CEs which form CPs = $f_{nCPCE} = n_{CPCE}(T)/n_{CE} = 0.01878$. The fraction of CPs, $f_{nCP} = \frac{1}{2} f_{nCPCE}$.

We first consider the effect of CEs ($n'_{CE}$) on the energy emission and absorption. For $T \sim 92K$, $T/T_C \sim 0.42$. $\therefore H_W(T)/H_W(0) = 0.96$ giving $H_W = 3144.2$ kOe. $H_{dip} = 0.841$ kOe and $H_{CE} = 132.5$ kOe. $\tau = 0.5 \times 10^{-4}$s and $\tau' = 2 \times 10^{-3}$s, being larger than the T = 220K value due to decreased T [81-83, 158]. This gives, the fraction of the energy which is emitted per ion per second by the ion pair 1, 2 due to its interaction with the CEs, $E_{e,CE} = 2.2964 \times 10^{-12} \times f_{n'CE}$ erg = $2.2532 \times 10^{-12}$ erg and the fraction of the energy which is absorbed per ion per second by the ion pair 1, 2 due to interaction with CEs, $E_{a,CE} = 0.5504 \times 10^{-13} x f_{n'CE}$ erg = $0.54 \times 10^{-13}$ erg.

We now consider the effect of CP CEs on the energy emission and absorption. As above, $H_{dip} = 0.841$ kOe, $H_W = 3144.2$ kOe, $\tau = 1.5 \times 10^{-4}$s due to smaller $v_{CP}$ (CP velocity) and $\tau' = 2 \times 10^{-3}$s. The magnetic field produced by the CP CEs, $H_{CP} = \frac{2ev_{CP}}{r^2}$ where $v_{CP}$ = CP velocity = $5.505 \times 10^6$ cm/s for $T \sim 92K$ since CP energy, $E_{CP} = k_BT$. $\therefore H_{CP} = 44.1$ kOe for r = 0.2Å. This gives, the fraction of the energy which is emitted per ion per second by the ion pair 1,2 due to its interaction with the CP CEs, $E_{e,CP} = 2.5038 \times 10^{-13} \times f_{nCP}$ erg = $2.3511 \times 10^{-15}$ erg, and the fraction of the energy absorbed, $E_{a,CP} = 1.7387 \times 10^{-16}$ erg. $\therefore$ Total energy emitted by the ion pair 1,2 per ion per second, $E_{e,T} = E_{e,CE} + E_{e,CP} = 22.5556 \times 10^{-13}$ erg and the total energy absorbed from the lattice by the ion pair 1,2 per ion per second, $E_{a,T} = E_{a,CE} + E_{a,CP} = 5.4175 \times 10^{-14}$ erg. Channeling data [89] shows $\Delta\theta_D = 220K = 303.71 \times 10^{-16}$ erg at $T_c$. Thus out of $E_{e,T}$, the energy which goes to the lattice per ion per second, $E_{e,T,L} = (541.7544 - 303.71) \times 10^{-16}$ erg = $238.044 \times 10^{-16}$ erg. $\therefore E_{e,T,L}/E_{e,T} = 1.06\%$. This quantity ($E_{e,L}/E_e$) is 1.49% at $T \sim 220K$. The smaller ratio at $T_c$ arises due to $\sim$ zero contribution to $E_{e,T,L}$ from $E_{e,CP}$ and decreased CE-lattice ion and CE-CE collisions owing to smaller T and CPs' presence. Further, the relation $H_{CP} = (2e)v_{CP}/r^2$ assumes shortest CP diameter case where both the CP CEs (CP partners) affect the ion 1 or ion 2 simultaneously. For the other case where the two CP partners affect two separate ions (1,1′ or 2,2′) at any instant, different possibilities exist like ions 1,1′(or 2,2′) having parallel spins or antiparallel spins. We have done calculations for all these cases and the results obtained are the same as given above.

**(b) Calculation for Cluster Boundary Ions**

(i) **T ~ 220 K**

For cluster boundary ions, $H_W = 0$ (paramagnetic state (Sec. II.5)) and $H_{dip}$, produced on a $Cu^{2+}$ ion by its singlet partner, = 0.32 kOe for average $r_{Cu2+-Cu2+}$ = 3.85Å. $H_{CE} = 132.5$ kOe for r = 0.2Å and $\tau = 0.5 \times 10^{-4}$ s, $\tau' = 10^{-3}$ s. In this case, $E_e = 19.89 \times 10^{-13}$ erg = 14411.5 K and $E_a = 1.2 \times 10^{-16}$ erg = 0.87 K which is a negligible number as compared to $E_e$ and $\Delta\theta_D$ = 80 K. Same result is obtained if other neighbouring ions are considered for $H_{dip}$ calculation producing broadening of the central $Cu^{2+}$ ion's levels.

(ii) **T~$T_c$**

For this case, $H_W = 0$, $H_{dip} = 0.32$ kOe, $H_{CE}$ (r = 0.2Å) = 132.5 kOe, $H_{CP}$ (r = 0.2Å) = 44.1 kOe, $\tau = 0.5 \times 10^{-4}$s for $E_{e,CE}$ and $1.5 \times 10^{-4}$s for $E_{e,CP}$ due to smaller $v_{CP}$, $\tau' = 2 \times 10^{-3}$s, $f_{n'CE} = 0.98121$ and $f_{nCP}$ =

0.01878/2. $\therefore$ $E_{e,T}$ = (237.274 $f_{n'CE}$ + 8.807 $f_{nCP}$) x $10^{-14}$ erg =(232.8156+ 0.0827) x $10^{-14}$ erg = 2.329 x $10^{-12}$ erg = 16870.58 K and $E_{a,T}$ = (0.0702 + 0.0002) x $10^{-15}$ erg = 0.0704 x $10^{-15}$ erg = 0.51 K which is again negligibly small compared to $E_{e,T}$ and $\Delta\theta_D$ = 220 K. Thus clusters' presence (i.e. a nonzero $H_W$ at the singlet coupled ions' site) is necessary for a nonzero $\Delta\theta_D$.

## APPENDIX 4

As has been mentioned in text (Sec. II.5), DOS calculations have been done for various $\Delta E_{el}$ vs. $E_{el}$ variations and same results have been obtained as the ones described in the text (Sec. II.5, III.2, IV.2) assuming a quick rising fast saturating exponential $\Delta E_{el}$ vs. $E_{el}$ variation. To give an example, we present here the DOS results obtained by assuming a linear dependence of $\Delta E_{el}$ on $E_{el}$ (Figs. A4.1, A4.2). For Fig. A4.1, all the descriptions (like the definition of dashed curve etc.) and parameter values are same as those of Fig. 32(a) except that $\Delta E_{el}$ has been assumed to have a linear variation with $E_{el}$ ($\Delta E_{el}$ =A$E_{el}$; $\Delta E_{el}(E_F)$ = 300 meV, $\Delta E_{el}(0)$ = 0). As can be seen, the results obtained in Figs. 32(a) and A4.1 are same; the $D_{fr}(E_{el})$ curves of both the cases give same pseudogap dip in the tunneling conductance curve. For Fig.

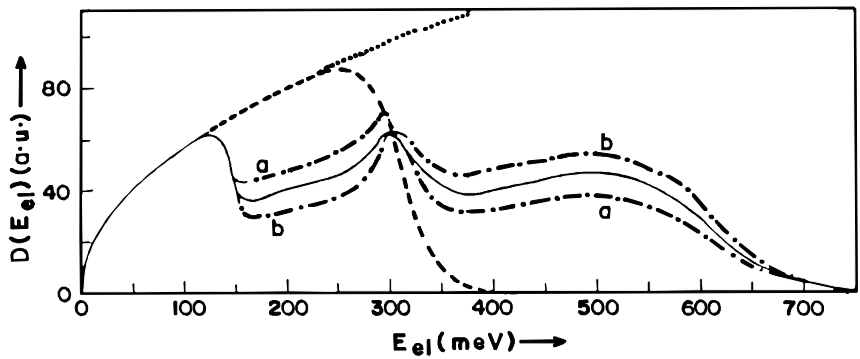

Fig.A4.1. Same as Fig. 32(a) except that $\Delta E_{el}$ vs. $E_{el}$ has been assumed to be linear here. Details are described in the text.

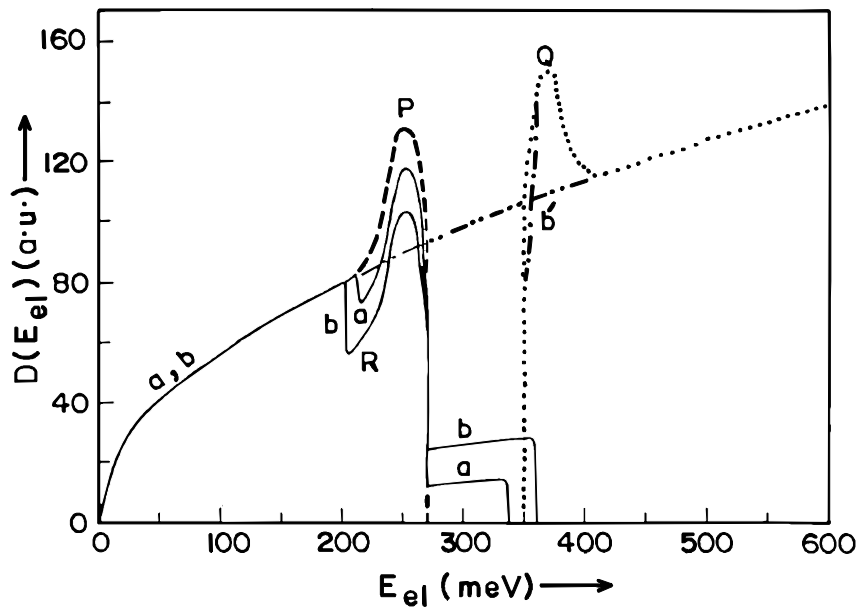

Fig.A4.2. Same as Fig. 33(b) except that $\Delta E_{el}$ vs. $E_{el}$ has been assumed to be linear here. Details are described in the text.

A4.2 (T < $T_c$ case) all the descriptions and parameter values are same as those of Fig. 33(b) except $\Delta E_{el}$ =A$E_{el}$ with $\Delta E_{el}(E_F)$ = 80 meV, $\Delta E_{el}(0)$ = 0 for curve a and $\Delta E_{el}(E_F)$ = 100 meV, $\Delta E_{el}(0)$ = 0 for curve b. Again the results obtained in Figs. A4.2 and 33(b) are same. In Fig. 33, and so in Fig. A4.2 here, thermal filling, which only reduces the gap amplitude (depth), has not been shown for clarity.

## APPENDIX 5

For T ~ $T_c$ case, where $\Delta E_{CP}$ scattering is also present (Part II) [194, 196], we have for the region below the gap ($E_{el} < (E_F - \Delta)$),

$D_{fr}(E_{el}) = D_f(E_{el}) - N_P D_f{'}(E_{el}) + N_P D_f{'}(E_{el} - \Delta E_{el}) - (N_P)_{CP} [D_f(E_{el}) - D_f{'}(E_{el})] + (N_P)_{CP} \times$
$[D_f(E_{el} - \Delta E_{el}) - D_f{'}(E_{el} - \Delta E_{el})].$ (A5.1)

The third term of the above equation will be nonzero only in very few cases when some empty states are present in $D_f{'}(E_{el})$ below the gap, like, for example, Fig. 33(a) case of Part II [194] where the temperature spread of $D_f(E_{el})$ curve's tail portion is slightly greater than the gap value. Similarly for the region inside the gap we have,

$D_{fr}(E_{el}) = N_P D_f{'}(E_{el} - \Delta E_{el}) + D_f(E_{el}).$ (A5.2)

The second term in the above equation contributes only if there are some filled states in the gap; generally this is zero or very small. Finally, for the region above the gap,

$D_{fr}(E_{el}) = N_P D_f{'}(E_{el} - \Delta E_{el}) + D_f(E_{el}) + (N_P)_{CP} [D_f(E_{el} - \Delta E_{el}) - D_f{'}(E_{el} - \Delta E_{el})].$ (A5.3)

As in Eq.(A5.2), the second term of Eq.(A5.3) is zero or very small. The third term of Eq.(A5.3) contributes only if some excited CPs are present above the gap.

---